\documentclass{aa}
\usepackage{txfonts}
\usepackage{graphicx}
\usepackage{color}

\usepackage{natbib,twoopt}
\usepackage[breaklinks=true]{hyperref} 

\hypersetup{
	pdftex=true, 
	colorlinks=true, 												
	breaklinks=true, 
	linkcolor=blue,          												
	citecolor=blue,        												
	filecolor=blue,     												
	pagecolor=blue, 
	urlcolor=blue,
	bookmarks=true,        					 						
	unicode=false,          												
	pdftoolbar=true,       												
	pdfmenubar=true,      											
	pdffitwindow=false,    											
	pdfstartview={Fit},   												
	pdftitle={Pulsating Stars in NGC 6231},    								
	pdfauthor={Stefan Meingast},    									
	pdfsubject={Frequency analysis and mode identification near the main sequence},   					
	pdfcreator={Stefan Meingast},   									
	pdfproducer={TeXLive 2012, TeXShop 3.11, OS X 10.8.2}, 					
	pdfkeywords={Stars: individual: NGC 6231, Stars: oscillations, Stars: variables: general, Methods: observational, Techniques: photometric}, 
	pdfnewwindow=true,      											
	pdfdisplaydoctitle=true,											
	citebordercolor=1 1 1,											
	linkbordercolor=1 1 1,
	filebordercolor=1 1 1,
	urlbordercolor=1 1 1
}

\bibpunct{(}{)}{;}{a}{}{,} 
\newcommandtwoopt{\citeads}[3][][]{\href{http://adsabs.harvard.edu/abs/#3}%
{\citealp[#1][#2]{#3}}} 
\newcommandtwoopt{\citepads}[3][][]{\href{http://adsabs.harvard.edu/abs/#3}%
{\citep[#1][#2]{#3}}} 
\newcommandtwoopt{\citetads}[3][][]{\href{http://adsabs.harvard.edu/abs/#3}%
{\citet[#1][#2]{#3}}} 
\newcommandtwoopt{\citeyearads}[3][][]%
{\href{http://adsabs.harvard.edu/abs/#3}{\citeyear[#1][#2]{#3}}}

\begin{document}

\title{Pulsating stars in NGC 6231}


\subtitle{Frequency analysis and photometric mode identification near the main sequence}
\author{Stefan Meingast\inst{1}
	\and Gerald Handler\inst{2}
	\and Robert R. Shobbrook\inst{3}}
\institute{Institut f\"ur Astrophysik, Universit\"at Wien, T\"urkenschanzstrasse 17, 1180 Wien, Austria
	\and Nicolaus Copernicus Astronomical Center, Bartycka 18, 00-716, Warsaw, Poland
  	\and Research School of Astronomy and Astrophysics, Australian National University, Canberra, ACT, Australia}
\date{}
\abstract
{}
{We used Johnson $UBV$ photometric CCD observations to identify pulsating and other variable stars in the young open cluster NGC 6231. The multi-color information was used to classify pulsating variables, perform frequency analysis, and - where possible - to compare observed to theoretical amplitude ratios for mode identification.}
{The CCD data were used to investigate a total of 473 stars in the field. The data reduction was performed with standard IRAF tools, where the extraction of light curves has been done with combined aperture and point-spread-function (PSF) photometry routines delivered with the DAOPHOT package. Differential light curves have been obtained by identifying a set of suitable comparison stars and the frequency analysis was then conducted on the basis of Fourier methods. Our classification of pulsating stars was based on the time scales and amplitudes of the variability with respect to the different filters and stellar parameters as calculated from published Str\"omgren and Geneva photometry. Attempts to set constraints on the pulsation mode were performed for stars with sufficiently high amplitude based on the significant dependence of amplitudes on wavelength.}
{We identified 32 variable stars in the field of the cluster out of which 21 are confirmed members of NGC 6231 and twelve are newly detected variable stars. Ten stars were classified as Slowly Pulsating B (SPB) stars in NGC 6231 out of which seven are new discoveries. We also analyzed six previously reported $\beta$ Cephei variables in more detail. One of them may be a hybrid $\beta$ Cephei/SPB pulsator. In addition, we investigated five more previously suspected pulsators of this group which we cannot convincingly confirm due to their small pulsation amplitudes. The remaining eleven variable stars are either not members of NGC 6231 or the membership status is questionable. Among them are three previously known $\delta$ Scuti stars, two newly detected pulsators of this class, one new and two already known eclipsing binaries, one new SPB variable, one possible Pre-Main-Sequence (PMS) pulsator and another new variable star for which we cannot present a classification. With more than 20 main sequence pulsators of spectral type B, NGC 6231 becomes the open cluster with the largest population of such pulsating stars known.}
{}
\maketitle
\section{Introduction}

NGC 6231 represents one of the youngest currently known open clusters in our galaxy with age estimates ranging from $3-5 \; \mathrm{Myr}$  \citepads{Baume99}, hereinafter referred to as BVF99, to $7 \; \mathrm{Myr}$ \citepads{1998AJ....115..734S}, hereinafter SBL98. It is located in the Southern Hemisphere, represents a part of the Scorpius OB I association and is characterized by an exceptionally rich population of very hot early type stars. Clusters themselves offer unique scientific possibilities in the study of asteroseismology and stellar evolution since the stars contained in them share very important physical properties. Stars in such aggregates are thought to originate from the very same interstellar cloud, thus having in common not only their distance to us, but also can be treated as equal in metallicity and in age with a certain intrinsic spread attributed to the star formation history in this region. 
One of the most remarkable characteristics of NGC 6231 is its richness in variable stars, harboring pulsators across the entire main sequence including $\beta$ Cephei, SPB, $\delta$ Scuti, and possibly even $\gamma$ Dor type variables. As a first step towards a successful asteroseismic model variable stars need to be identified and monitored over a large time period to ensure an unbiased and accurate understanding of their frequency spectra. Previous studies of NGC 6231 often concentrated on the bright $\beta$ Cephei population so that up to this point 6 stars are confirmed members of this class and 5 more are considered candidates \citepads{2005ApJS..158..193S}. These studies include \citetads{1979MNRAS.189..571S}, who was the first to classify a pulsating star in the cluster, \citetads{1983MNRAS.203.1041B}, hereinafter referred to as Ba83, who organized a multi-site campaign, \citetads{1983MNRAS.205..309B}, hereinafter referred to as BS83, and \citetads{BE85} (BE85). The most recent and also most extensive search for variable stars has been carried out by \citetads{Arentoft:2001gg}, hereinafter referred to as ASK01. Their discussion is based on large data sets featuring CCD observations covering bright variables in a wide field, as well as fainter pulsators in a smaller field. They reported 17 new variable stars in total, where for 11 of these they presented a classification. In addition they gave tentative frequency solutions for previously known pulsating stars in the field. ASK01 also were the first to confirm the presence of pulsators other than $\beta$ Cephei stars in NGC 6231. Many of these variables, however, lack confirmation of their variability and a verification of the involved time scales. Moreover, the analysis by ASK01 is focused on just one passband in the Str\"omgren photometric system and their classifications are mainly based on the time scales of the variability and broadband color information. Important indicators for pulsations, however, are amplitude and phase differences in distinct passbands, which furthermore allow a photometric determination of the spherical degree of the underlying pulsation modes.

Apart from the variability studies, thorough photometric observations were conducted by SBL98 who performed $UBVRI$ and H$\alpha$ photometry to study the PMS population, the initial-mass-function, and photometric properties of NGC 6231; BVF99 obtained $UBVI$ measurements with a lower detection limit compared to SBL98 to also determine parameters like reddening, distance, and age, and \citetads{1995MNRAS.276..627B}, hereinafter BL95, obtained intermediate and narrowband photometric data. A detailed compilation of older data often used in our work was presented by \citetads{1991A&AS...90..195P}, hereinafter PHC91, which is a part in a series of papers concerning large regions of the entire OB association including NGC 6231. In their extensive work they gathered information from various previous studies concerning broad-, intermediate-, and narrowband photometry and information on the spectral type of many stars. Spectroscopic studies include \citetads{LM83} (LM83), \citetads{Ra96}, hereinafter referred to as Ra96, and \citetads{GM01}, who investigated the binary nature of cluster members.

\begin{figure}
	\resizebox{\hsize}{!}{\includegraphics{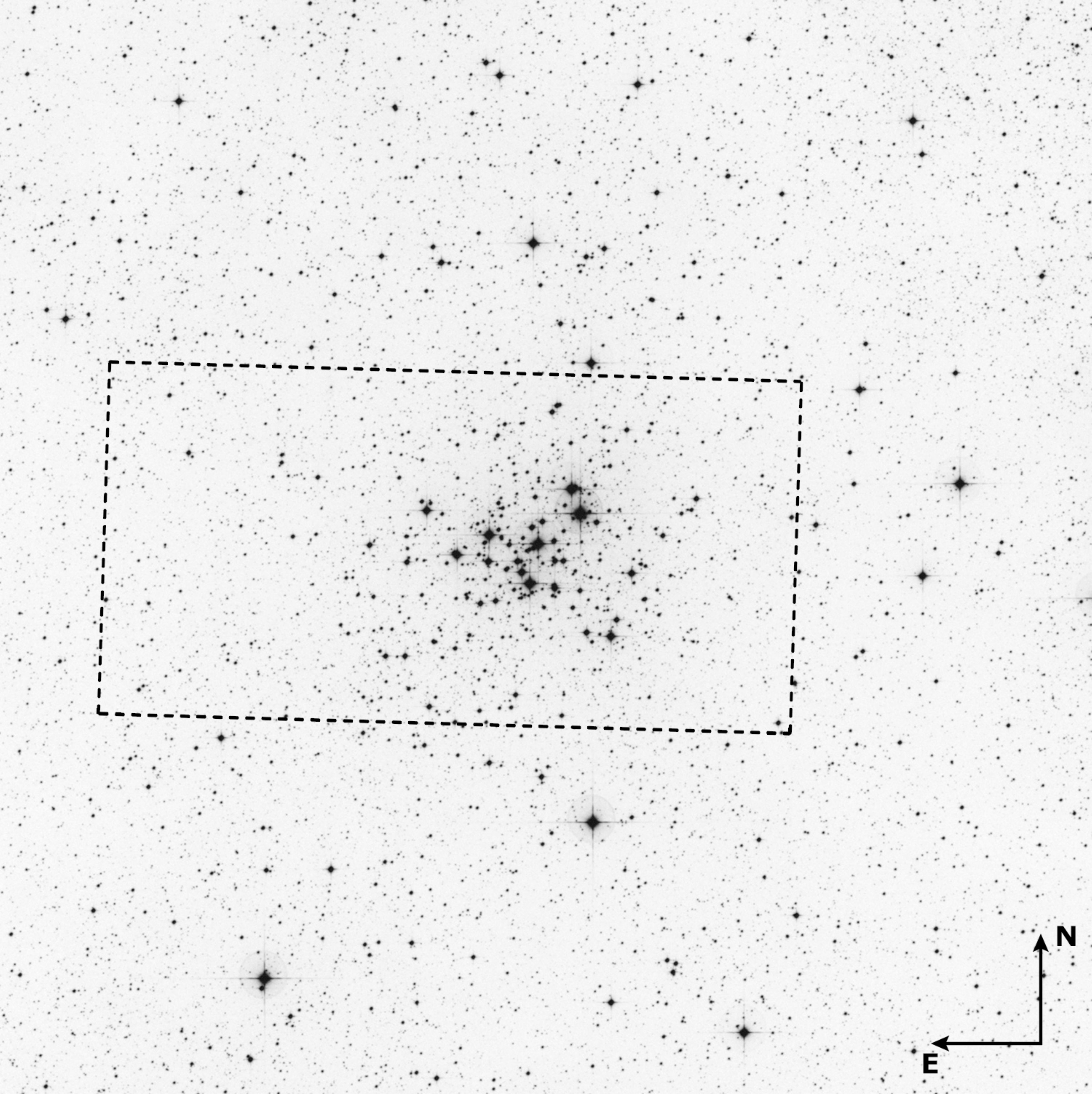}}
	\caption[Field of view of the observations]{The dashed frame shows the FOV of the observations on top of a \textit{Digitized Sky Survey} image of NGC 6231. Our reference field has a size of about $25.5 \times 12.7\; \mathrm{arcmin}^{2}$, whereas the background image is centered on $\alpha = 16\; \mathrm{h} \; 54 \; \mathrm{m} \;  8.6 \; \mathrm{s}$, $\delta=-41^{\circ} \; 49' \; 36.4''$ and has a size of $40 \times 40 \; \mathrm{arcmin}^{2}$.}  
	\label{img:FOV}
\end{figure}

The photometric results agree reasonably well with each other where important results refer to the cluster's distance and age. SBL98 find a color excess of $E_{\left( B-V \right)} = 0.466\, \pm \,0.054 \; \mathrm{mag}$ with $R \equiv A_{V} / E_{\left( B-V \right)} = 3.3\, \pm \,0.1$ and finally the distance modulus $V_{0} - M_{V} = 11.0\, \pm \,0.07 \; \mathrm{mag}$, where $V_{0}$ is the dereddened $V$ magnitude, i.e. $V_{0} = V - A_{V} = V - R \cdot E_{\left( B-V \right)}$. The distance to NGC 6231 then is $1590\, \pm \,50 \; \mathrm{pc}$. As an indicator of the systematic uncertainty in these values, other results can be considered: BVF99 find a distance modulus of $V_{0} - M_{V} = 11.5\, \pm \,0.25 \; \mathrm{mag}$ or $1990\, \pm \,200 \; \mathrm{pc}$. In light of these results it is also very important to mention here, that all photometric studies found large differential reddening in the field of the cluster ranging from values slightly below the adopted one and up to $E_{\left( B-V \right)} = 0.7 \; \mathrm{mag}$ for individual stars with a trend towards a larger excess in the southern parts. To estimate the reddening for stars in NGC 6231 we adopted the results of SBL98. With $E_{(U-B)} / E_{(B-V)} = 0.72 + 0.05 \cdot E_{(B-V)}$ we find $E_{(U-B)} \approx 0.346$ on average but also here one has to keep the differential reddening in mind.


Concerning the age determination by SBL98 and BFV99 it is important to note that their estimates point towards the conclusion that low mass stars still need to reach the stage of hydrogen core burning, giving rise to a potential rich PMS population. The cluster membership information for individual stars is available from BVF99 and \citetads{RCB97}, hereinafter referred to as RCB97, both of which base their results on photometric data.

In this paper we present new time-series CCD photometric observations of NGC 6231 covering a large field-of-view (FOV). All stars in the field up to a magnitude limit of $m_V = 15 \; \mathrm{mag}$ (SBL98 photometry) together with some individually chosen targets were investigated in detail allowing us to give an overview of the parameters of all bright pulsators in this cluster. The data are then used to determine accurate frequency solutions and corresponding amplitude values in three distinct passbands (Johnson $UBV$). For the identification of the pulsation modes it is imperative to set constraints on the stellar physical parameters including mass, surface gravity, and effective temperature. These values have been derived by means of Str\"omgren photometry and a comparison with stellar evolutionary tracks.

The outline of the paper is as follows: Section \ref{sec:DR} describes the data characteristics and the reduction procedure to obtain differential light curves. Section \ref{sec:DA} outlines the methods used for data analysis, Sect. \ref{sec:Results} gives a very thorough overview of the results for each individual star. Since this discussion is lengthy, this section is only available online, where we also present detailed light curves, frequency spectra and attempt mode identifications. A summary of our results is given in Sect. \ref{sec:Summary} where we also present an outlook and suggestions for further steps towards a complete picture of pulsators in NGC 6231.

\section{Observations and data reduction}
\label{sec:DR}

The data we present in this study were obtained at the Siding Spring Observatory using the (now decommissioned) $40 \; \mathrm{inch}$ telescope and one $4\mathrm{k} \times 2\mathrm{k}$ CCD of the Wide Field Imager with a pixel scale varying from $0.377 \; \mathrm{arcsec/pix}$ near the optical axis to $0.345 \; \mathrm{arcsec/pix}$ at the edge of the frame translating to a FOV of about $25.65 \times 12.81 \; \mathrm{arcmin}^2$. The FOV of the observations which varied from night to night is approximately sketched in Fig. \ref{img:FOV}. The entire observing campaign covered about two months and was split up into three distinct shorter runs carried out in May, June, and July 2007. The observing sequence was carried out by cycling through the filters with exposure times around $15 \; $s in $B$ and $V$, and either $60 \;$s or $80 \;$s in $U$ to compensate for the lower atmospheric transmission and quantum efficiency of the detector in this passband.

After cleaning the set from bad data (blurred images, cloudy nights, large offsets) a total of about $90 \; \mathrm{h}$ of time series photometry in the Johnson $UBV$ filters remained. A short overview of these observing runs is given in Table \ref{tab:runs} which lists all frames that passed through this cleaning step. The quality of the data is characterized by different aspects and varies from star to star. Especially the data in $U$ showed several spontaneous jumps in the light curves during some nights which might either be attributed to the detector itself, imperfect flat fielding in the $U$ band, or changing atmospheric conditions. In order not to include such corrupted parts in the determination of frequencies and amplitudes each light curve of a candidate variable star was checked in all three bands for similar features. If the variation in $U$ was not consistent with the results in $B$ and $V$ the concerning parts of the $U$ light curve were excluded from our further analysis. Since this was quite often the case only for some of the variable stars a reliable set in $U$ was produced. In cases with low $U$ data quality it was not possible to constrain the spherical degree of the pulsation. The numbers in Table \ref{tab:runs}, however, represent all data and do not include the second cleaning iteration in the $U$ band as described above since the final numbers vary from star to star depending on the $U$ data quality. The total number of usable measurements in all bands then varied from a lower limit of about 1800 (with only 400 in U) to about 2150 (with more than 600 in U).

\begin{table}
	\centering
		\begin{tabular}[t]{l c c c c }
			\hline\hline
			Run  & Nights & Period &images & total length \\
			& (\#) & JD + 2454200& (\#) & (h) \\
			\hline
			May 2007 & 4 & 28.0 - 34.3 & 830 & 32.0 \\
			June 2007 & 5 &60.9 - 65.9 & 825 & 34.5 \\
			July 2007 & 3 & 90.9 - 92.8 & 627 & 23.4 \\
			\hline
			Total & 12 & 28.0 - 92.8 & 2282 &89.9h	\\
			\hline
		\end{tabular}
	\caption[Summary of the observed data collected during the entire run]{Summary of the data collected during the entire run. Only the images used in the data analysis are listed where the numbers represent observations in all three filters.}
	\label{tab:runs}
\end{table}

From the included calibration frames we estimated the readout noise to be $5.5 \; \mathrm{e}^{-}$, the gain to be $1.5 \; \mathrm{e}^{-}/\mathrm{ADU}$, and the dark current to be $0.015 \; \mathrm{e}^{-}/\mathrm{s}$. The data reduction steps included bias and dark current subtraction and a correction of the gain variations using flat fields. Since only few calibration frames were present in the data set due to the rather long CCD readout time of about $75 \; \mathrm{s}$, we created master frames for each month separately. Only the master flat fields showed a significant difference for each month attributed to different ambient conditions, dust on the filters and possibly telescope maintenance where the setup was altered. All data reduction tasks have been performed with standard IRAF tools.

The data themselves not only suffered from large pointing offsets (up to several arcmin), but also from changing positions of the camera relative to the telescope focal plane. This was necessary to adjust for unstable atmospheric transmission in order not to saturate some of the potential bright variable stars. Here we preferred to keep the longer exposure times and a defocused camera to have the largest possible duty cycle considering the long readout time of the detector. This led to shape and size variations of the PSF (and therefore also changed the curve of growth) which made it difficult to choose an ideal aperture for the photometry. In addition, the increasing stellar density towards the center of the cluster required a crowded field photometry approach. To avoid photometric contamination and to achieve the desired millimag precision the light curve extraction was done in two steps. At first PSF photometry has been performed using the DAOPHOT package in IRAF. Since in most images the shape of the PSF significantly differed from a gaussian or lorentzian profile, non-negligible residuals were visible in the PSF-subtracted images. In order to deal with these residuals we used subsequent aperture photometry on these PSF-subtracted images to apply a suitable correction to the PSF photometry. 


To compute differential magnitudes a selected subset of stars was combined into an artificial comparison star. To this end we employed the algorithm developed by \citetads{2005AN....326..134B}, which calculates a weighted mean of all selected comparison stars on the basis of the standard deviation in their own differential light curves. To get an estimate on the single data point error we examined the light curves of the comparison stars which typically showed a standard deviation of about $4 \; \mathrm{mmag}$ for the brightest objects.

We then searched all frames for variable stars by visual inspection of the light curves and by computing frequency spectra. We chose the upper limit in frequency to be half of the average sampling rate of the observations while omitting larger gaps. For the $U$ measurements this upper limit corresponds to about $300 \; \mathrm{d}^{-1}$. Since the data were optimized (in terms of integration time) for the bright $\beta$ Cephei variables in NGC 6231 many stars which would potentially fall into the $\delta$ Scuti instability region could not be studied in detail due to their low signal-to-noise ratio (SNR) for these much fainter pulsators. 

As a consequence of the optimization for the bright variables and the defocused state of the camera our data set included a total number of 473 stars in spite of the large field. For 32 stars, independent of their cluster membership, we were able to confirm intrinsic variability. These objects were subject to a more thorough analysis. To this end we adopted the numbering scheme of SBL98 who analyzed all but one variable star in our field for which we then adopted the UCAC3 identification number \citepads{UCAC3}. The cross identification with BVF99 and UCAC3 is given in Table \ref{tab:par} among other results obtained with the methods described in the following section.

\section{Methods}
\label{sec:DA}

\begin{table*}
	\caption{Basic data, including positions and photometry, cross identification of SBL98, BVF99, and UCAC3 and spectral type information for all detected variable stars in the FOV of our observations, as well as the results of our calculation of stellar parameters for all stars in the FOV for which the required measurements were available. Stars which supposedly are a cluster member of NGC 6231 have their SBL98 IDs printed in boldface. SBL464 has not been identified as a member so far, but our analysis supports this idea.} 
	\label{tab:par}
	\centering
	\begin{tabular}{cccccccccccc}
		\hline\hline
SBL98	&	BVF99	&	UCAC3	&	RA\tablefootmark{a}	&	DEC\tablefootmark{a}	&	$m_{V}$\tablefootmark{b}	&	$M_{V}$\tablefootmark{c}	&	SpT\tablefootmark{d}	&	$E_{(b-y)}$\tablefootmark{c}	&	$T_{eff}$\tablefootmark{c}	&	log $g$\tablefootmark{c}	&	log $(L/L_{\odot})$\tablefootmark{e}	\\
ID			&	ID	&	ID			&	(hh:mm:ss)	&	(dd:mm:ss)	&	(mag)	&(mag)	&			&	(mag)	&	$(kK)$	&	(dex)	&	(dex)				\\
\hline
\textbf{113}	&		&	097-238816	&	16:53:39.1	&	-41:47:48.1	&	9.729	&		&	B1.5Ve	&			&	{\it 26.0}	& {\it 3.3}	&			\\
\textbf{164}	&		&	097-238918	&	16:53:47.0	&	-41:48:55.1	&	12.361	&	0.18	&	B6-8V	&	0.304	&	13.3		&	4.6	&	2.2	\\
\textbf{210}	&		&	097-239017	&	16:53:52.1	&	-41:48:54.0	&	12.011	&	-0.76	&	Ap		&	0.272	&	16.7		&	4.5	&	2.8	\\
\textbf{226}	&	20	&	097-239064	&	16:53:54.6	&	-41:52:15.2	&	9.58		&	-3.85	&	B1V		&	0.319	&	26.8		&	3.5	&	4.5	\\
\textbf{268}	&	26	&	097-239145	&	16:53:58.6	&	-41:48:41.8	&	9.752	&	-2.94	&	B2V		&	0.306	&	22.9		&	3.8	&	4.0	\\
\textbf{275}	&		&	097-239158	&	16:53:59.3	&	-41:53:04.2	&	11.888	&	-0.97	&	B4-5V	&	0.303	&	16.3		&	4.3	&	2.9	\\
\textbf{283}	&	106	&	097-239173	&	16:53:60.0	&	-41:46:26.4	&	12.452	&	-0.74	&	B		&	0.309	&	14.5		&	4.1	&	2.7	\\
\textbf{303}	&	37	&	097-239216	&	16:54:01.8 	&	-41:51:11.9	&	10.281	&	-3.31	&	B2IV-V	&	0.336	&	21.8		&	3.4	&	4.1	\\
306				&		&	097-239222	&	16:54:01.9 	&	-41:53:24.0	&	13.987	&		&			&			&			&		&		\\
\textbf{317}	&	21	&	097-239323	&	16:54:06.9 	&	-41:49:23.5	&	10.914	&	-1.92	&	B2-3V	&	0.31		&	20.5		&	4.3	&	3.5	\\
\textbf{353}	&	41	&	097-239258	&	16:54:03.2 	&	-41:51:49.0	&	9.51		&	-3.38	&	B0.5V	&	0.286	&	28.6		&	4.1	&	4.4	\\
\textbf{394}	&	39	&	097-239403	&	16:54:10.7	&	-41:47:47.4	&	10.599	&	-1.92	&	B2IV-V	&	0.294	&	20.8		&	4.4	&	3.5	\\
417				&	279	&	097-239450	&	16:54:12.9	&	-41:52:28.9	&	14.618	&		&			&			&			&		&		\\
\textbf{437}	&	38	&	097-239486	&	16:54:14.3	&	-41:55:01.2	&	10.299	&	-3.16	&	B0Vn	&	0.415	&	25.3		&	3.9	&	4.1	\\
444				&	113	&	097-239499	&	16:54:14.9	&	-41:46:58.4	&	12.602	&		&			&			&			&		&		\\
455				&	178	&	097-239520	&	16:54:15.7	&	-41:51:40.0	&	13.455	&		&	A5III		&			&			&		&		\\
\textbf{456}	&	36	&	097-239519	&	16:54:15.7	&	-41:49:32.2	&	10.174	&	-2.79	&	B1Vn	&	0.332	&	22.3		&	3.8	&	3.9	\\
\textbf{461}	&	40	&	097-239532	&	16:54:16.3	&	-41:50:26.5	&	10.745	&	-1.79	&	B1.5V	&	0.338	&	20		&	4.4	&	3.4	\\
464				&	56	&	097-239539	&	16:54:16.8	&	-41:45:08.3	&	11.554	&		&	B8-9V	&			&{\it 13.3}		&{\it 4.5}	&		\\
\textbf{480}	&	44	&	097-239568	&	16:54:17.9	&	-41:48:01.4	&	11.059	&		&	B2IVn	&			&	{\it 19.2}	& {\it 4.7}	&		\\
\textbf{482}	&	148	&	097-239573	&	16:54:18.1	&	-41:50:16.4	&	13.062	&	-0.15	&	B9V		&	0.373	&	13.8		&	4.4	&	2.4	\\
\textbf{486}	&	22	&	097-239578	&	16:54:18.3	&	-41:51:35.6	&	9.603	&	-3.47	&	B0.5V	&	0.311	&	28.9		&	4.0	&	4.4	\\
\textbf{515}	&	49	&	097-239633	&	16:54:20.6	&	-41:49:28.6	&	11.198	&	-2.83	&			&	0.335	&	18.6		&	3.3	&	3.7	\\
521				&	25	&	097-239643	&	16:54:21.3	&	-41:51:41.8	&	9.715	&		&	B0-2V	&			&			&		&		\\
\textbf{528}	&	66	&	097-239654	&	16:54:21.7	&	-41:49:54.5	&	11.719	&	-0.99	&	B6Vn	&	0.343	&	15.8		&	4.2	&	2.8	\\
574				&	101	&	097-239760	&	16:54:26.1	&	-41:51:31.3	&	12.462	&		&			&			&			&		&		\\
612				&	30	&	097-239858	&	16:54:31.2	&	-41:55:28.6	&	9.945	&		&	F0III		&			&			&		&		\\
\textbf{653}	&	28	&	097-239952	&	16:54:36.0	&	-41:53:38.0	&	9.803	&	-3.61	&	B1V		&	0.352	&	30		&	4.0	&	4.5	\\
705				&		&	097-240058	&	16:54:42.4	&	-41:47:16.4	&	14.426	&		&			&			&			&		&		\\
\textbf{712}	&	23	&	097-240079	&	16:54:43.2	&	-41:49:35.4	&	9.601	&	-3.61	&	B1V		&	0.328	&	27.8		&	3.8	&	4.4	\\
752				&		&	097-240182	&	16:54:48.7	&	-41:52:30.7	&	12.239	&		&			&			& {\it 8.2}		& {\it 3.7}	&		\\
				&		&	097-240594	&	16:55:18.1	&	-41:46:24.2	&			&		&			&			&			&		&		\\
	\hline
	\end{tabular}
\tablefoot{
   \tablefoottext{a}{Coordinates from the USNO CCD Astrograph Catalog 3 (UCAC3).}
   \tablefoottext{b}{Photometry from SBL98.}
   \tablefoottext{c}{Calculated using published Str\"omgren photometry, supplemented by Geneva photometry (values in italics).}
   \tablefoottext{d}{Spectral type information from RCB97, PHC91, or the catalogue of stellar spectral classifications \citepads{2009yCat....1.2023S}.}
   \tablefoottext{e}{Calculated from the $M_{V}$ value and interpolated bolometric corrections.}  
}
\end{table*}

\subsection{Data analysis}

To study the selected variable stars in more detail we carried out frequency analysis of the differential light curves with Period04 \citepads{2005CoAst.146...53L}, a program which allows frequency search on the basis of the Discrete Fourier Transform and subsequent least-squares fitting of the determined multi-periodic function to obtain accurate values for the amplitudes and phases of potential pulsation modes and their individual contributing components. For stars where the light curves showed zero-point offsets on a daily (or monthly) basis an adjustment was performed in such a way that the mean differential magnitude for each day (or month) amounted 0. For stars which showed variability on time-scales of several days this adjustment was performed for each month in order not to corrupt the amplitude information. If the periods of the variability were much shorter than one day (like for the $\beta$ Cephei stars) the adjustment was carried out for each observed night. 

As a criterion for significance of a frequency we calculated the signal-to-noise ratio (SNR) by a relating the signal strength to the average residual signal (i.e. after prewhitening) calculated in a box centred on the frequency of interest with a width of $2 \; \mathrm{d}^{-1}$. For values below $1 \; \mathrm{d}^{-1}$ the software accounts for the reflection at frequency 0. The errors in the frequencies and amplitudes are determined by Period04 via the least-squares fit and therefore represent lower limits (see e.g. \citeads{1999DSSN...13...28M}.)

Following \citetads{1993A&A...271..482B}, if the SNR was greater or equal 4, a frequency was classified as significant and was attributed to an intrinsic change of the stellar properties. Concerning this criterion, however, it is important to note that the actual significance levels may vary from object to object or even within the same light curve (see e.g. \citeads{2010Ap&SS.329..267K}).

Signals deemed to be significant were then subtracted from the light curves, a process called prewhitening, whereupon the residuals then again were searched for significant frequencies. We repeated these steps until no further signal could be found with a SNR above the adopted criterion. In cases where light curves were dominated by noise and a frequency did not exceed the SNR threshold but came reasonably close to it and the visual inspection clearly revealed variability, a signal could also be included if its SNR was lower than 4. This, if applicable, will be explicitly mentioned for each star in Sect. \ref{sec:Results}.

\begin{figure*}
	\resizebox{\hsize}{!}{\includegraphics{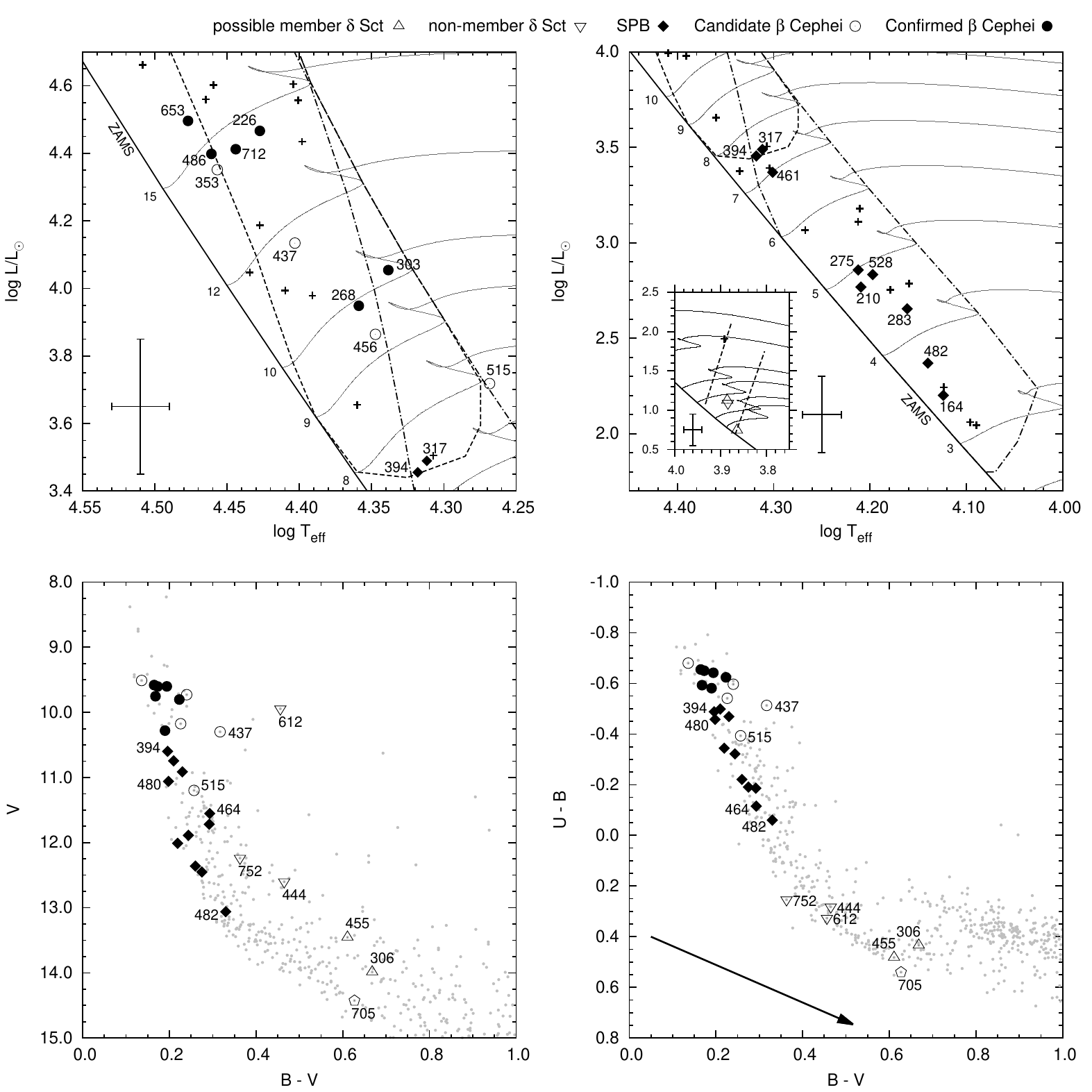}}
	\caption[Photometric data]{The four panels show the positions of variable stars in our FOV in theoretical HR diagrams (top row), a color-magnitude (bottom left), and a color-color diagram (bottom right). The number labels next to the data points represent the IDs of SBL98 such as we use them throughout our analysis. The HR diagrams also show the ZAMS (slanted full line), evolutionary tracks for different stellar masses where the mass is given in solar masses at their origin at the ZAMS, the $\beta$ Cephei instability strip (dashed line), the SPB instability strip (dash-dotted line), the estimated error in $\mathrm{T}_{eff}$ and $\mathrm{log} \; (L/L_{\odot})$ indicated in the bottom left corner of the individual plots and further cluster members which were not determined to be variable as crosses. Both plots in the top row are theoretical HR diagrams but show different sections in the parameters space for a better visualization of the results. The top right plot also contains an inlay where a third region in the HR diagram, covering the $\delta$ Sculti instability domain (marked with dashed lines), is displayed. Here the stars SBL306, SBL455, and SBL612 are plotted, where the first two are possible members of NGC 6231 and SBL612 is a foreground star. The bottom row shows the results of SBL98's photometry. Only stars for which the position is important in the individual discussions are labeled here. In addition the reddening vector is indicated in the color-color diagram where the different groups of pulsators are very well separated. Only for SBL705 pulsations may not be the origin of its variability.}
	\label{img:HRD}
\end{figure*}

\subsection{Identification of pulsations}
\label{sec:IDpuls}
It is important to not only rely on the frequency information of time-series data for the identification pulsating stars. Other effects such as surface inhomogeneities (starspots) or binarity can cause sinusoidal-shaped light curve variations which may misleadingly be interpreted as pulsations. Clear indications, however, for a pulsational origin of variability are multi-periodic light curves and amplitude differences depending on wavelength. In order to assign a star to a distinct class of pulsator we also performed a calculation of the effective temperature and its luminosity using Str\"omgren photometry from either the compilation of PHC91, or - if not available in the former study - measurements of BL95 or \citetads{Sh83}. To this end we used the algorithm developed by \citetads{1993A&A...268..653N} in combination with the bolometric corrections of \citetads{1996ApJ...469..355F} and an absolute magnitude of the Sun of $M_{V} = 4.74 \; \mathrm{mag}$ \citepads{2000asqu.book.....C} to calculate stellar luminosities. With this procedure we also obtained a star's surface gravity which was later used for the attempt to identify pulsation modes (see Sect.\ \ref{sec:MI}). The results of this step are listed in Table \ref{tab:par} for all (non-peculiar) variable stars for which the required photometric measurements were available. Moreover, we calculated these stellar parameters for all remaining bright stars where photometric and spectral type information was available to check for non-pulsating stars in the instability strips within our detection limit. For four stars with no available Str\"omgren photometry, we retrieved their Geneva colors from The General Catalogue of Photometric Data (GCPD; \citeads{1997A&AS..124..349M}) and determined their effective temperatures and surface gravities using the calibrations by \citetads{1997A&AS..122...51K}. In the top row of Fig. \ref{img:HRD} these stars are displayed in a theoretical HR diagram also showing instability strips and evolutionary tracks.

These were computed with the Warsaw-New Jersey stellar evolution code described in \citetads{1998A&A...333..141P} which uses OP opacities \citepads{2005MNRAS.362L...1S} and the \citetads{2004A&A...417..751A} chemical element mixture. The parameters of the tracks included an overall metal abundance of $\mathrm{Z} = 0.025$ and a hydrogen abundance of $\mathrm{X} = 0.7$. The high metallicity was chosen because NGC 6231 is a young cluster and was found to be metal-rich by \citetads{2010A&A...517A..32P}. Further constraints feature a rotational velocity of 100 km/s at the ZAMS whereas convective core overshooting was not included. In all cases Kurucz atmospheres  were used in combination with pulsation models based on OP opacities.

Since no single measurement errors for the photometry were available and also since it was not possible to include the systematic errors of the calibrations, the errors were estimated by calculating the stellar parameters with different sets of photometric indices. Hence, for all stars, for which narrow- and intermediate band photometry from at least three different sources was available, the calibration was performed and the standard deviation of the results was taken as a first estimation of the errors. They amount $\sigma_{\mathrm{log} \, T_{\mathrm{eff}}} = 0.019 \; \mathrm{dex}$, $\sigma_{\mathrm{log} (L/L_{\odot})} = 0.17 \; \mathrm{dex}$, and $\sigma_{\mathrm{log} \, g} = 0.32 \; \mathrm{dex}$. Hoping to include also systematic errors, the values were rounded up and finally were set to $\sigma_{\mathrm{log} \, T_{\mathrm{eff}}} = 0.02 \; \mathrm{dex}$, $\sigma_{\mathrm{log} (L/L_{\odot})} = 0.2 \; \mathrm{dex}$, and $\sigma_{\mathrm{log} \, g} = 0.35 \; \mathrm{dex}$. For the Geneva photometry results, $\sigma_{\mathrm{log} \, g}$ may be as large as 0.5 dex. To support our classifications, we also consider previous studies of pulsating stars in NGC 6231, in addition to the time scales of the variability and the stellar parameters derived as just described.

\subsection{Mode identification}
\label{sec:MI}

In order to attempt a comparison between observed and theoretical amplitude ratios mode identification was performed with the software FAMIAS (Frequency Analysis and Mode Identification for ASteroseismology) developed by \citetads{FAMIAS}. FAMIAS allows to extract theoretical amplitude ratios for various pre-defined stellar models which can be selected by choosing appropriate parameters including $T_{eff}$, log $g$, mass, metallicity and further atmospheric parameters. Effective temperature and surface gravity were derived via the photometric calibrations, the mass was estimated by plotting each star in a theoretical HR diagram with overlaid evolutionary tracks for different stellar masses (Fig. \ref{img:HRD}). We included all models which passed through the observational error box of $\mathrm{log} \, g$ and $T_{\mathrm{eff}}$ according to the errors as given in Sect. \ref{sec:IDpuls}. Moreover, we also allowed for a variation in mass estimated from the error box in the HR diagram. To this end we searched the grids in the ($T_{eff}$, log $g$) error box corresponding to incremental mass steps. In general and consistent with our obtained errors in the HR diagram for the $\beta$ Cephei stars we allowed for an uncertainty of $\pm 1 \; M_{\odot}$ and for the SPB stars $\pm 0.5 \; M_{\odot}$. The errors of the observed amplitude ratios were calculated using standard error propagation. All metallicities were set to $[\mathrm{m/h}] = 0.3 \; (\approx Z = 0.025)$, consistent with the evolutionary tracks used. It is important to note here, that many of our variable stars may indeed be binaries (as suggested by previous studies) and therefore their actual pulsation amplitudes would also be different from the observed ones. However, amplitude {\it ratios} are generally hardly affected because this would require a companion of substantial luminosity and marked different color.

Photometric mode identification only allows a determination of the spherical degree $l$ of a pulsation mode \citepads{1988Ap&SS.140..255W}. Only amplitude ratios for $l \le 4$ have been calculated since higher degrees are not likely observed due to cancellation effects. In most cases, however, the amplitudes of the oscillations are only of the order a few mmag, giving rise to large relative errors in the amplitude ratios so that only constraints could be set and no definite classification can be given. The large number of models which passed through our error box was another contributor to ambiguities in the mode identification process. Whenever it was possible to obtain an identification, the classification of an oscillation was based on a visual comparison of observed amplitude ratios and theoretical predictions.

\begin{table*}[t]
	\caption{Determined frequencies and suggested spherical degree $l$ of the corresponding pulsation mode. Only stars where frequency analysis resulted in at least one significant signal are listed. The frequencies are sorted by decreasing SNR and listed are all possible pulsation modes according to our analysis. The $A_{V, f_{1}}$ column lists the amplitude of the frequency in the $V$ band with largest SNR, i.e. $f_{1}$. The first  horizontal line in the list separates the confirmed members of NGC 6231 (above) from possible member stars and targets below the second horizontal line represent non-members. Typical errors are of the order of a few $10^{-4} \; \mathrm{d}^{-1}$ for the frequencies and about $0.2 \; \mathrm{mmag}$ for the $V$ band amplitude. More detailed error estimates will be given in Sect. \ref{sec:Results} where each star is discussed individually.}
\label{tab:results}
	\centering
	\begin{tabular}{c c c c c c c c c c c c c}
	\hline\hline
		SBL98& Type & $f_{1}$ & $l_{1}$ & $f_{2}$ & $l_{2}$ & $f_{3}$ & $l_{3}$ & $f_{4}$ & $f_{5}$ & $f_{6}$ & $A_{V, f_{1}}$ & Comment \\
		ID &&$(\mathrm{d}^{-1})$&&$(\mathrm{d}^{-1})$&&$(\mathrm{d}^{-1})$&&$(\mathrm{d}^{-1})$&$(\mathrm{d}^{-1})$&$(\mathrm{d}^{-1})$ & (mmag) & \\
	\hline
		164 & SPB 			& 1.4183 	&  	3		&&&&&&&& 19.5 & new \\
		210 & SPB 			& 0.9615 	& 2	& 0.8332 &  & 0.5004 &  & 1.9223 &&& 28.4 & $f_{4}$ = $2f_{1}$, new\\
		226 & $\beta$ Cephei 	& 0.7017 	&  			& 14.9098 & & 2.4242 & & 2.6579 & 17.2502 &  & 7.5 & hybrid? \\
		268 & $\beta$ Cephei 	& 8.3847 	&  			& 12.9891 &  & 7.2748 &&&&& 4.7 & \\
		275 & SPB 			& 1.6686 	&			& 1.7662 & & 0.9093 &  & 2.0437 &  && 13.0 & \\
		283 & SPB 			& 1.6694 	& 1 	& 0.249 &  & 1.4688 &  & 2.9952 &&& 11.3 & new \\
		303 & $\beta$ Cephei 	& 10.1225& & 10.991 & & 10.5106 & 4 & 11.6904 & 12.045  & 11.902  & 11.1 & \\
		317 & SPB 			& 0.6133 	&  			& 1.3381 &  & 3.4903 &  & 4.9252 &&& 7.9 & new \\
		437 & $\beta$ Cephei	& 7.461 	&			&&&&&&&& 2.0 & candidate \\
		456 & $\beta$ Cephei 	& 12.6701&			&&&&&&&& 2.0 & candidate\\
		461 & SPB 			& 0.3597 	& 1 	&&&&&&&& 4.4 & \\
		480 & SPB 			& 2.2981 	&  			& 4.5224 &&&&&&& 4.2 & new, candidate \\
		482 & SPB 			& 1.9461 	& 		& 1.3466 & 1,3 &&&&&& 7.4 & new \\
		486 & $\beta$ Cephei 	& 11.3791&  			& 13.8258 &  & 18.058 &  & (0.38) &&& 2.1 & \\
		515 & $\beta$ Cephei 	& 3.5221	&			& 5.162	&&&&&&& 4.0 & candidate\\
		528 & SPB 			& 1.1959 	& 		& 0.8962 &  & 1.3335 &  &&&& 12.7 & new \\
		653 & $\beta$ Cephei 	& 9.2647 	& 0 			& 13.334 &&&&&&& 4.5 & \\
		712 & $\beta$ Cephei 	& 9.8481 	&  			& 9.2947 &  &&&&&& 5.7 & \\
		\hline
		306 & $\delta$ Scuti 		& 23.7713	&  			& 9.4853 &  & 20.3195 &  & 9.7797 & 10.5421 && 8.9 &   \\
		455 & $\delta$ Scuti 		& 23.7651&  			& 19.8906 &  & 20.869 &  & 18.4495 &  && 10.0 & \\
		464 & SPB 			& 1.1001	&  			& 1.3389 &&&&&&& 4.7 & new  \\
		\hline
		444 & $\delta$ Scuti 		& 15.0539 &  			& 20.8483 &  & 17.9737 &  & 23.0557 &&& 2.3 & new   \\
		612 & $\delta$ Scuti 		& 11.1654 &  			& 10.7018 &  & 10.4427 &  & 9.2525 &&& 11.5 &  \\
		752 & $\delta$ Scuti 		& 15.2413 &  			& 15.3989 &  & 17.1006 &  & 15.0161  & 18.7892 && 2.7 & new \\
	\hline
	\end{tabular}
\end{table*}
 
\section{Results for individual stars}
\label{sec:Results}
This section is available only in electronic form via http://www.aanda.org and includes a detailed analysis of each individual variable star.

\section{Summary and conclusions}
\label{sec:Summary}

Our analysis of the data set revealed variability for 32 out of 473 stars of which 21 are - according to previous photometric studies - confirmed members of the cluster NGC 6231. Those 21 members include six confirmed $\beta$ Cephei stars and five more candidates of this class and ten Slowly Pulsating B stars, out of which only three were candidates from the literature; we could confirm them all. Therefore we report the discovery of seven new hot pulsating stars in NGC 6231. Our results generally comply very well with earlier studies, but especially the good agreement with the determined frequencies of BS83 who organized a multi-site campaign (and therefore suffered much less from aliasing), demonstrates that the quality of our dataset is very high.

Of the remaining eleven variables only one star is definitely classified as a nonmember of NGC 6231 (SBL612) where for the other stars a possible membership is discussed on the basis of color-excess and broadband photometry if available.

In the entire set we were able to identify 74 independent oscillations modes and one combination frequency. 50 of these were found in confirmed cluster member stars. Attempts to identify pulsation modes were only performed for those confirmed members where reliable amplitudes in $U$ could be determined and where the comparison to models produced reasonable results. However, as a consequence of our large observational error box we were only able to interpret very few oscillations. The results of the frequency analysis and the mode identification are listed in Table \ref{tab:results} for all stars for which at least one significant frequency could be determined. To support our classifications we also calculated stellar parameters on the basis of Str\"omgren photometry and checked whether the resulting position would put a target into a theoretically determined instability strip. 

All six known $\beta$ Cephei stars (SBL226, SBL268, SBL303, SBL486, SBL653, SBL712) could be confirmed and we find excellent agreement with previous results concerning the frequencies. For SBL226 we even find evidence for hybrid $\beta$ Cephei/SPB pulsation. The candidate $\beta$ Cephei stars (SBL113, SBL353, SBL437, SBL456, SBL515), however, are characterized by extremely low amplitudes in their variability. Additionally, the data quality in $U$ was very low for these stars so that they could not be undoubtedly confirmed by us and therefore must remain candidate pulsators for now. Nevertheless, these stars show evidence of variability in our data and we also were able to make suggestions with respect to the timescales of their potential pulsations. These, if they indeed are pulsations, together with the derived stellar parameters favor a classification as a $\beta$ Cephei star. Furthermore, all these stars for which we could calculate stellar parameters, fall into the $\beta$ Cephei instability strip within the errors as readily shown in Fig. \ref{img:HRD}. Here SBL437 and SBL515 show similar values in the CMD and two-colour diagram. However, their calculated temperatures are found to be very different. We found that $c_1$, which is the primary temperature indicator for hot stars, is very different for these objects. Since NGC6231 shows large differential reddening and $c_1$ is less affected by reddening as, e.g., $B-V$, the positions in the diagrams do not reflect temperature effects very well. Additionally, the position of SBL515 in Fig. \ref{img:HRD} relative to the other $\beta$ Cephei stars and the different time scales of the pulsations indicate that it might not be a cluster member as suggested by BVF99.
Since the amplitudes of these candidates typically are found below the mmag regime, future observations should carefully aim towards a well calibrated data set and also must rely on stable instrumentation in order to unambiguously determine their frequencies and amplitudes to finally classify these stars as pulsating variables. Mode identification was only performed for three stars in this group (SBL226, SBL303, and SBL653). Here only $f_{1}$ of SBL653 and $f_{3}$ of SBL303 allow a careful interpretation and SBL226 serves as an example of the difficulties involved in the mode identification procedure. While the analysed oscillation of SBL653 seems to be a radial mode, our analysis of $f_{3}$ of SBL303 points toward an $l = 4$ mode.

The data quality with respect to the frequency analysis and relative photometric amplitude errors concerning the SPB stars in NGC 6231 is generally much better when compared to the $\beta$ Cephei stars which can be explained by their in general much larger intrinsic pulsation amplitudes. ASK01 identified three SPB candidates in NGC 6231 (SBL275, SBL394, and SBL461) whose nature we managed to confirm. For SBL461 we were also able to interpret pulsation modes with $f_{1}$ pointing towards a dipole mode. SBL461 is blended with SBL459 for which we also find evidence of variability at around $2.4 \; \mathrm{d}^{-1}$ and $14.8 \; \mathrm{d}^{-1}$ which, however, cannot be confirmed unambiguously.

On top of these three candidate SPB stars, seven new pulsators of this class were found (SBL164, SBL210, SBL283, SBL317, SBL480, SBL482, SBL528), all of which are also confirmed members of NGC 6231 and only SBL480 retains a candidate status. We also attempted mode identification for four of the SPB variables (SBL164, SBL210, SBL283, and SBL482). For SBL164 the observations of $f_{1}$ are compatible with an $l=3$ mode. For SBL210 we conclude that $f_{1}$ is an $l=2$ mode. SBL283 seems to drive a dipole mode while for $f_{2}$ of SBL482 we find compatibility with $l=1$ and $l=3$. Concerning the positions in the HR diagram, the same holds true as for the $\beta$ Cephei stars. All stars for which we could calculate stellar parameters fall into the SPB instability strip (Fig. \ref{img:HRD}). We detected one more previously unknown pulsating star in NGC 6231, SBL752, which resembles most likely a $\delta$ Scuti type variable. However, our analysis leads to the conclusion that this star has been misclassified as a member and is in fact a foreground star.

We also found variability in stars where the literature concerning membership status does not give a consistent picture. These are SBL306, SBL417, SBL455, SBL464, and SBL705. SBL306 and SBL455 show a frequency spectrum typical for $\delta$ Scuti stars and broadband photometry suggests that these stars are members of NGC 6231. We classify SBL464 as a new SPB variable and here also the photometry and a comparison with other cluster members suggests that it is part of NGC 6231.

SBL705 shows variability on time scales much longer than the entire observing run, but is also characterized by variations on shorter time scales as demonstrated in the light curve (see Sect. \ref{sec:Results}). However, no periodicities could not be derived reliably. No statement concerning a classification of SBL705 could be given within the limits of our analysis. SBL417 was speculated to be a PMS pulsator, however, we were not able to confirm its variability without doubt.

We also investigated the time scales of the variability of one confirmed nonmember of NGC 6231, SBL612, which is a bright foreground $\delta$ Scuti pulsator. For this variable we only give a tentative frequency solution. In addition, we classify SBL444 as a newly discovered $\delta$ Scuti star which does also not seem to be a member of the cluster. We also found evidence for two eclipsing binaries, SBL574, which is an already known binary and another system not included in the observations of SBL98. Its designation is 097-240594 in the UCAC3 catalog and seems to be an Algol type binary.

Despite the presence of a large number of early-type variable stars in NGC 6231, the cluster contains further B-type stars which we did not discover to be variable within the limits of our analysis. The effective temperatures and luminosities for those 30 stars for which sufficient data was available place them into the theoretical instability regions in the HR Diagram. This suggests that the cluster either contains even more pulsating stars than we currently know of, or that there are constant stars within the instability strips. 

Given that all these stars should have the same metallicity, there must be at least one other parameter that affects the amplitudes of the pulsators. One obvious candidate is rotation. For $\beta$ Cephei stars, there is a tendency that the amplitude decreases with increasing rotation rate \citepads{2005ApJS..158..193S}. Theory \citepads{2007MNRAS.377..645S} predicts that collective saturation of the driving mechanism by several pulsation modes produces amplitudes of the generally observed sizes, and implies a mass dependence of the amplitude. NGC 6231 is an ideal laboratory to test these dependencies and theories, besides its obvious potential to carry out ensemble asteroseismology. To this end, extensive time-resolved multicolour photometry, but also high-resolution spectroscopy, is required. Due to the potentially small amplitudes, however, new measurements must be very precise and well calibrated to improve on our results.

\begin{acknowledgements}
This research is supported by the Austrian Fonds zur F\"orderung der wissenschaftlichen Forschung under grants P18339-N08 and P20526-N16. We also acknowledge support by the Polish NCN grant 2011/01/B/ST9/05448. This research has made use of the VizieR catalogue access tool, CDS, Strasbourg, France. We would like to thank the referee, Chris Koen, for his invaluable comments and suggestions which helped to improve our methods and interpretations as well as the overall quality of this paper.
\end{acknowledgements}


\bibliography{references}

\begin{thebibliography}{40}
\expandafter\ifx\csname natexlab\endcsname\relax\def\natexlab#1{#1}\fi

\bibitem[{{Arentoft} {et~al.}(2001){Arentoft}, {Sterken}, {Knudsen},
  {Freyhammer}, {Duerbeck}, {Pompei}, {Delahodde}, \&
  {Clasen}}]{Arentoft:2001gg}
{Arentoft}, T., {Sterken}, C., {Knudsen}, M.~R., {et~al.} 2001, \aap, 380, 599

\bibitem[{{Asplund} {et~al.}(2004){Asplund}, {Grevesse}, {Sauval}, {Allende
  Prieto}, \& {Kiselman}}]{2004A&A...417..751A}
{Asplund}, M., {Grevesse}, N., {Sauval}, A.~J., {Allende Prieto}, C., \&
  {Kiselman}, D. 2004, \aap, 417, 751

\bibitem[{{Balona}(1983)}]{1983MNRAS.203.1041B}
{Balona}, L.~A. 1983, \mnras, 203, 1041

\bibitem[{{Balona} \& {Engelbrecht}(1985)}]{BE85}
{Balona}, L.~A. \& {Engelbrecht}, C.~A. 1985, \mnras, 212, 889

\bibitem[{{Balona} \& {Laney}(1995)}]{1995MNRAS.276..627B}
{Balona}, L.~A. \& {Laney}, C.~D. 1995, \mnras, 276, 627

\bibitem[{{Balona} \& {Shobbrook}(1983)}]{1983MNRAS.205..309B}
{Balona}, L.~A. \& {Shobbrook}, R.~R. 1983, \mnras, 205, 309

\bibitem[{{Baume} {et~al.}(1999){Baume}, {V{\'a}zquez}, \&
  {Feinstein}}]{Baume99}
{Baume}, G., {V{\'a}zquez}, R.~A., \& {Feinstein}, A. 1999, \aap Supplement,
  137, 233

\bibitem[{{Boehm-Vitense}(1981)}]{1981ARA&A..19..295B}
{Boehm-Vitense}, E. 1981, \araa, 19, 295

\bibitem[{{Breger} {et~al.}(1993){Breger}, {Stich}, {Garrido}, {Martin},
  {Jiang}, {Li}, {Hube}, {Ostermann}, {Paparo}, \&
  {Scheck}}]{1993A&A...271..482B}
{Breger}, M., {Stich}, J., {Garrido}, R., {et~al.} 1993, \aap, 271, 482

\bibitem[{{Broeg} {et~al.}(2005){Broeg}, {Fern{\'a}ndez}, \&
  {Neuh{\"a}user}}]{2005AN....326..134B}
{Broeg}, C., {Fern{\'a}ndez}, M., \& {Neuh{\"a}user}, R. 2005, Astronomische
  Nachrichten, 326, 134

\bibitem[{{Cox}(2000)}]{2000asqu.book.....C}
{Cox}, A.~N. 2000, {Allen's astrophysical quantities} (Springer; 4th ed.
  edition)

\bibitem[{{Flower}(1996)}]{1996ApJ...469..355F}
{Flower}, P.~J. 1996, \apj, 469, 355

\bibitem[{{Garc{\'{\i}}a} \& {Mermilliod}(2001)}]{GM01}
{Garc{\'{\i}}a}, B. \& {Mermilliod}, J.~C. 2001, \aap, 368, 122

\bibitem[{{Garrison} \& {Schild}(1979)}]{1979AJ.....84.1020G}
{Garrison}, R.~F. \& {Schild}, R.~E. 1979, \aj, 84, 1020

\bibitem[{{Handler} {et~al.}(2012){Handler}, {Shobbrook}, {Uytterhoeven},
  {Briquet}, {Neiner}, {Tshenye}, {Ngwato}, {van Winckel}, {Guggenberger},
  {Raskin}, {Rodr{\'{\i}}guez}, {Mazumdar}, {Barban}, {Lorenz},
  {Vandenbussche}, {{\c S}ahin}, {Medupe}, \& {Aerts}}]{2012MNRAS.424.2380H}
{Handler}, G., {Shobbrook}, R.~R., {Uytterhoeven}, K., {et~al.} 2012, \mnras,
  424, 2380

\bibitem[{{Koen}(2010)}]{2010Ap&SS.329..267K}
{Koen}, C. 2010, \apss, 329, 267

\bibitem[{{Kunzli} {et~al.}(1997){Kunzli}, {North}, {Kurucz}, \&
  {Nicolet}}]{1997A&AS..122...51K}
{Kunzli}, M., {North}, P., {Kurucz}, R.~L., \& {Nicolet}, B. 1997, \aaps, 122,
  51

\bibitem[{{Lenz} \& {Breger}(2005)}]{2005CoAst.146...53L}
{Lenz}, P. \& {Breger}, M. 2005, Communications in Asteroseismology, 146, 53

\bibitem[{{Levato} \& {Morrell}(1983)}]{LM83}
{Levato}, H. \& {Morrell}, N. 1983, \aplett, 23, 183

\bibitem[{{McSwain} \& {Gies}(2005)}]{SG05}
{McSwain}, M.~V. \& {Gies}, D.~R. 2005, \apjs, 161, 118

\bibitem[{{Mermilliod} {et~al.}(1997){Mermilliod}, {Mermilliod}, \&
  {Hauck}}]{1997A&AS..124..349M}
{Mermilliod}, J.-C., {Mermilliod}, M., \& {Hauck}, B. 1997, \aaps, 124, 349

\bibitem[{{Montgomery} \& {O'donoghue}(1999)}]{1999DSSN...13...28M}
{Montgomery}, M.~H. \& {O'donoghue}, D. 1999, Delta Scuti Star Newsletter, 13,
  28

\bibitem[{{Napiwotzki} {et~al.}(1993){Napiwotzki}, {Schoenberner}, \&
  {Wenske}}]{1993A&A...268..653N}
{Napiwotzki}, R., {Schoenberner}, D., \& {Wenske}, V. 1993, \aap, 268, 653

\bibitem[{{Pamyatnykh} {et~al.}(1998){Pamyatnykh}, {Dziembowski}, {Handler}, \&
  {Pikall}}]{1998A&A...333..141P}
{Pamyatnykh}, A.~A., {Dziembowski}, W.~A., {Handler}, G., \& {Pikall}, H. 1998,
  \aap, 333, 141

\bibitem[{{Paunzen} {et~al.}(2010){Paunzen}, {Heiter}, {Netopil}, \&
  {Soubiran}}]{2010A&A...517A..32P}
{Paunzen}, E., {Heiter}, U., {Netopil}, M., \& {Soubiran}, C. 2010, \aap, 517,
  A32

\bibitem[{{Perry} {et~al.}(1991){Perry}, {Hill}, \&
  {Christodoulou}}]{1991A&AS...90..195P}
{Perry}, C.~L., {Hill}, G., \& {Christodoulou}, D.~M. 1991, \aaps, 90, 195

\bibitem[{{Raboud}(1996)}]{Ra96}
{Raboud}, D. 1996, \aap, 315, 384

\bibitem[{{Raboud} {et~al.}(1997){Raboud}, {Cramer}, \& {Bernasconi}}]{RCB97}
{Raboud}, D., {Cramer}, N., \& {Bernasconi}, P.~A. 1997, \aap, 325, 167

\bibitem[{{Raboud} \& {Mermilliod}(1998)}]{RM98}
{Raboud}, D. \& {Mermilliod}, J.-C. 1998, \aap, 333, 897

\bibitem[{{Seaton}(2005)}]{2005MNRAS.362L...1S}
{Seaton}, M.~J. 2005, \mnras, 362, L1

\bibitem[{{Shobbrook}(1979)}]{1979MNRAS.189..571S}
{Shobbrook}, R.~R. 1979, \mnras, 189, 571

\bibitem[{{Shobbrook}(1983)}]{Sh83}
{Shobbrook}, R.~R. 1983, \mnras, 205, 1215

\bibitem[{{Skiff}(2009)}]{2009yCat....1.2023S}
{Skiff}, B.~A. 2009, VizieR Online Data Catalog, 1, 2023

\bibitem[{{Smolec} \& {Moskalik}(2007)}]{2007MNRAS.377..645S}
{Smolec}, R. \& {Moskalik}, P. 2007, \mnras, 377, 645

\bibitem[{{Stankov} \& {Handler}(2005)}]{2005ApJS..158..193S}
{Stankov}, A. \& {Handler}, G. 2005, \apjs, 158, 193

\bibitem[{{Sterken} \& {Bouzid}(2004)}]{2004RMxAC..20...79S}
{Sterken}, C. \& {Bouzid}, M.~Y. 2004, in Revista Mexicana de Astronomia y
  Astrofisica, vol. 27, Vol.~20, Revista Mexicana de Astronomia y Astrofisica
  Conference Series, ed. G.~{Tovmassian} \& E.~{Sion}, 79--80

\bibitem[{{Sung} {et~al.}(1998){Sung}, {Bessell}, \&
  {Lee}}]{1998AJ....115..734S}
{Sung}, H., {Bessell}, M.~S., \& {Lee}, S.-W. 1998, \aj, 115, 734

\bibitem[{{Watson}(1988)}]{1988Ap&SS.140..255W}
{Watson}, R.~D. 1988, \apss, 140, 255

\bibitem[{{Zacharias} {et~al.}(2009){Zacharias}, {Finch}, {Girard}, {Hambly},
  {Wycoff}, {Zacharias}, {Castillo}, {Corbin}, {Divittorio}, {Dutta}, {Gaume},
  {Gauss}, {Germain}, {Hall}, {Hartkopf}, {Hsu}, {Holdenried}, {Makarov},
  {Martinez}, {Mason}, {Monet}, {Rafferty}, {Rhodes}, {Siemers}, {Smith},
  {Tilleman}, {Urban}, {Wieder}, {Winter}, \& {Young}}]{UCAC3}
{Zacharias}, N., {Finch}, C., {Girard}, T., {et~al.} 2009, VizieR Online Data
  Catalog, 1315, 0

\bibitem[{{Zima}(2008)}]{FAMIAS}
{Zima}, W. 2008, Communications in Asteroseismology, 155, 17

\end{thebibliography}

\clearpage

\Online

\setcounter{section}{3}
\section{Results for individual stars}
\label{sec:Resultsindiv}

In this section of our study we present detailed results and discussions for each star individually. Our analysis is split into five different groups, reviewing $\beta$ Cephei and SPB stars in NGC 6231 separately, followed by a discussion of further variables including possible members, nonmembers, and eclipsing binaries. However, a detailed analysis was only performed for the stars which are situated in NGC 6231. For the remaining objects we only carried out a coarse study, giving e.g. only pulsation frequencies.

\begin{figure}
	\resizebox{\hsize}{!}{\includegraphics{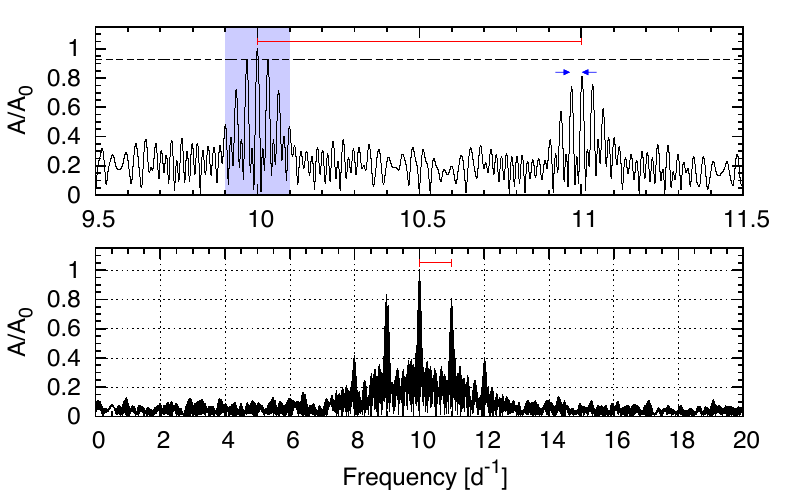}}
	\caption[Spectral Window of the observations]{Spectral window of the observations. The bottom panel shows the overall structure of the window with daily 
aliases indicated by the red marker where the $\pm 1 \; \mathrm{d}^{-1}$ peaks reach about $80\%$ of the strongest signal. The top panel displays a closer view revealing the monthly aliases (blue arrows) with a maximum signal at about $93\%$ marked as a dashed horizontal line. The region marked in blue denotes the size of the envelope of the monthly aliases which represents the length of a run in one month (which is about 5 days).}
	\label{img:SW}
\end{figure}

The individual spectral windows of the light curves are very similar to each other since all data were extracted from one set and NGC 6231 was observed in all 3 filters during each night. The only noteworthy distinction is created when large parts of the light curves in $U$ had to be deleted due to corrupted data, but even then the general shape of the spectral window was maintained. An example is displayed in Fig. \ref{img:SW} where the spectral window for the data of SBL303 was folded with an artificial frequency at $10 \; \mathrm{d}^{-1}$. Main features are the daily alias $(\pm 1 \; \mathrm{d^{-1}})$ created by the observing gaps during the days and monthly aliases $(\pm 0.03 \; \mathrm{d^{-1}})$ created by the observations in May, June, and July which are separated by about a month.

\subsection{$\beta$ Cephei stars}

\begin{figure}
	\resizebox{\hsize}{!}{\includegraphics{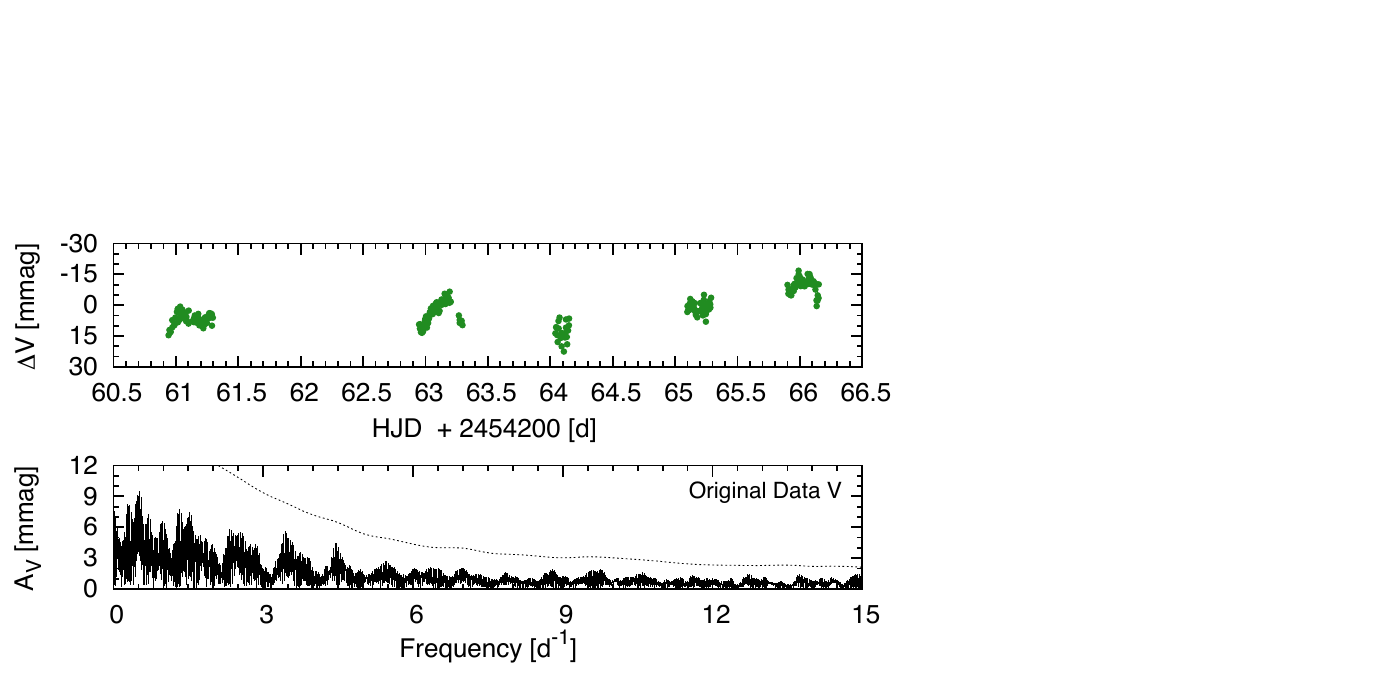}}
	\caption[SBL113]{Light curve and frequency spectrum of the $\beta$ Cephei candidate SBL113. No frequency was significant with our detection criterion of $\mathrm{SNR} \ge 4$ which is displayed as dotted line in the lower panel.}
	\label{img:SBL113}
\end{figure}

\textbf{SBL113} is classified as a possible $\beta$ Cephei pulsator by BE85 and according to RCB97 is a member of NGC 6231. This target is listed as a B1.5V star by LM83 who also imply the possible presence of a dust shell around SBL113. PHC91 find a spectral type of B0.5IV without peculiarities. Photometric studies of this star revealed an unusually low $\beta$ index in the Str\"omgren photometry pointing towards H$\beta$ emission. \citetads{SG05} carried out a search for Be stars in southern open clusters and do not classify SBL113 as such a star. It is also known that the Be phenomenon can occur only temporally. In light of the uncertain nature of this object it was excluded from our calibration of the Str\"omgren photometry. So far variability studies have been carried out only by BE85 since SBL113 lies outside the FOV of the ASK01 observations. BE85 found variability in the range of $11 - 11.5 \; \mathrm{d}^{-1}$ with amplitudes barely exceeding $1 \; \mathrm{mmag}$ but were not able to determine individual frequencies. Moreover, they reported a slower superimposed variation pattern but could not find corresponding periodicities. Based on these results BE85 considered this star to be a probable $\beta$ Cephei type pulsator. Unfortunately, also our frequency analysis did not result in a single significant peak. However, both light curve and periodogram, as shown in Fig. \ref{img:SBL113}, clearly reveal variability with frequencies around $3 \; \mathrm{d}^{-1}$ where we find the largest amplitude at $3.44 \; \mathrm{d}^{-1}$. Furthermore, even variability on time scales with periods longer than $1 \; \mathrm{d}$ is indicated. However, no evidence was found that this object shows faster $\beta$ Cephei type pulsation as suggested by BE85 which could be explained by the small amplitudes of the indicated pulsation compared to our noise level which is considerably higher. As a consequence this star must remain a candidate.

\begin{figure*}
	\resizebox{\hsize}{!}{\includegraphics{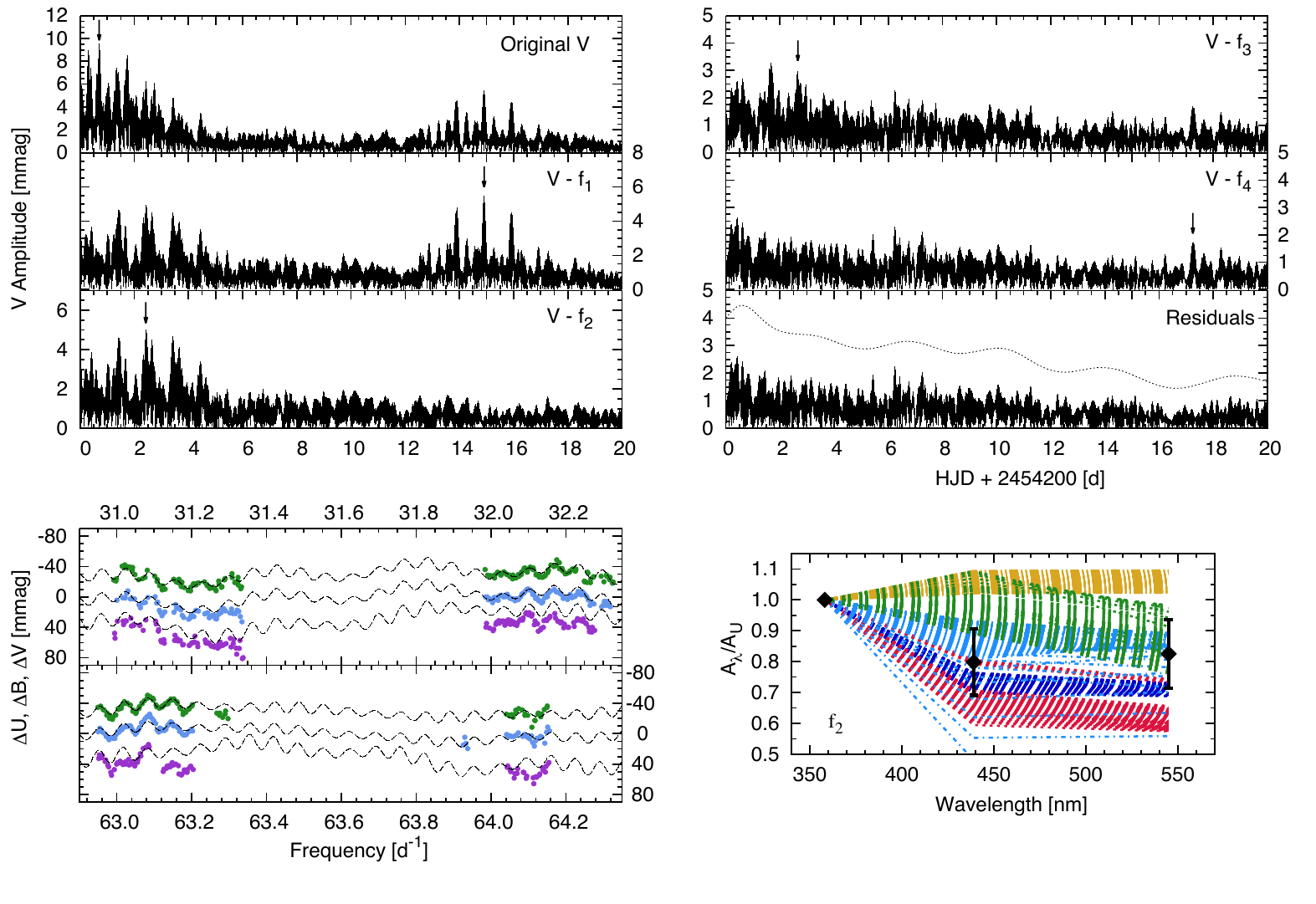}}
	\caption[SBL226]{The frequency spectra of the $\beta$ Cephei star SBL226 for the unadjusted data in V show different stages of the analysis where each of the panels in the top figures displays consecutive prewhitening steps. The last panel shows the residuals after subtracting five frequencies and the $\mathrm{SNR} = 4$ threshold as dotted line. The arrows in the individual panels mark each significant frequency throughout the analysis procedure. The bottom left panels show exemplary light curves for all bands where purple represents data in $U$, blue in $B$, and green in $V$. The dashed lines represent the calculated multi-periodic fits. The graph in the bottom right corner shows the mode identification where both the predicted and observed amplitude ratios are shown. We searched the grids for stars with masses ranging from 14 to $16 \; M_{\odot}$. The various colours of the dotted lines represent theoretical pulsation modes ranging from $l=0$ to $l=4$. Dark blue corresponds to $l=0$, red to $l=1$, light blue to $l=2$, green to $l=3$, and yellow to $l=4$. The black squares and their errorbars show the observed values. The determined frequencies take the values $f_{1} = 0.7017 \; \mathrm{d^{-1}}$, $f_{2} = 14.9098 \; \mathrm{d^{-1}}$, $f_{3} = 2.4242 \; \mathrm{d^{-1}}$, $f_{4} = 2.6579 \; \mathrm{d^{-1}}$, and $f_{5} = 17.2502 \; \mathrm{d^{-1}}$ with errors ranging from $\sigma_{f_{1}} = 0.0003 \; \mathrm{d^{-1}}$ to $\sigma_{f_{5}} = 0.0013 \; \mathrm{d^{-1}}$.}
	\label{img:SBL226}
\end{figure*}

\textbf{SBL226} was reported to be a $\beta$ Cephei pulsator by Ba83 with a period of $0.07 \; \mathrm{d}$ and according to RCB97 and BVF99 is a cluster member. PHC91 and 
GM01 list this star as a binary with equal B1V components, \citetads{RM98} (RM98) find it to be a single star. Dedicated variability studies have been carried out by BS83 and ASK01. Both find three frequencies but of these only one matches exactly and another one can be explained as an alias. In contrast to BS83, who find a remarkably simple frequency spectrum, our analysis reveals a different picture. We found five significant frequencies in total of which two are dominating the variability at $f_{1} = 0.7017 \; \mathrm{d}^{-1}$ and $f_{2} = 14.9098 \; \mathrm{d}^{-1}$. After analyzing many light curves of different stars we found typical frequencies in our unadjusted data representing drifts of the zero-point. However, $0.7 \; \mathrm{d}^{-1}$ was not one of them. Therefore $f_{1}$ is interpreted as intrinsic variability. In addition clear amplitude differences in all three passbands rule out possible rotation and binary effects. $f_{2}$ matches with the results of ASK01 and BS83 perfectly. For the remaining detected oscillations we found significant aliasing and also differences between $V$ and $B$ data. For $f_3$ an alias of $2f_1$ was considered but since there is no signal at the exact position and the amplitude is much larger at the given value, $f_3$ is considered as an independent oscillation. We attempted mode identification only for $f_{1}$ and $f_{2}$ since only for these frequencies reliable amplitudes in $U$ could be derived. To this end we calculated amplitude ratios for models with a mass between $14 - 16 \; M_{\odot}$, as indicated by its position in the theoretical HR diagram relative to the evolutionary tracks. For  $f_{1}$ no match was found since the models showed too many simultaneously excited modes. However, considering the oscillation frequency $(f_{1} = 0.7 \; \mathrm{d^{-1}})$ only a non-radial pulsation is possible since $l=0$ modes altogether exhibit faster variations. For $f_{2}$ we also found too many excited modes for a decisive conclusion. In Fig. \ref{img:SBL226} we show frequency spectra, sample light curves and, the mode identification for $f_{2}$ as an example for the following discussions. Given this result we can only exclude an $l=4$ mode for $f_{2}$. In light of the results from the frequency analysis and the calibration of effective temperature and luminosity (including errors), SBL226 can also be considered as a candidate for a $\beta$ Cephei/SPB hybrid. The multiperiodic variability in both the low and high frequency domain in combination with the bandpass-dependent amplitudes are very strong evidence for this interpretation.

\begin{figure*}
	\resizebox{\hsize}{!}{\includegraphics{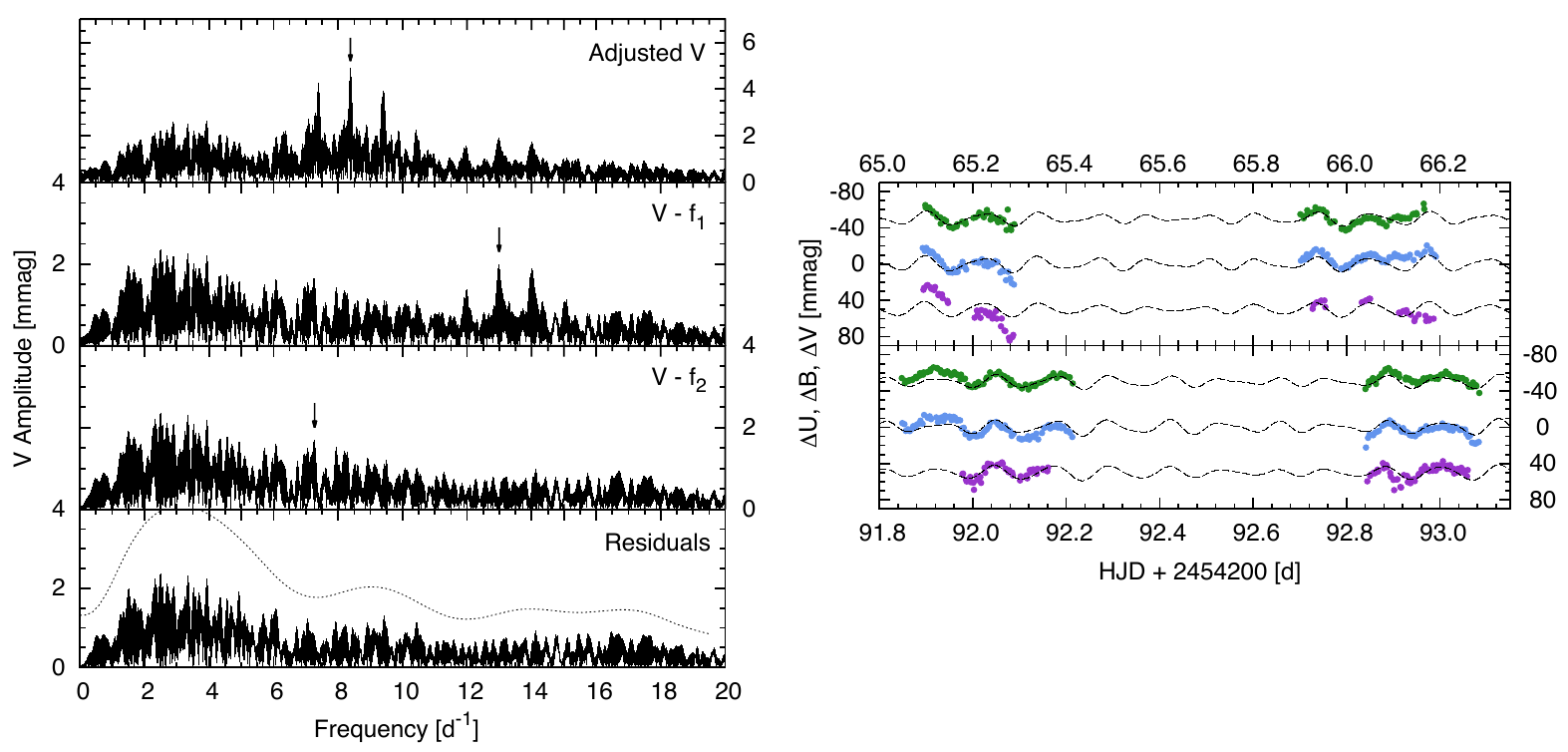}}
	\caption[SBL268]{Frequency spectra and selected light curves for the $\beta$ Cephei star SBL268. The analysis was performed with daily adjusted data, suppressing low frequency variability. The notations and markings are the same as in Fig. \ref{img:SBL226} where applicable. $f_{1} = 8.3847 \; \mathrm{d^{-1}}$, $f_{2} = 12.9891 \; \mathrm{d^{-1}}$, and $f_{3} = 7.2748 \; \mathrm{d^{-1}}$.  The errors range from $\sigma_{f_{1}} = 0.0004 \; \mathrm{d^{-1}}$ to $\sigma_{f_{3}} = 0.001 \; \mathrm{d^{-1}}$ for the frequencies.} 
	\label{img:SBL268}
\end{figure*}

\textbf{SBL268} was initially suspected to be a $\beta$ Cephei variable by Ba83 which was later confirmed by BE85. LM83 list this star as a double-lined spectroscopic binary consisting of two equal B2V components and RM98 classify it as a single object. Both BVF99 and RCB97 arrive at the conclusion that SBL268 is a member of NGC 6231. Variability studies have been carried out by BE85 and ASK01. The data from BE85 yielded evidence for three significant frequencies and they also report the possible presence of long period variability. ASK01 found just one match with the older investigation at $8.4 \; \mathrm{d}^{-1}$ and an additional frequency at $2.6 \; \mathrm{d}^{-1}$.  Our analysis is based on daily adjusted data since the light curves were characterized by large zero-point offsets. The results are displayed in Fig. \ref{img:SBL268}. We were able to detect three frequencies above the SNR threshold, two of them ($f_{1} = 8.3847 \; \mathrm{d^{-1}}$ and $f_{2} = 12.9891 \; \mathrm{d^{-1}}$) are an exact match with BE85. Apart from the (still not confirmed) slower variability, the third frequency of BE85 is also present as an alias at $f_{3} = 7.2748 \; \mathrm{d^{-1}}$ but only reaches significance if one looks at a daily adjusted data set in $V$. The multi-periodicity and amplitude difference clearly point towards pulsations. To search for low frequency variability as mentioned by BE85, the more rapid $\beta$ Cephei signals were subtracted from the original unadjusted data. This led to a residual light curve which was dominated by variations of the order of the length of the run making it very difficult to detect any low frequency modulation. However, a peak in the unadjusted spectrum (and therefore not in Fig. \ref{img:SBL268}) is visible at $0.38 \; \mathrm{d^{-1}}$ where BE85 suspected a signal but it is not significant. Furthermore, the residual frequency spectrum suggests the presence of additional variability at about $10 \; \mathrm{d^{-1}}$ with an amplitude in the mmag regime. In $U$ the data are extremely unreliable for large parts, only $f_{1}$ with an amplitude $> 5 \; \mathrm{mmag}$ comes close to the SNR threshold, but does not exceed it. Due to these large uncertainties we did not attempt mode identification of the individual oscillations. Clearly SBL268 shows oscillations and in addition it is located in the $\beta$ Cephei instability strip. Therefore we can confirm its membership in this class of variables.

\begin{figure*}
	\resizebox{\hsize}{!}{\includegraphics{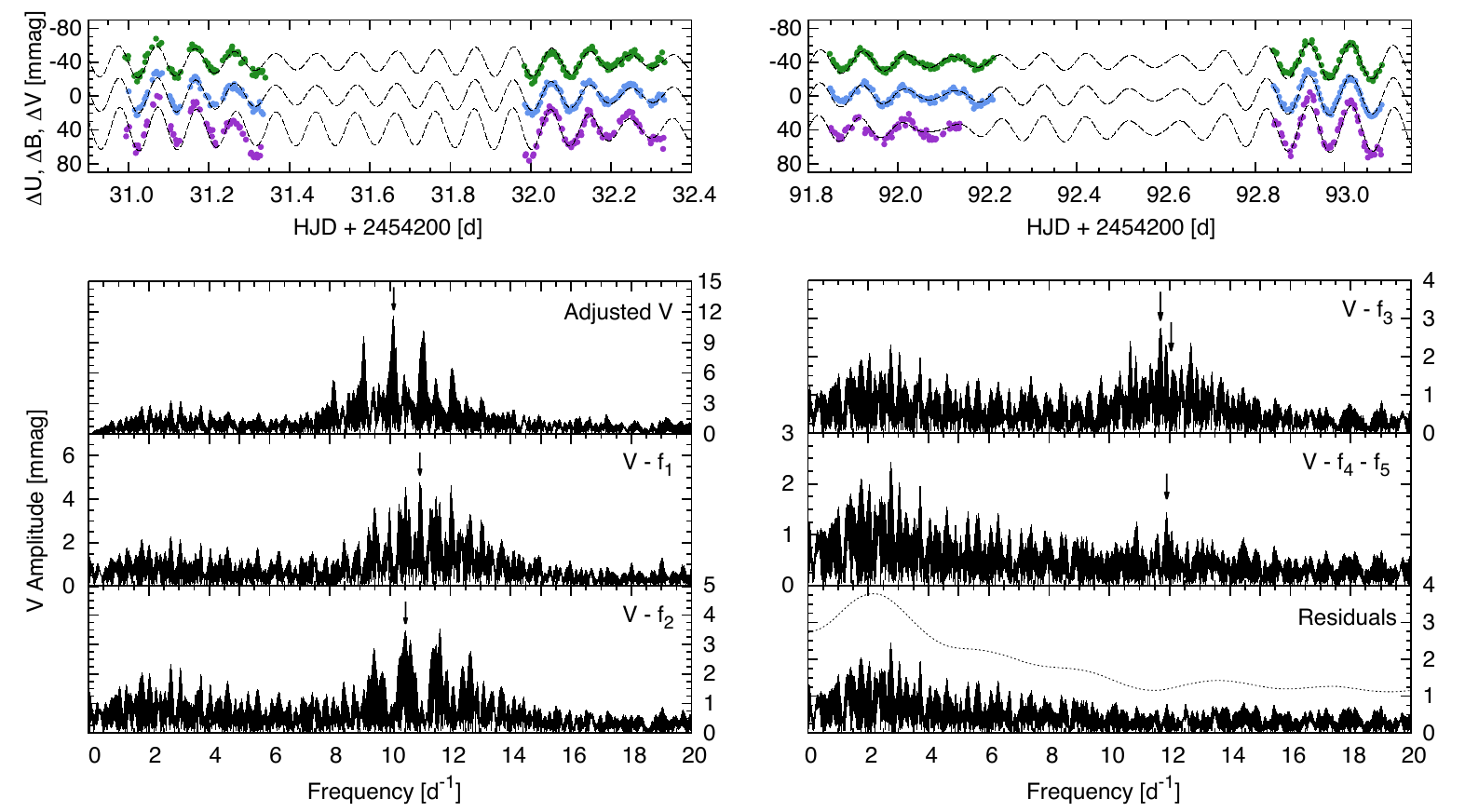}}
	\caption[SBL303]{The frequency spectra of the $\beta$ Cephei variable SBL303 in the bottom panels show five steps in our data evaluation. The top panels show light curves for all bands. Since too many models passed through our error box, the mode identification panel is not displayed here. However, $f_{3}$ shows increasing amplitude with wavelength which is only seen in $l=4$ modes. All notations are the same as in Fig. \ref{img:SBL226}. The frequencies take the values: $f_{1} = 10.1225 \; \mathrm{d^{-1}}$, $f_{2} = 10.991 \; \mathrm{d^{-1}}$, $f_{3} = 10.5106 \; \mathrm{d^{-1}}$, $f_{4} = 11.6904 \; \mathrm{d^{-1}}$, $f_{5} = 12.045 \; \mathrm{d^{-1}}$, and $f_{6} = 11.902 \; \mathrm{d^{-1}}$. Their errors range from $\sigma_{f_{1}} = 0.0002 \; \mathrm{d^{-1}}$ to $\sigma_{f_{6}} = 0.001 \; \mathrm{d^{-1}}$.} 
	\label{img:SBL303}
\end{figure*}

\textbf{SBL303} has been classified as a $\beta$ Cephei pulsator as early as 1983 by Ba83. Its spectral type is listed as B2IV-V by PHC91 and it is classified as a member by both BVF99 and RCB97. Ra96, RM98, and GM01 consider this star as a binary but no definite conclusion has been drawn so far since too few spectroscopic measurements are available. Furthermore, RM98 suggest two components with masses of $8.8$ and $1.7 \; M_{\odot}$ respectively. In light of this large difference in mass, the combined stellar light would be strongly dominated by the more massive component. Therefore the Str\"omgren photometry is fully representative of the primary star and can be used to determine its effective temperature and surface gravity. This calibration puts the star into the $\beta$ Cephei instability domain. SBL303 has been subject to variability studies by BS83 and ASK01. Comparing their work, the rapid $\beta$ Cephei pulsation frequencies match relatively well, however, both find evidence for superimposed slower variability, but here the results differ. Our data reveal a more complex picture: The multi-periodic oscillation spectrum at higher frequencies is recovered and also clear signs of variability on longer time scales are present. The data, however, suffer from zero point offsets between the individual nights and months. No peak in the regime $< 1 \; \mathrm{d^{-1}}$ is deemed significant with the adopted SNR threshold criterion. Therefore no statement can be made whether this is due to zero point variations or indeed intrinsic. To determine reliable amplitudes which do not suffer from offsets and extrinsic long term variability, we applied an adjustment on a daily basis by subtracting the mean magnitude for each day to study the more rapid $\beta$ Cephei pulsations. Our analysis resulted in six independent oscillations. Comparing our results to BS83, all frequencies except the long period variation are reproduced either exactly ($f_{1} = 10.1225 \; \mathrm{d^{-1}}, f_{4} = 11.6904 \; \mathrm{d^{-1}}, f_{5} = 12.045 \; \mathrm{d^{-1}}$) or can be explained with a monthly $\pm 0.03 \; \mathrm{d^{-1}}$ alias ($f_{2} = 10.991 \; \mathrm{d^{-1}}, f_{3} = 10.5106 \; \mathrm{d^{-1}}$). In the latter cases the aliases were examined and rejected because (a) each of them had a significantly lower amplitude and (b) the residuals of the fit also increased when these were chosen. The same is true for a comparison with the results of ASK01: low-frequency variations were not matched but the others match exactly $(f_{1}, f_{2})$ or can be explained with a daily alias $(f_{3}, f_{4})$. Within the errors also the amplitudes match with those from BS83 except for $f_{5}$. The weakest signal we detected $(f_{6})$ at $11.902 \; \mathrm{d^{-1}}$, which reached a $\mathrm{SNR} > 4$ only in V, has not been found before. With all evidence taken together, the analysis clearly confirms the $\beta$ Cephei nature of SBL303. Despite the good data quality for SBL303 our attempts to characterise pulsation modes were not successful since too many models passed through our error box. However, one conclusion can be drawn from our tests. $f_{3}$ shows a trend towards larger amplitudes with increasing wavelength. When calculating models for a mass range of $9 - 11 \; M_{\odot}$, the only spherical degree that is compatible with this behavior is $l=4$. This may at first sight be surprising because of geometrical cancellation for such a high-degree mode. However, it has been suggested that higher-l modes tend to occur more frequently in the more rapidly rotating $\beta$~Cephei stars \citepads{2012MNRAS.424.2380H}; SBL 303 rotates at $v \sin i = 140$ kms$^{-1}$ \citepads{GM01}. Frequency spectra and light curves are displayed in Fig. \ref{img:SBL303}.

\begin{figure}
	\resizebox{\hsize}{!}{\includegraphics{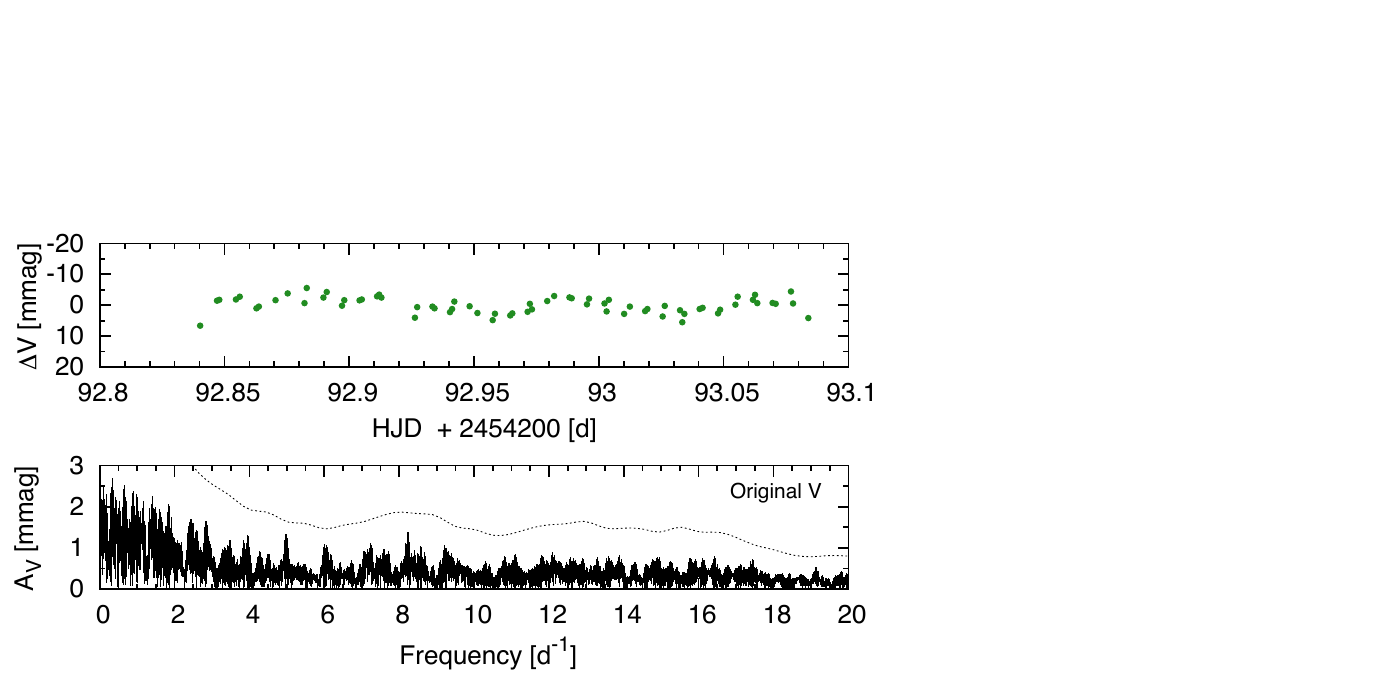}}
	\caption[SBL353]{Light curve and frequency spectrum of the $\beta$ Cephei candidate SBL353. No signal reaches our $\mathrm{SNR}=4$ threshold but for some nights a visual inspection suggests the presence of intrinsic variability.} 
	\label{img:SBL353}
\end{figure}

\textbf{SBL353} was first suggested to be a $\beta$ Cephei candidate by BE85 and both RCB97 and BVF99 classify it as a member of the cluster. PHC91 list spectral type 
B0.5V. According to LM83 it is a spectroscopic binary, RM98 however conclude that it is a single star. Str\"omgren photometry places this star near the other very bright $\beta$ Cephei variables on the theoretical HR diagram in NGC 6231, but one has to keep in mind its possible binary nature. No one so far managed to detect a single significant frequency for this object: BE85 mention that there might be oscillations at $0.3 \; \mathrm{d^{-1}}$ or at $10.6 \; (11.6) \; \mathrm{d^{-1}}$ but do not arrive at a conclusion. ASK01 also analyzed this star and found possible oscillations in the $10-15 \; \mathrm{d^{-1}}$ range. Our attempts to detect significant frequencies with the new data did not change this situation. One reason why it is so difficult to extract high quality light curves from the CCD data is the fact that SBL353 is situated next to a very bright O star. This star produced a bright halo around itself in the images and depending whether the instrument was in focus or not, led to serious contamination in the data. As a consequence the given statements should be considered with some caution. Our analysis resulted in individual significant frequencies, but these depend on the filter under consideration: Two frequencies were found in $V$ at $8.24 \; \mathrm{d^{-1}}$ and $4.98 \; \mathrm{d^{-1}}$ which is not consistent with the previous studies and in $B$ not even a small signal was visible in these regions. On the other hand, the $B$ data showed excess power at $3 \; \mathrm{d^{-1}}$, which in turn was not visible in $V$. This kind of variability might be attributed to the contamination of the bright O star. However, a insignificant signal was present at $0.3 \; \mathrm{d^{-1}}$ which agrees with the suggested period of BE85 and a visual inspection of the light curve showed indeed higher frequency variation for some nights. As a consequence this star must remain a $\beta$ Cephei candidate for now. A light curve sample and the frequency spectrum of SBL353 are displayed in Fig. \ref{img:SBL353}. Here the light curve spans only one night where one can see rapid variations which might be $\beta$ Cephei oscillations.

\begin{figure}
	\resizebox{\hsize}{!}{\includegraphics{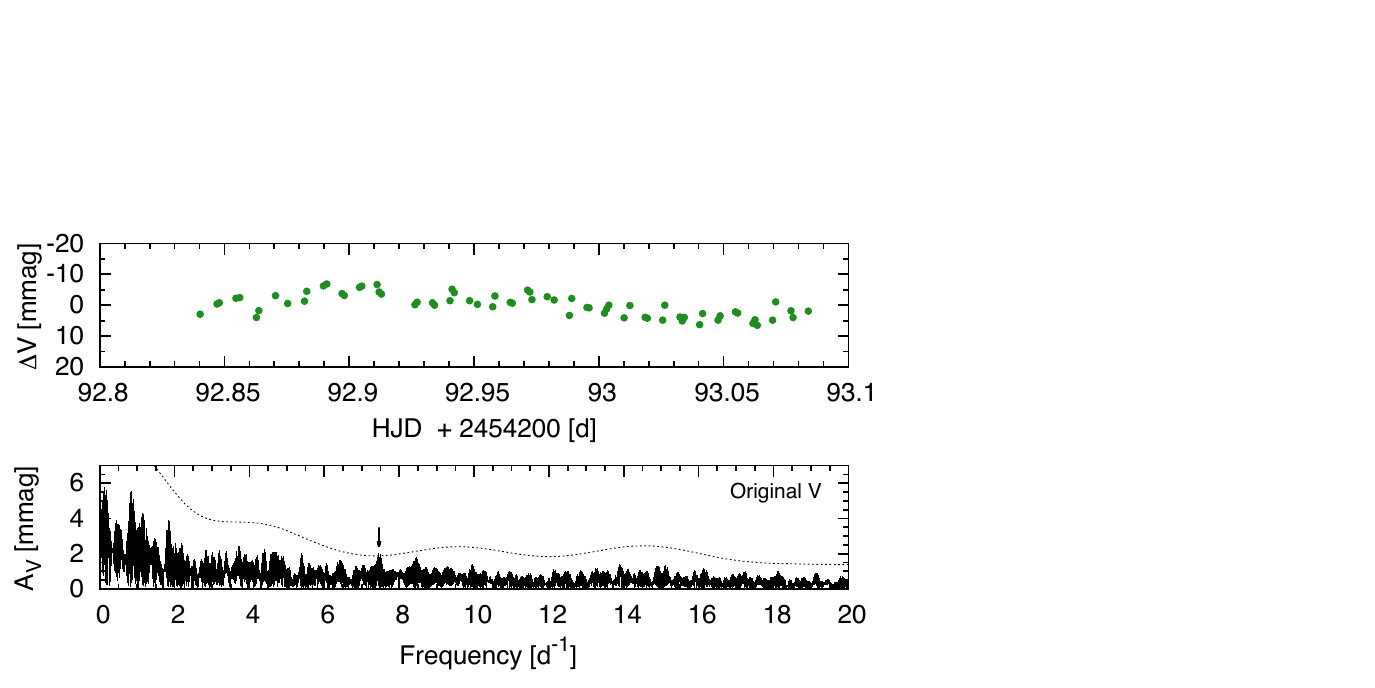}}
	\caption[SBL437]{Example light curve and frequency spectrum of the $\beta$ Cephei candidate SBL437. Our analysis resulted in one significant frequency at $f_{1} = 7.461 \pm 0.001 \; \mathrm{d^{-1}}$.} 
	\label{img:SBL437}
\end{figure}

\textbf{SBL437} is a $\beta$ Cephei candidate according to BE85 and ASK01. BVF99 and RCB97 determined it to be a cluster member and PHC91 list it with spectral type B0Vn. Both variability studies did not result in the detection of any frequency above the SNR threshold, but still indications were present that this star is indeed variable. BE85 reported possible frequencies around $9.1 \; \mathrm{d^{-1}}$ and $14 \; \mathrm{d^{-1}}$ and even a long term variation with a period of the order of $30 \; \mathrm{d}$. ASK01 mention possible frequencies around $15 \; \mathrm{d^{-1}}$. Our data set supports the picture of a $\beta$ Cephei candidate and we were able to detect one oscillation above the SNR threshold at $f_{1} = 7.461 \; \mathrm{d^{-1}}$. Further analysis reveals some interesting features: Visually inspecting the light curves and periodograms led to the conclusion that the long term variation is not caused by zero point offsets as is the case with many other stars. Most stars in our data set show extrinsic variability of the order of the length of the run due to zero-point offsets at  $0.016 \; \mathrm{d^{-1}}$.
In this case the strongest but still insignificant signal which would mark a long period variation can be found at $0.14 \; \mathrm{d^{-1}}$ with a strong alias at $0.86 \; \mathrm{d^{-1}}$. Going up in the frequency range another potential frequency can be found at $3.73 \; \mathrm{d^{-1}}$ which depicts $f_{1}/2$. At even higher frequencies (and consistent with the results of BE85 and ASK01) possible signals can be found around $14 \; \mathrm{d^{-1}}$ and $15 \; \mathrm{d^{-1}}$ both of which do not seem to be a harmonic of $f_{1}$. The data quality in U is worse than average for this star and we could not even find a sign of the strongest frequency found in the other filters which can be attributed to its small amplitude. Therefore no attempt to identify the pulsation mode was made. Figure \ref{img:SBL437} shows data for one night in $V$ and our frequency spectrum of SBL437. Clearly additional and even more important, more precise data are needed to detect additional intrinsic variability and settle for a definite conclusion whether this star belongs to the $\beta$ Cephei class or not.

\begin{figure}
	\resizebox{\hsize}{!}{\includegraphics{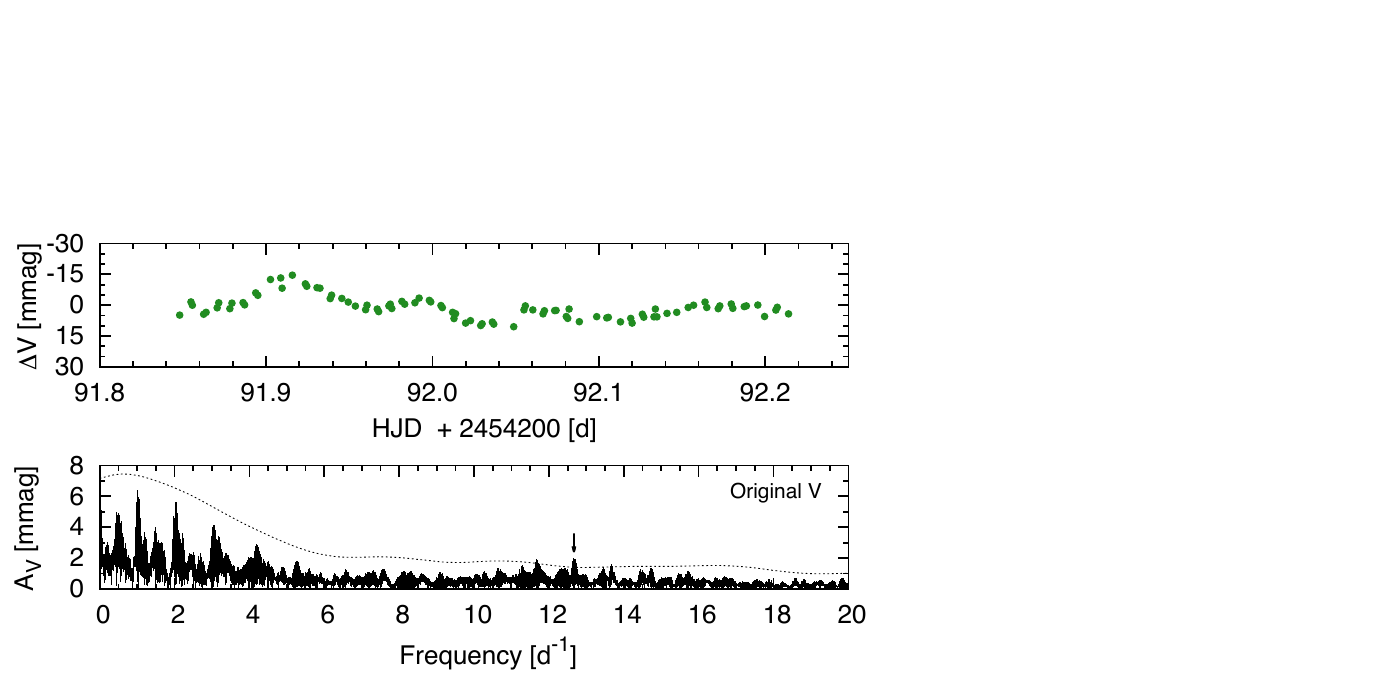}}
	\caption[SBL456]{Sample data and frequency spectrum of the $\beta$ Cephei candidate SBL456. Notations and markings are the same as in Fig. \ref{img:SBL226}. Only one frequency at $f_{1} = 12.670 \pm 0.001 \; \mathrm{d^{-1}}$ was found to be significant.} 
	\label{img:SBL456}
\end{figure}

\textbf{SBL456} is a $\beta$ Cephei candidate as proposed by ASK01 with spectral type B1Vn as listed by PHC91 which still lacks confirmation. According to BVF99 it is likely a member of NGC 6231. The variability study of ASK01 resulted in the detection of three frequencies at $3.1 \; \mathrm{d^{-1}}$, $3.9 \; \mathrm{d^{-1}}$, and $0.1 \; \mathrm{d^{-1}}$ respectively and they suggested that oscillations with even shorter periods are likely present but their data did not allow a detection in this range.  Our data and significance criterion allowed us to only detect one frequency at $f_{1} = 12.670 \; \mathrm{d^{-1}}$ together with its aliases. Examining the light curves and the periodogram after prewhitening shows that variations at longer time scales are present as well (excluding the often seen signal at $0.016 \; \mathrm{d^{-1}}$ which is introduced by the length of the entire observing run). However, no signal in this regime exceeds the SNR threshold but probable oscillations can be found at $3.05 \; \mathrm{d^{-1}}$ (which agrees with the result of ASK01) and a second peak can be found at about $0.5 \; \mathrm{d^{-1}}$. Since the variability at $0.5 \; \mathrm{d^{-1}}$ is sometimes seen in other stars as well, it is also possible that it is caused by zero-point offsets. If it is intrinsic, there is no way of telling which peak represents the true oscillation since in this range the periodogram has a complex structure and is very crowded due to aliasing. However, none of the mentioned possible oscillations reached a $\mathrm{SNR} > 4$. This holds true for both $B$ and $V$ light curves. The $U$ data set is characterized by very bad quality, not even $f_{1}$ could be detected unambiguously. If one takes the oscillation spectrum and the position in the HR diagram (which places the star into the $\beta$ Cephei instability domain) as indicators for the class of pulsating star, the conclusion is that this indeed is a $\beta$ Cephei type variable. To resolve the suggested frequencies clearly more data are required that do not suffer from offsets, allow the detection of signals below $1 \; \mathrm{mmag}$, and ideally feature a better spectral window. An example light curve and the frequency spectrum of the $V$ data are displayed in Fig. \ref{img:SBL456}.

\begin{figure*}
	\resizebox{\hsize}{!}{\includegraphics{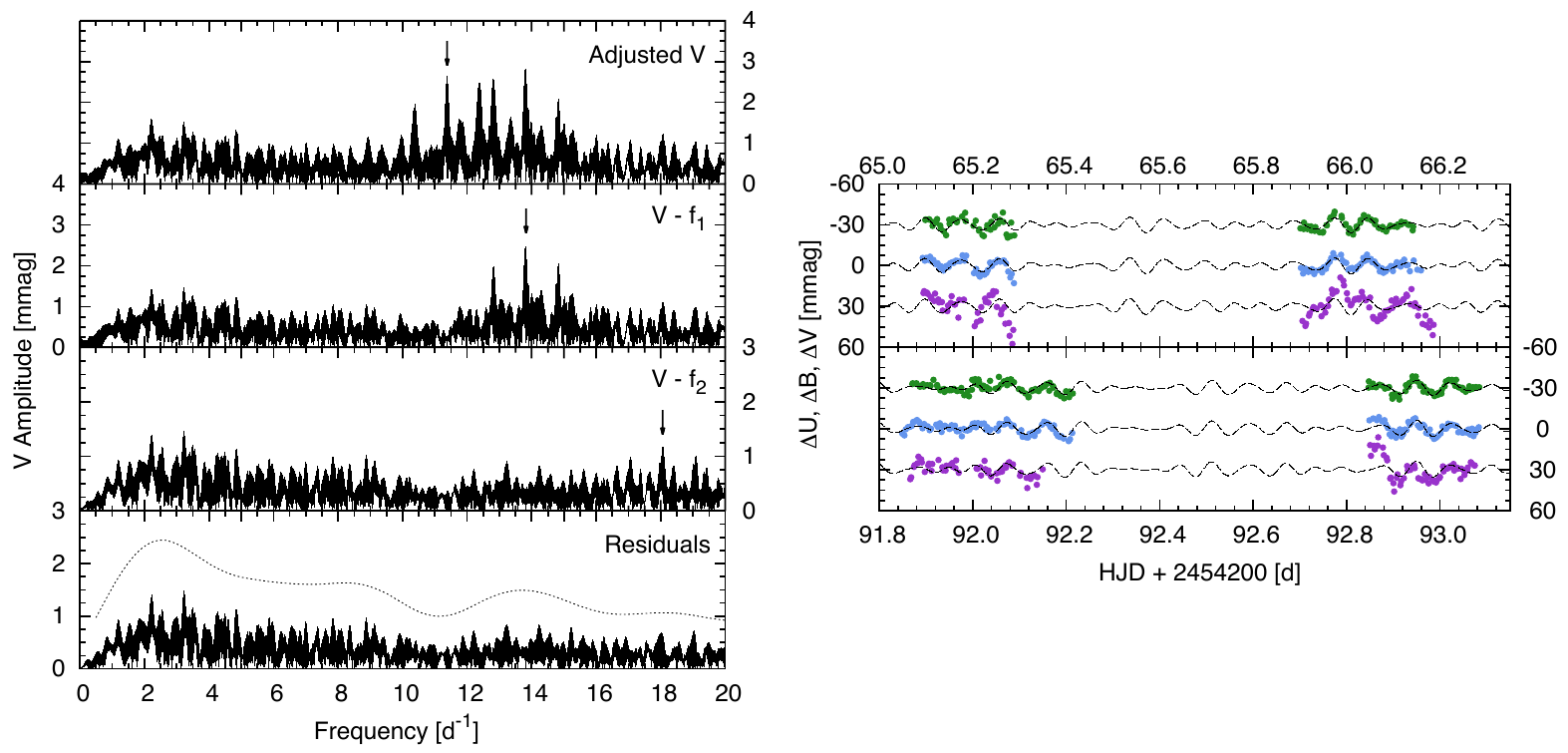}}
	\caption[SBL486]{The prewhitening steps in our frequency analysis of the $\beta$ Cephei star SBL486 are shown, as well as sample light curves with the multi-periodic fit. No mode identification could be performed due to unreliable data in $U$. $f_{1} = 11.3791 \; \mathrm{d^{-1}}$, $f_{2} = 13.8258 \; \mathrm{d^{-1}}$, $f_{3} = 18.058 \; \mathrm{d^{-1}}$. The errors range from $\sigma_{f_{1}}= 0.0007 \; \mathrm{d^{-1}}$ to $\sigma_{f_{3}}= 0.001 \; \mathrm{d^{-1}}$. All notations and markings are the same as in Fig. \ref{img:SBL226}.} 
	\label{img:SBL486} 
\end{figure*}

\textbf{SBL486} is listed as a single B0.5V star by PHC91 and RM98 and is a member of NGC 6231 according to both BVF99 and RCB97. It is classified as a $\beta$ Cephei pulsator and was examined by BE85 who found three significant frequencies, but they also mentioned that they could not exclude errors due to aliasing. SBL486 was also in the field of view in the ASK01 observations. They found five significant oscillations, two of which match (considering aliasing) the BE85 data. We were able to detect three significant frequencies on the basis of a daily adjusted set. Figure \ref{img:SBL486} shows our data analysis steps and some representative light curves. $f_{1} = 11.3791 \; \mathrm{d^{-1}}$ and $f_{3} = 18.058 \; \mathrm{d^{-1}}$ match the results of BE85, $f_{2} = 13.8258 \; \mathrm{d^{-1}}$ can be explained by an alias. The matching alias frequency of $f_{2}$ of BE85 ($f = 14.798$, separated by one daily and one monthly alias) was checked with the current data and resulted in a significantly lower peak in the Fourier spectrum and also produced larger residuals when fit together with the other oscillations to the entire light curve in all bands. Taking this into account the frequency found by BE85 was rejected. The data were again dominated by offsets on time scales spanning the entire run making it very difficult to determine low frequency variability. Nevertheless, visual inspection of the unadjusted light curves suggested a period larger than $2 \; \mathrm{d}$ which was found in the unadjusted (and not displayed) frequency spectrum at $f_{4} = 0.38 \; \mathrm{d^{-1}}$ with an amplitude of $3 \; \mathrm{mmag}$. This matches one of the low frequency variations detected by ASK01. However, the signal was not significant when fit to the whole run. When investigated on the basis of just a single month it exceeded the SNR threshold and the amplitude of this variation varied by a factor of three (from about two mmag up to six mmag) for the individual months. Furthermore, the second data set of the observations of BE85 in 1984 covered eleven nights and they claim that the residuals of the fit can be explained by single data point errors ($\sigma = 2.8 \; \mathrm{mmag}$) leaving no place for a low frequency variation with larger amplitudes. Consequently this frequency was not included in the final multi-periodic fit but is certainly worth mentioning. The data quality in U was particularly bad for this star, so that no frequency was found to be above the SNR threshold. The multi-periodicity, amplitude differences and also the calibrated Str\"omgren photometry confirm the $\beta$ Cephei nature of SBL486. In light of the poor data quality in $U$ where no single frequency was determined to exceed our criterion we did not attempt to identify pulsation modes.

\begin{figure}
	\resizebox{\hsize}{!}{\includegraphics{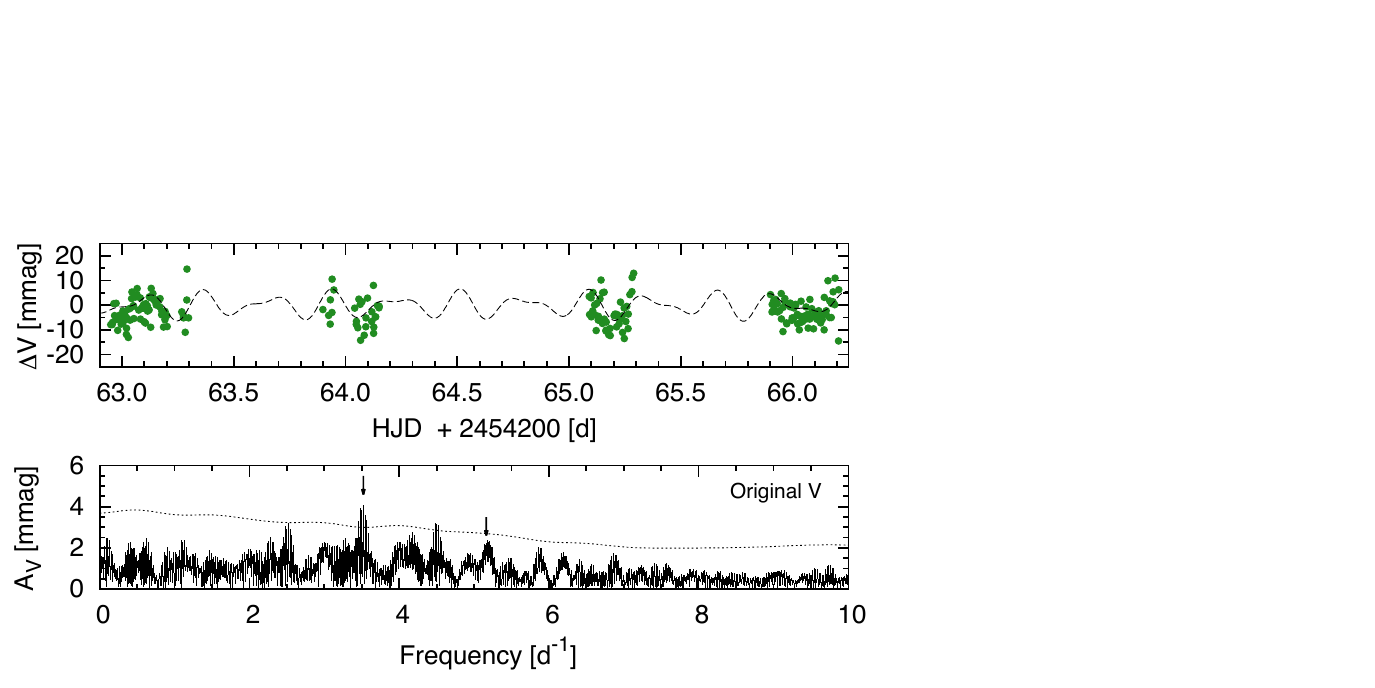}}
	\caption[SBL515]{Light curve in $V$ and frequency spectrum of the candidate $\beta$ Cephei star SBL515. Two frequencies were included in our results; $f_{1} = 3.5221 \; \mathrm{d^{-1}}$ and $f_{2} = 5.162 \; \mathrm{d^{-1}}$ with the errors $\sigma_{f_{1}}= 0.0006\; \mathrm{d^{-1}}$ and $\sigma_{f_{2}}= 0.001 \; \mathrm{d^{-1}}$.} 
	\label{img:SBL515}
\end{figure}

\begin{figure*}
	\resizebox{\hsize}{!}{\includegraphics{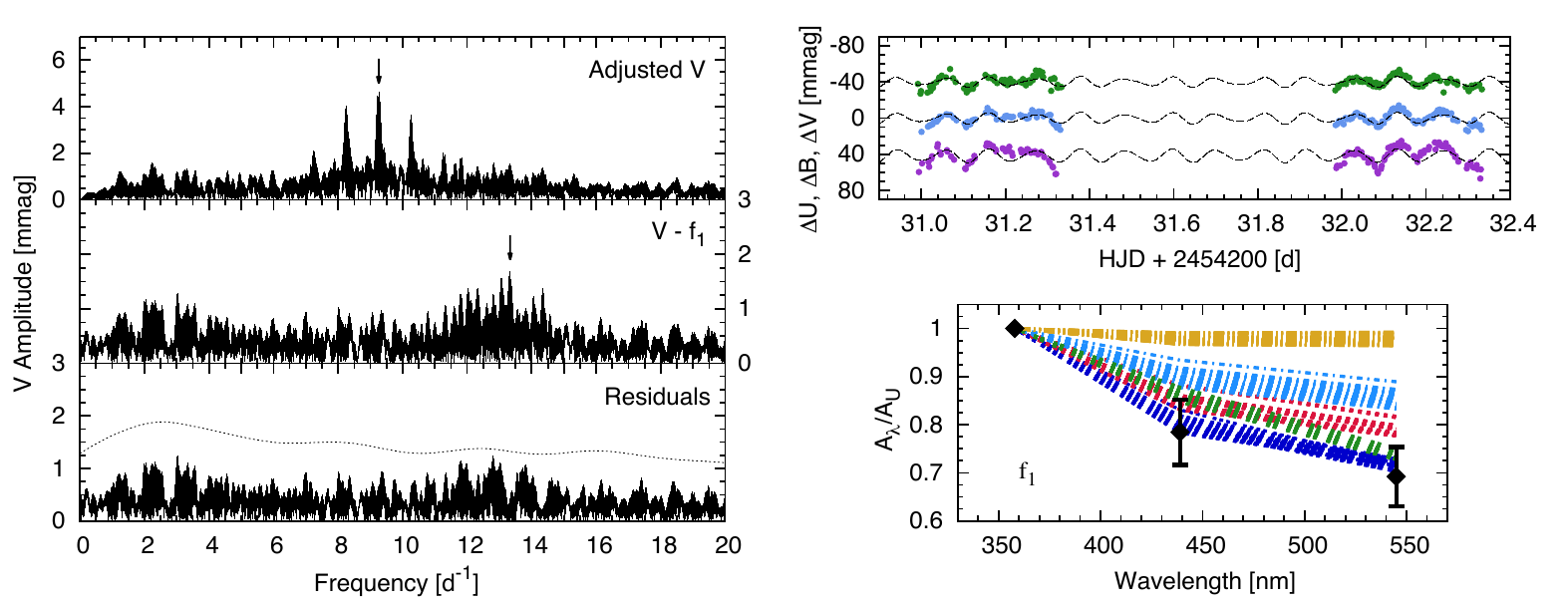}}
	\caption[SBL653]{Results for the $\beta$ Cephei star SBL653 with frequency spectra, light curves and an attempt to identify the pulsation mode of $f_{1} = 9.2647 \; \mathrm{d^{-1}}$. $f_{2} = 13.334 \; \mathrm{d^{-1}}$ with the errors $\sigma_{f_{1}}= 0.0003 \; \mathrm{d^{-1}}$ and $\sigma_{f_{2}}= 0.0008 \; \mathrm{d^{-1}}$. The amplitude ratios were calculated for masses from 15 to $16 \; M_{\odot}$. All notations and markings are the same as in Fig. \ref{img:SBL226}.} 
	\label{img:SBL653}
\end{figure*}

According to ASK01 \textbf{SBL515} is suspected to be a $\beta$ Cephei pulsator. These authors found several frequencies but were unable to give a final frequency solution due to aliasing. Furthermore they mentioned that this star is much fainter than other $\beta$ Cephei stars in NGC 6231 (see Fig. \ref{img:HRD}), implying that it is not a cluster member even though it is classified as a member by BVF99. 
Even though the measurements of this star are affected by one of the most luminous stars in the cluster and its PSF is merged with SBL512 we were able to determine two frequencies in the $V$ passband. $f_1 = 3.5221 \; \mathrm{d^{-1}}$ is a very good match with the ASK01 analysis; $f_2 = 5.162 \; \mathrm{d^{-1}}$ reaches $\mathrm{SNR} = 3.97$ and therefore does not reach our detection criterion. However, since this value also agrees with the time scales of the ASK01 data we consider it as an intrinsic oscillation. Figure \ref{img:SBL515} shows parts of the light curve and the frequency spectrum with the two detections marked as arrows. SBL515 is also the only star among the (confirmed and candidate) $\beta$ Cephei stars for which the determination of the stellar parameters was done with photometry from BL95. Its position in the HR diagram is shown in Fig. \ref{img:HRD}. The position which is significantly different from the other stars can be interpreted as an indicator that this star is not member of NGC 6231, or that it is not a $\beta$ Cephei star. In addition, the time scales of the variability of SBL 515 reported by ASK01 are considerably larger than those of the other $\beta$ Cephei stars in the cluster, suggesting a more evolved evolutionary state, strengthening the interpretation that it is not a cluster member.

\begin{figure*}
	\resizebox{\hsize}{!}{\includegraphics{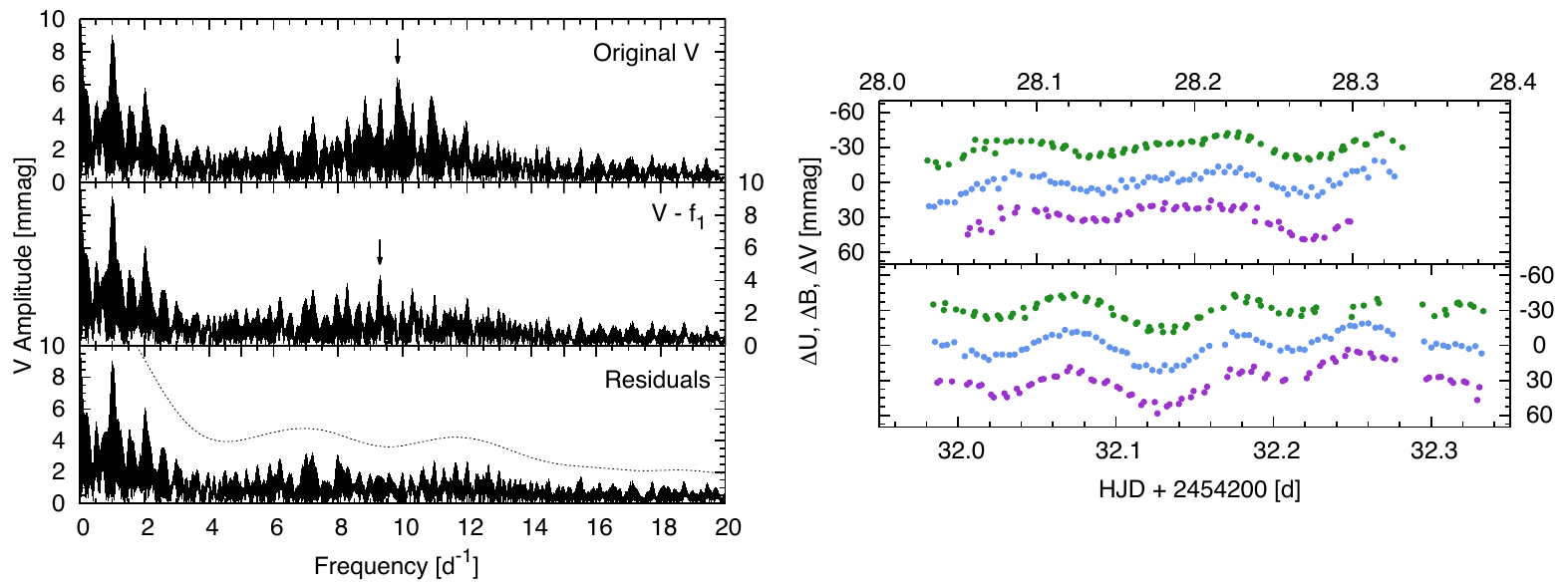}}
	\caption[SBL712]{Same as Fig. \ref{img:SBL226} where applicable, but for the $\beta$ Cephei star SBL712. No fit to the data is shown since zero-point offsets dominate the light curve. $f_{1} = 9.8481 \pm 0.0006 \; \mathrm{d^{-1}}$, $f_{2} = 9.2947 \pm 0.0008 \; \mathrm{d^{-1}}$.} 
	\label{img:SBL712}
\end{figure*}

\textbf{SBL653} was discovered to be a $\beta$ Cephei star by Ba83 and later its variability was further investigated by BS83 and ASK01. Both RCB97 and BVF99 considered it as a member of NGC 6231 within their adopted photometric criteria. PHC91 list it as a single star with spectral type B1V. The variability study of BS83 resulted in a simple frequency spectrum with only two dominant oscillations ($f_{1,BS83} = 9.27 \; \mathrm{d^{-1}}$, $f_{2,BS83} = 16.405 \; \mathrm{d^{-1}}$). ASK01 found a daily alias of the strongest signal, $f_{1,BS83}$, and also evidence for superimposed slower variability with a time scale of the order of $1 \; \mathrm{d}$. They claimed that this slower variation might not be intrinsic since fitting aliases for the strongest frequency results in the disappearance of this slow variability. 
With our data set we were able to detect two frequencies: $f_{1} = 9.2647 \; \mathrm{d^{-1}}$, which matches the signals in the other data sets and $f_{2} = 13.334 \; \mathrm{d^{-1}}$. The latter is above the SNR threshold in $B$, and close to it in $V$. Consequently it is considered as detected. No sign of slower variability is found and the second frequency listed by BS83 is not visible either. However, the residuals after prewhitening are not clean and suggest additional frequencies in the range $11 - 14 \; \mathrm{d^{-1}}$, where the strongest signal is found at $12.8 \; \mathrm{d^{-1}}$. No peak in this range has a SNR greater than 4. To determine more reliable amplitudes the light curves have been adjusted on a daily basis to eliminate zero point offsets. Only for $f_{1}$ a reliable amplitude could be determined in $U$ for which then it was possible to investigate the nature of the oscillation. Amplitude ratios were computed for masses ranging from 15 to $16 \; M_{\odot}$ as suggested by the evolutionary tracks in the HR diagram. The results favor a radial mode but the relative errors are too large to arrive at a final conclusion. In addition, stellar models with lower masses were also checked since RM98 find $13.13 \; M_{\odot}$. All models passing through our error box with masses ranging from $13$ to $16 \; M_{\odot}$ show the same behaviour regarding the amplitude ratios also suggesting an $l=0$ pulsation. Figure \ref{img:SBL653} displays a light curve sample, frequency spectra, and the mode identification.


\begin{figure*}
	\resizebox{\hsize}{!}{\includegraphics{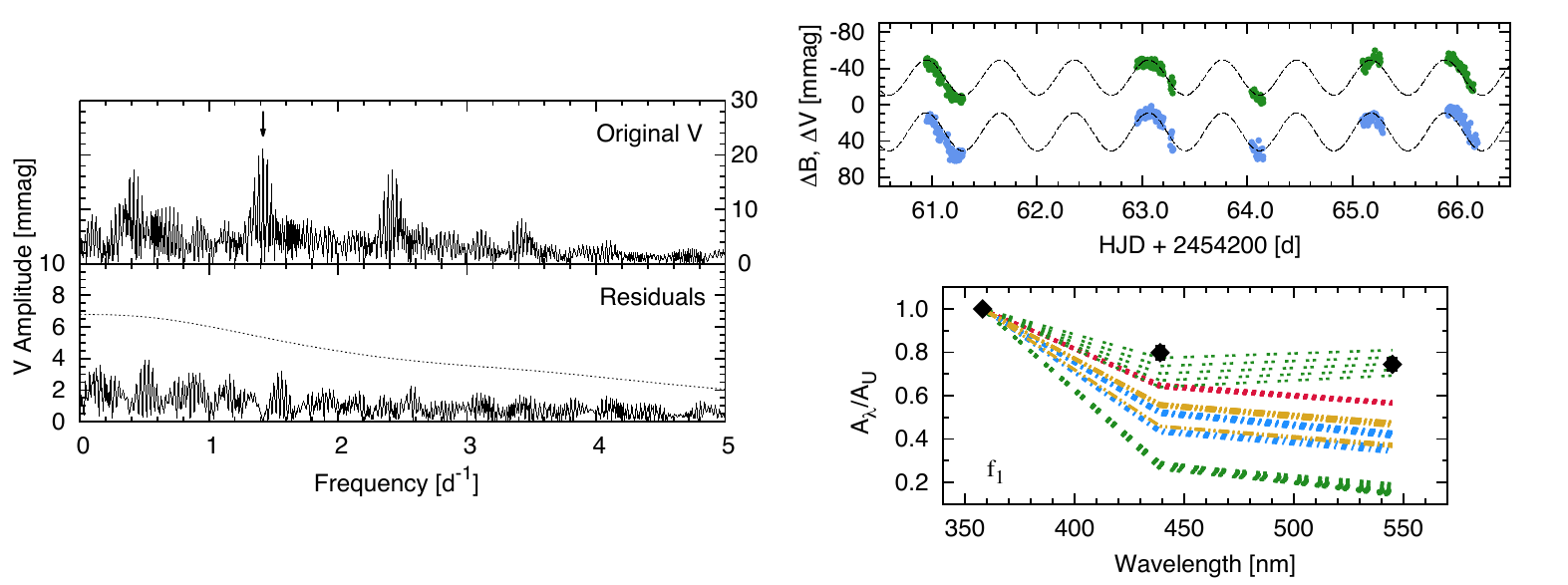}}
	\caption[SBL164]{Same as Fig. \ref{img:SBL226}, but for the newly discovered SPB variable SBL164. $f_{1} = 1.4183 \pm 0.0001 \; \mathrm{d^{-1}}$. The theoretical amplitude ratios display in the bottom right panel were computed for stellar masses between 3.5 and $4 \; M_{\odot}$. }
	\label{img:SBL164}
\end{figure*}

\textbf{SBL712} was the first discovered pulsating star in NGC 6231 by \cite{1979MNRAS.189..571S} who analyzed observations taken in 1972 and 1973. It is also an already classified $\beta$ Cephei type pulsator for which these early results indicate that the periods are not stable since the frequency spectra exhibited significant change during this time. It is listed as a single star with spectral type B1V by PHC91 who also determined it to be a member of NGC 6231. BS83 found similarities with the older data and detected six independent oscillations. Apart from aliasing problems the frequencies determined by ASK01 match well with the results of BS83. Our analysis reveals a very complex frequency spectrum, but only two frequencies are detected above the adopted SNR threshold ($f_{1} = 9.8481 \; \mathrm{d^{-1}}$ and $f_{2} = 9.2947 \; \mathrm{d^{-1}}$). The two frequencies we could detect match very well the older results. In the residual frequency spectrum evidence for additional frequencies is present but only one additional signal detected by BS83 ($f = 10.98 \; \mathrm{d^{-1}}$) comes close to $\mathrm{SNR} = 4$. For other frequencies mentioned by BS83 such as $f = 9.62 \; \mathrm{d^{-1}}$ and $f = 8.23 \; \mathrm{d^{-1}}$ a daily adjusted data set helps to improve the quality but still these frequencies do not exceed the SNR threshold. Moreover, the low-frequency variation found by BS83 at $1.44 \; \mathrm{d^{-1}}$ does not appear at all in our analysis. Adjusting the data has been considered to improve the reliability of the analysis and possibly detect more frequencies but no further signal was detected. In $U$ the situation is worse: no frequency is deemed significant and for the two present signals in the other two bands, daily aliases have a larger amplitude. This is attributed to large amounts of corrupted data in $U$ which had to be deleted before the analysis to guarantee a reliable result. Figure \ref{img:SBL712} shows the frequency spectra for an unadjusted data set along with sample light curves of SBL712. Since the data are dominated by zero-point offsets the fit to the original data does not match the light curves well and is not displayed here. No attempt to constrain the spherical degrees of the pulsation modes has been made.

\subsection{Slowly pulsating B stars}

\begin{figure*}
	\resizebox{\hsize}{!}{\includegraphics{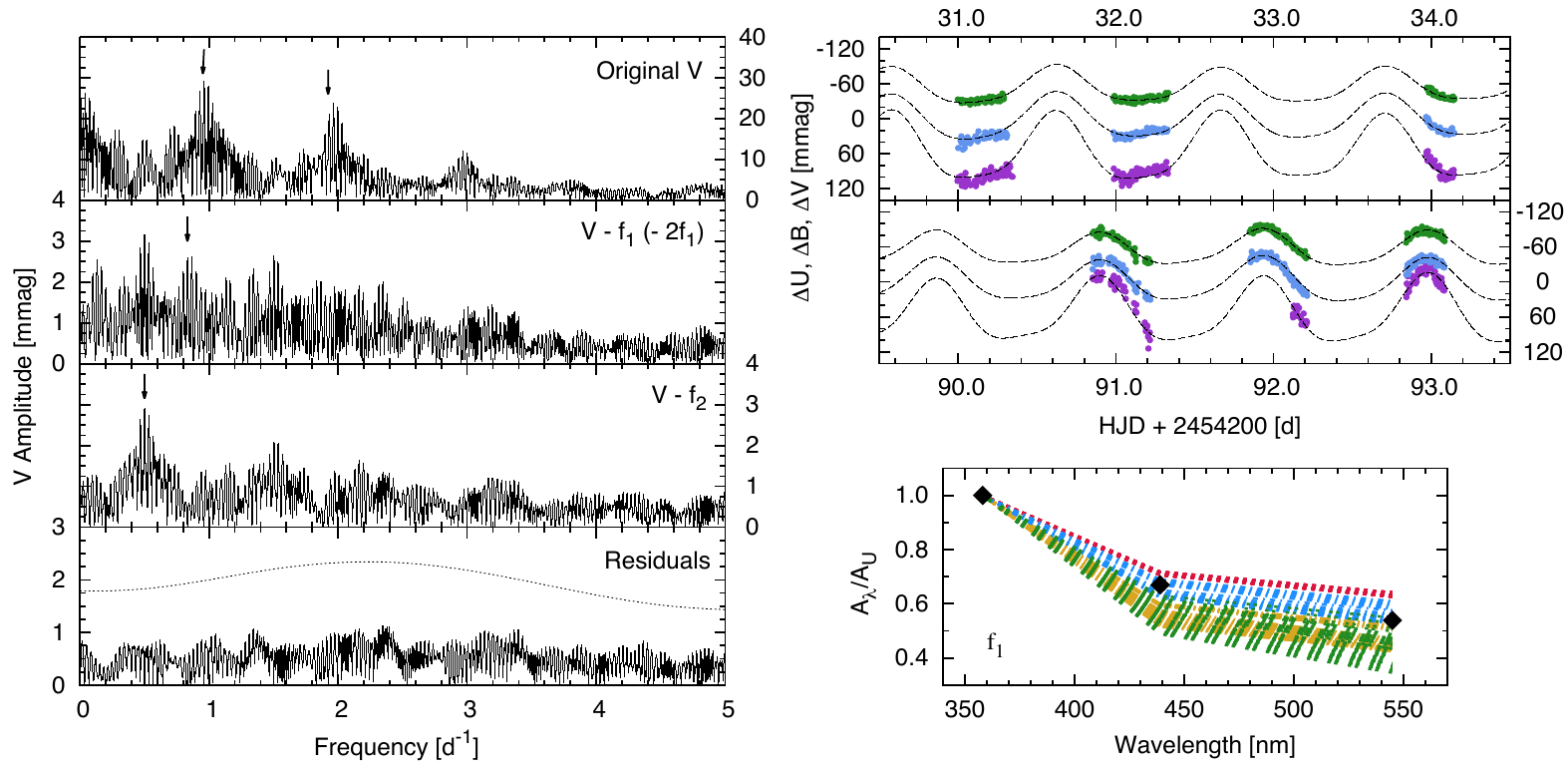}}
	\caption[SBL210]{Frequency spectra, light curves and amplitude ratios for the newly discovered SPB variable SBL210. Notations and markings are the same as in Fig. \ref{img:SBL226}. The amplitude ratios were calculated for masses between 4.5 and $5.5 \; M_{\odot}$. The detected frequencies are: $f_{1} = 0.9615 \; \mathrm{d^{-1}}$,  $f_{2} = 0.8332 \; \mathrm{d^{-1}}$,  $f_{3} = 0.5004 \; \mathrm{d^{-1}}$, and $2f_{1}$. The errors in the frequencies range from $\sigma_{f_{1}} = 0.00005 \; \mathrm{d^{-1}}$ to $\sigma_{f_{3}} = 0.0006 \; \mathrm{d^{-1}}$.} 
	\label{img:SBL210}
\end{figure*}

\textbf{SBL164} is a newly discovered variable star. It is classified as a member of NGC 6231 by RCB97 who also photometrically derived spectral type B6-8V. RM98 found SBL164 to be a single star. Our frequency analysis showed that the light curve is dominated by a single frequency at $f_{1} = 1.4183 \; \mathrm{d^{-1}}$, but we also suspect additional oscillations in the low frequency range with amplitudes up to $4 \; \mathrm{mmag}$ due to the shape of the residual frequency spectrum and the imperfect (but very good) fit of the single frequency solution to the light curves. Since multi-periodicity is an important indicator for pulsations as the origin of the light output variations here one has to rely on the photometric amplitude ratios, the time scales of the variability and the position in the HR diagram. These facts point towards the conclusion that SBL164 is a new SPB star in NGC 6231. The mode identification turned out to be difficult for various reasons. The calibration of the Str\"omgren photometry puts SBL164 below the ZAMS in the $T_{eff} - \mathrm{log} \, g$ diagram, which is due to an unphysically large log g value. For this reason - and only for this star - the error in surface gravity has been increased to $\sigma_{\mathrm{log} \, g} = 0.5 \; \mathrm{dex}$. This also suggests that the position in the HR diagram might be not correct, pointing towards a different stellar mass. RM98 find a mass of $4.07 \; M_{\odot}$ whereas our comparison of Str\"omgren photometry to the evolutionary tracks in the HR diagram suggests a mass of about $3.5 \; M_{\odot}$. In Fig. \ref{img:SBL164} we show frequency spectra, light curves and amplitude ratios for oscillations of stars with masses in the range from $3.5$ to $4 \; M_{\odot}$ (no excited modes were found for $3 \; M_{\odot}$). The two distinct branches observed for the $l > 1$ modes is explained by the incremental uncertainty in mass. Here the observations are only compatible with an $l=3$ mode of a $4 \; M_{\odot}$ star.

\begin{figure*}
	\resizebox{\hsize}{!}{\includegraphics{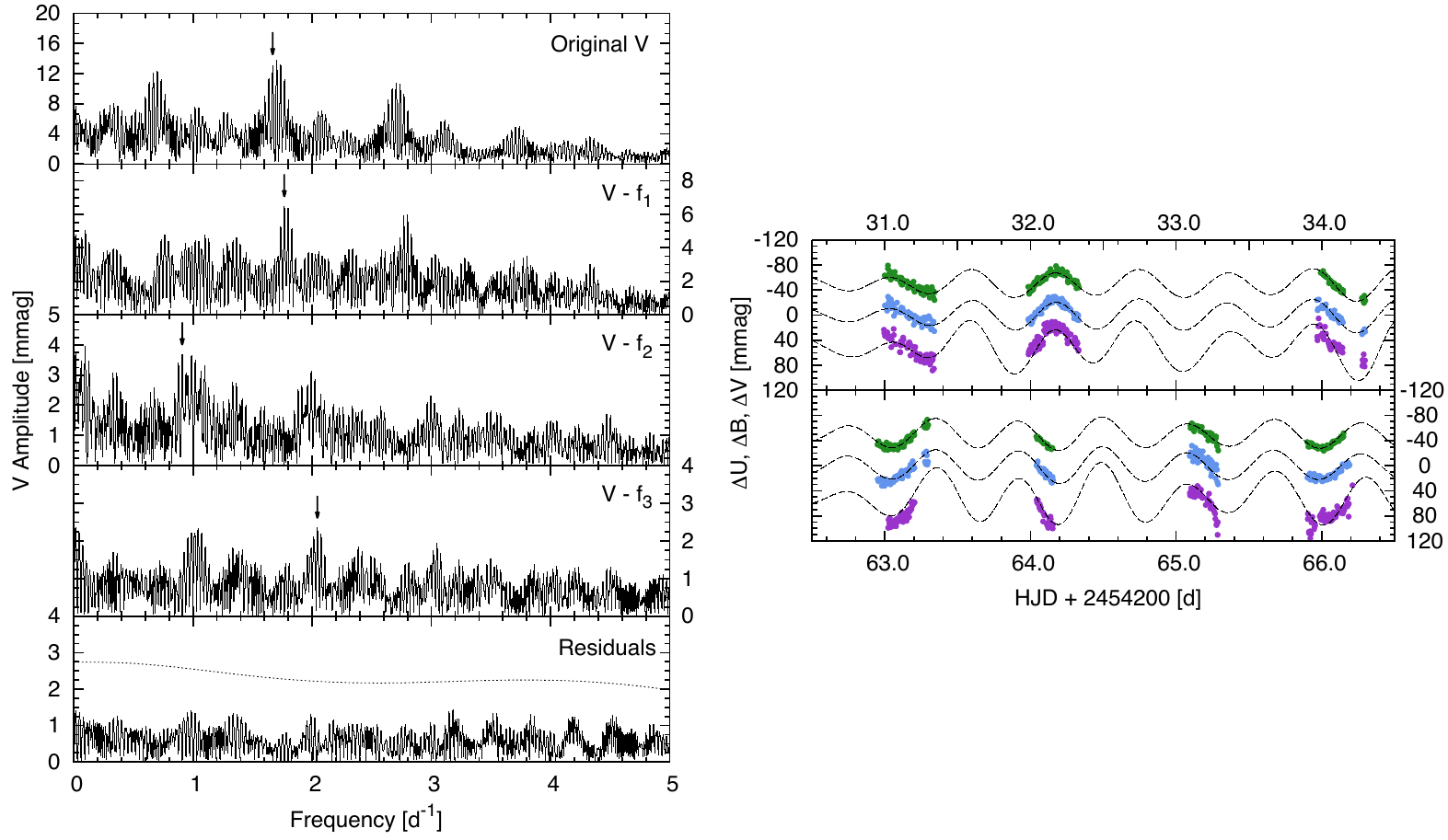}}
	\caption[SBL275]{Same as Fig. \ref{img:SBL226} where applicable, but for the SPB star SBL275. $f_{1} = 1.6686 \; \mathrm{d^{-1}}$, $f_{2}= 1.7662 \; \mathrm{d^{-1}}$, $f_{3} = 0.9093 \; \mathrm{d^{-1}}$, and $f_{4}= 2.0437 \; \mathrm{d^{-1}}$, with errors ranging from $\sigma_{f_{1}}= 0.0001 \; \mathrm{d^{-1}}$ to $\sigma_{f_{4}}= 0.0006 \; \mathrm{d^{-1}}$.} 
	\label{img:SBL275}
\end{figure*}

\textbf{SBL210} has not been subject to a variability study so far. It is classified as a member of NGC 6231 by RCB97 who also list it as a candidate Ap star on the basis of photometric boxes. However, such a classification is clearly inconsistent with its Str\"omgren colours that are typical for a B star. In addition, \citetads{SG05} also attribute spectral type B to SBL210. Therefore we treated it as a chemically normal B-type star for the determination of effective temperature and luminosity. Our analysis of the data revealed three independent frequencies and one harmonic ($f_{1} = 0.9615 \; \mathrm{d^{-1}}$, $f_{2} = 0.8332 \; \mathrm{d^{-1}}$, $f_{3} = 0.5004 \; \mathrm{d^{-1}}$, and $2f_{1}$). Only the strongest signal and its first harmonic are significant in all data sets. In the $V$ data two more frequencies are found to be above the SNR threshold. $f_{3}$ has to be considered with caution since we found this frequency in some other data sets as well and it might be explained by zero point offsets. However, this star does not show significant offsets during the entire two months period of the observing run, therefore we think that $f_{3}$ is an intrinsic variation. We found the data in $U$ to be much more reliable compared to other stars, because the data for this star showed less systematic deviations than for others and because its amplitudes are comparably high. Therefore the whole light curve could be analysed. In Fig. \ref{img:SBL210} examples of the light curves, the corresponding fits, as well as original and residual frequency spectra are shown. Ra98 derive a mass of $5.06 \; M_{\odot}$ for SBL210 which matches very well with calculated luminosity, effective temperature and overlaid evolutionary tracks as visible in Fig. \ref{img:HRD}. On the basis of multi-periodicity, amplitude differences in all bands, the time scales of the variability and its position in HR diagram, we classify SBL210 as an SPB variable. Only $f_{1}$ with its large amplitude could be examined closer due to unreliable data in $U$ (which affects the smaller amplitudes). The results of an attempt to identify the spherical degree $l$ of the associated mode are also visible in Fig. \ref{img:SBL210}. The grids for stars with masses between 4.5 and $5.5 \; M_{\odot}$ have been checked with the already mentioned parameters for NGC 6231 within the error boxes of the Str\"omgren photometry for the degrees $l=1$ to $4$. Again the uncertainty in mass makes it very difficult to determine the spherical degree. From a visual inspection we suggest that $f_{1}$ is an $l=2$ mode.

\begin{figure*}
	\resizebox{\hsize}{!}{\includegraphics{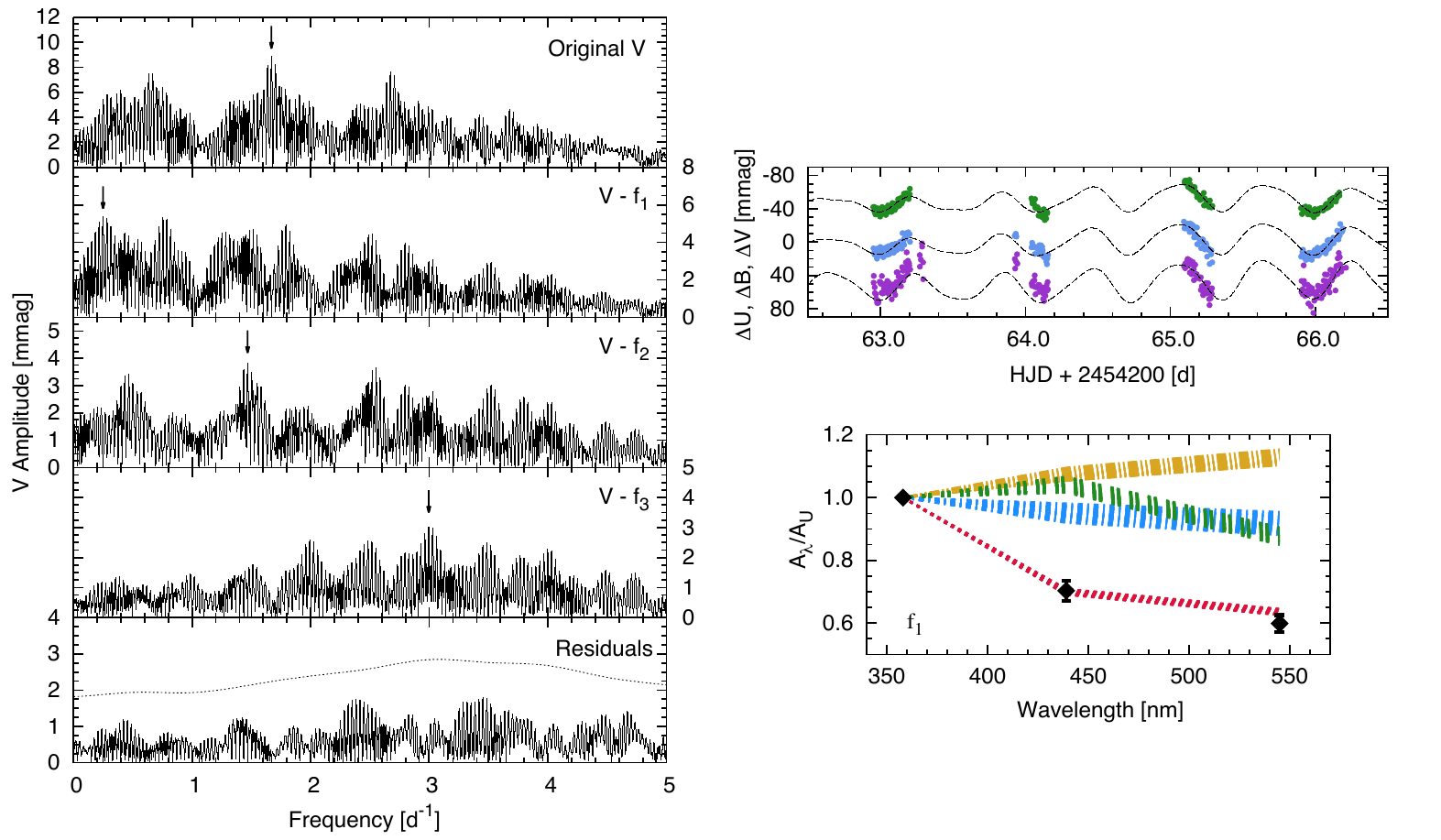}}
	\caption[SBL283]{Same as Fig. \ref{img:SBL226}, but for the newly discovered SPB star SBL283. The detected frequencies are: $f_{1} = 1.6694 \; \mathrm{d^{-1}}$, $f_{2}= 0.249 \; \mathrm{d^{-1}}$, $f_{3} = 1.4688 \; \mathrm{d^{-1}}$, and $f_{4}= 2.9952 \; \mathrm{d^{-1}}$, with errors from $\sigma_{f_{1}} = 0.0002 \; \mathrm{d^{-1}}$ to $\sigma_{f_{4}} = 0.0006 \; \mathrm{d^{-1}}$. The theoretical amplitude ratios shown here were calculated for a $5 \; M_{\odot}$ star. If one allows for an uncertainty in mass also only the $l=1$ modes match the observations well.}
	\label{img:SBL283}
\end{figure*}

\begin{figure*}
	\resizebox{\hsize}{!}{\includegraphics{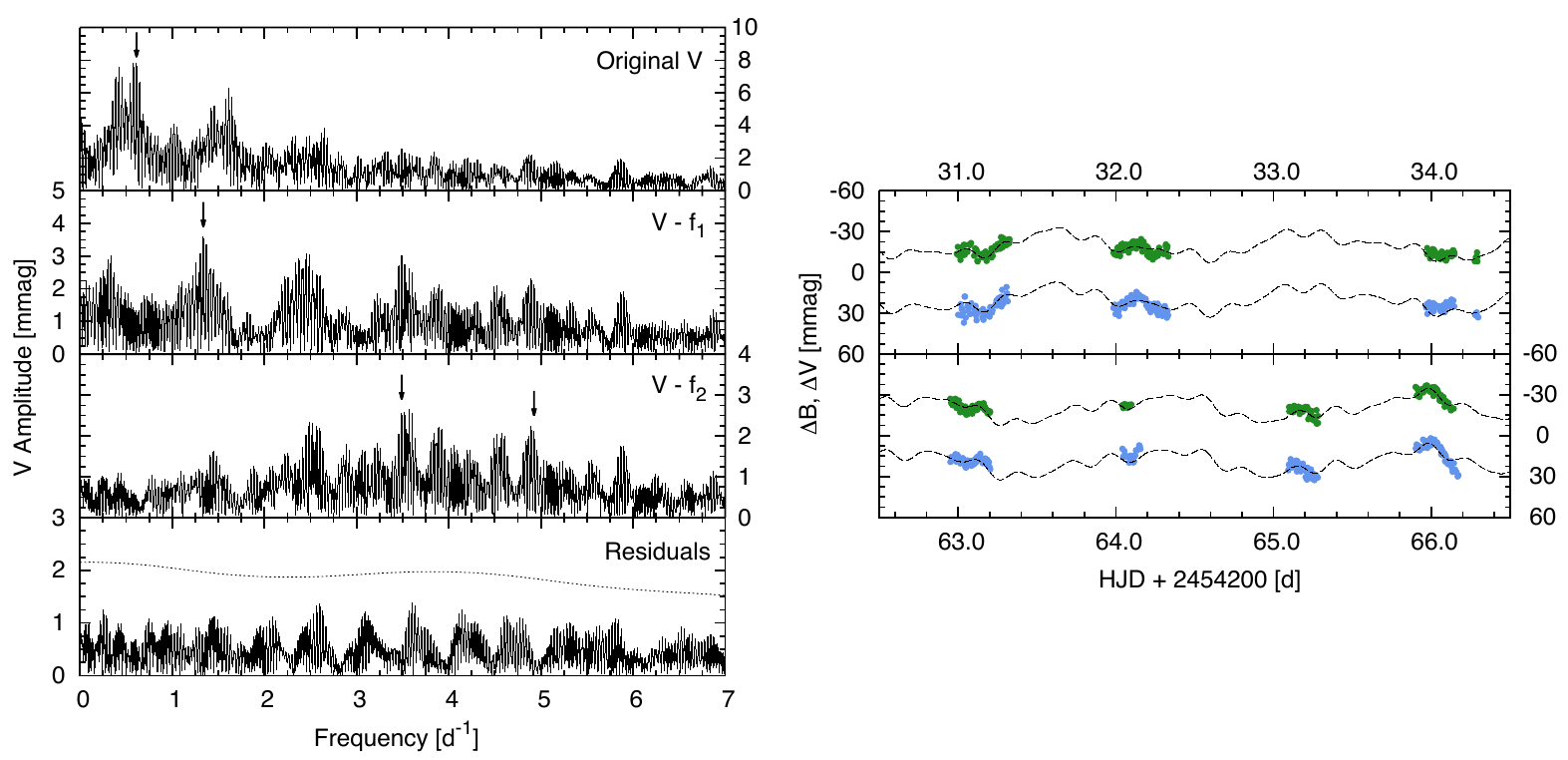}}
	\caption[SBL317]{Same as Fig. \ref{img:SBL226}, but for the newly discovered SPB star SBL317. $f_{1} = 0.6133 \; \mathrm{d^{-1}}$, $f_{2}= 1.3381 \; \mathrm{d^{-1}}$, $f_{3} = 3.4903 \; \mathrm{d^{-1}}$, and $f_{4}= 4.9252 \; \mathrm{d^{-1}}$, with errors ranging from $\sigma_{f_{1}} = 0.0002 \; \mathrm{d^{-1}}$ to $\sigma_{f_{4}} = 0.0007 \; \mathrm{d^{-1}}$.} 
	\label{img:SBL317}
\end{figure*}

\textbf{SBL275} was discovered to be a variable star by ASK01 who also suggested a possible SPB nature. According to RCB97 this star is a cluster member and has spectral type B4-5V as derived photometrically. The variability study by ASK01 resulted in two significant frequencies at $1.65 \; \mathrm{d^{-1}}$ and $0.74 \; \mathrm{d^{-1}}$. Our analysis revealed a more complex oscillation spectrum. The periodogram for the original data set in V showed maximum power at $1.7 \; \mathrm{d^{-1}}$. Prewhitening this oscillation and subsequent fitting of further significant frequencies did not result in a satisfactory fit to the light curve. Therefore a careful examination of the light curve was conducted which showed the possibility of beating of two close frequencies. For the determination of these frequencies which are responsible for the beating phenomenon, several different pairs and their aliases were checked. Those chosen in the end mark the pair which minimized the residuals. The best match was found at $f_{1} = 1.6686 \; \mathrm{d^{-1}}$ and $f_{2}= 1.7662 \; \mathrm{d^{-1}}$. Due to aliasing these two oscillations produced the largest peak in the frequency spectrum at $1.7 \; \mathrm{d^{-1}}$. This also matches well with the results of ASK01 when aliasing is taken into account. The search for additional oscillations was very difficult because the $V$ and $B$ light curves suggested different frequencies after prewhitening $f_{1}$ and $f_{2}$. In $V$ two more frequencies ($f_{3} = 0.9093 \; \mathrm{d^{-1}}$ and $f_{4}= 2.0437 \; \mathrm{d^{-1}}$) turned out to be significant which could not be found in $B$. The third strongest signal in $B$ is found at $1.33 \; \mathrm{d^{-1}}$ which in turn is not visible in $V$. Since no distinction is possible which of these frequencies is intrinsic only the frequencies in $V$ for the remaining signals are included in our final solution since the light curve in this filter exhibits the least scatter and seems to be most reliable throughout our entire data analysis. Also no indication of more rapid oscillations is present above the noise level of the data. A light curve sample and the individual steps in the frequency analysis are displayed in Fig. \ref{img:SBL275}. 
Together with the results from ASK01 it is clear that this star shows variability in the range typical for SPB stars. An important indicator to exclude rotational and binary effects is multi-periodicity which is clearly present here. Also amplitude differences are taken as an indicator for pulsations as origin of the variability. Furthermore calculating the stellar luminosity and effective temperature puts SBL275 into the instability domain of SPB type pulsators. The indicated mass of about $5 \; M_{\odot}$ compares well with the results from RM98 who find $5.17 \; M_{\odot}$. All this evidence taken together, we declare SBL275 to be an SPB type pulsator. Setting constraints on the underlying modes turned out to be very difficult for this star, since for the given effective temperature and surface gravity combined with the uncertainty in mass (4.5 to $5.5 \; M_{\odot}$) far too many model frequencies were excited to draw any conclusion.

\textbf{SBL283} has not been classified as a variable star so far and is listed as member of NGC 6231 by both BVF99 and RCB97. PHC91 give only spectral type B without further subclass or luminosity class. Our frequency analysis revealed the richest and most complex spectrum compared to the other SPB stars in the field giving rise to many individual oscillations in the low frequency regime. For this star prewhitening and subsequent fitting of frequencies with the largest amplitude in the periodogram led to an imperfect fit. For this reason the frequencies were selected manually. The searching, fitting and prewhitening after some steps led to eleven significant frequencies in the end with residuals far below $1 \; \mathrm{mmag}$ in the amplitude spectrum. Since the data quality is not unusually different from any other star with comparable apparent brightness, we fear that many of these formally significant frequencies are not intrinsic. Moreover, the residuals after prewhitening the eleven frequencies were below the single data point error, also suggesting an implausible result. Therefore only four frequencies were included in the final fit ($f_{1} = 1.6694 \; \mathrm{d^{-1}}$, $f_{2}= 0.249 \; \mathrm{d^{-1}}$, $f_{3} = 1.4688 \; \mathrm{d^{-1}}$, and $f_{4}= 2.9952 \; \mathrm{d^{-1}}$). The residual frequency spectrum and the still imperfect fit to the light curve can be interpreted as indications for additional variability. Figure \ref{img:SBL283} shows the frequency spectra for the original data set after prewhitening 1-3 frequencies and the residuals for the $V$ data. Note here, that the 11 frequencies mentioned above do not reach the $\mathrm{SNR} = 4$ limit in the plot of the residuals because they only become significant after pre-whitening. Also displayed are light curves in $U$, $B$, and $V$ with the corresponding calculated fit. SBL283 is clearly multiperiodic, shows different amplitudes in the three bands, is positioned in the SPB instability strip and shows variability at typical time scales for these stars. Thus we classify it as an SPB type variable. Reliable amplitude values could only be determined for $f_{1}$ in all three bands. For the mode identification we considered models with masses ranging from 4 to $5 \; M_{\odot}$. Also for this star a very large amount of excited model frequencies was found in the observed range. All $l=1$ modes showed consistent behaviour to which the observations are very well matched. For $l>1$ we observed the opposite behaviour suggesting a broad range of amplitude ratios. However, no mode with $l>1$ was able to reproduce the observed values. For demonstration purposes we only show the mode identification for a $5 \; M_{\odot}$ model in Fig. \ref{img:SBL283}.

\textbf{SBL317} has been suspected to be a variable star by BVF99 on the basis of the difference between their measurements and PHC91's photometry. RCB97 find it to be a member of NGC 6231, BVF99 however list the star as a non-member. The latter result is likely an error in the catalog since (a) this star is used in the discussion of their work as a member and (b) the calibration of the Str\"omgren photometry indicates a very similar color excess $E_{b-y}$ compared to other members. For these reasons it can safely be assumed that SBL317 is a member of NGC 6231. The spectral type is listed as B3 without luminosity class in PHC91 and RCB97 photometrically find B2-3V. Our variability analysis resulted in three significant frequencies. A fourth one has also been included in the final list since it is apparent in both $V$ and $B$ data, comes close to the SNR threshold and significantly improves the fit to the data. These are: $f_{1} = 0.6133 \; \mathrm{d^{-1}}$, $f_{2}= 1.3381 \; \mathrm{d^{-1}}$, $f_{3} = 3.4903 \; \mathrm{d^{-1}}$, and $f_{4}= 4.9252 \; \mathrm{d^{-1}}$. There was no need to adjust the data for this target because the monthly offsets were lower than the amplitudes of the oscillations and the offsets were found to be only slightly above the single data point error. Unfortunately, the data in $U$ resulted in one of the worst light curves with many unreliable parts so that not even a single frequency could be found in this passband. For this reason also no attempt to identify pulsation modes was made. Figure \ref{img:SBL317} shows example light curves and our frequency analysis steps in the periodograms. Since most $U$ data had to be deleted no light curve is shown representing this passband. This star is certainly a very interesting object. Not only does it show variability on different time scales, it is also located near the region where SPB and $\beta$ Cephei instability strips overlap in the HR diagram. On the basis of multi-periodicity and since the low frequency variations dominate the light curve and no definite conclusion can be made on the more rapid variability, we classify SBL317 as an SPB star.

\textbf{SBL394} has been reported as a possible SPB star by ASK01, but they were not able to arrive at a definite conclusion regarding the frequencies due to 
aliasing. \cite{1979AJ.....84.1020G} established its spectral type to be B2IV-V, BVF99 and RCB97 both classified it as a member of NGC 6231. Furthermore, it is a possible binary according to RM98. Since the amplitudes of the potential multiple periods are very small for this star, and significant offsets are present, the data have been adjusted by subtracting the mean magnitude for each month. In this way possible SPB variability is not corrupted and the effects of the offsets are minimized. Only one frequency could be determined to be significant at $f_{1} = 1.1298 \; \mathrm{d^{-1}}$ with an amplitude of about $3 \; \mathrm{mmag}$. Possible additional variability might be at $1.33 \; \mathrm{d^{-1}}$ which comes close to a value determined by ASK01. Altogether, only the data in $V$ revealed a significant frequency and it cannot be safely ruled out that this arises due to binarity. Visually inspecting the light curves also suggests the presence of more rapid variability, but we were not able to determine such pulsations with our frequency analysis. If such variability was present, it certainly would have an amplitude below $1 \; \mathrm{mmag}$. However, even though only one frequency in one filter was detected, this star is considered as an SPB variable since (a) it shows signs of multi-periodicity even though no additional oscillations could be detected (which can be blamed to the data offsets), (b) ASK01 detected multi-periodicity, and (c) it is found near other SPB stars in the theoretical HR and color-color diagrams.

\begin{figure}
	\resizebox{\hsize}{!}{\includegraphics{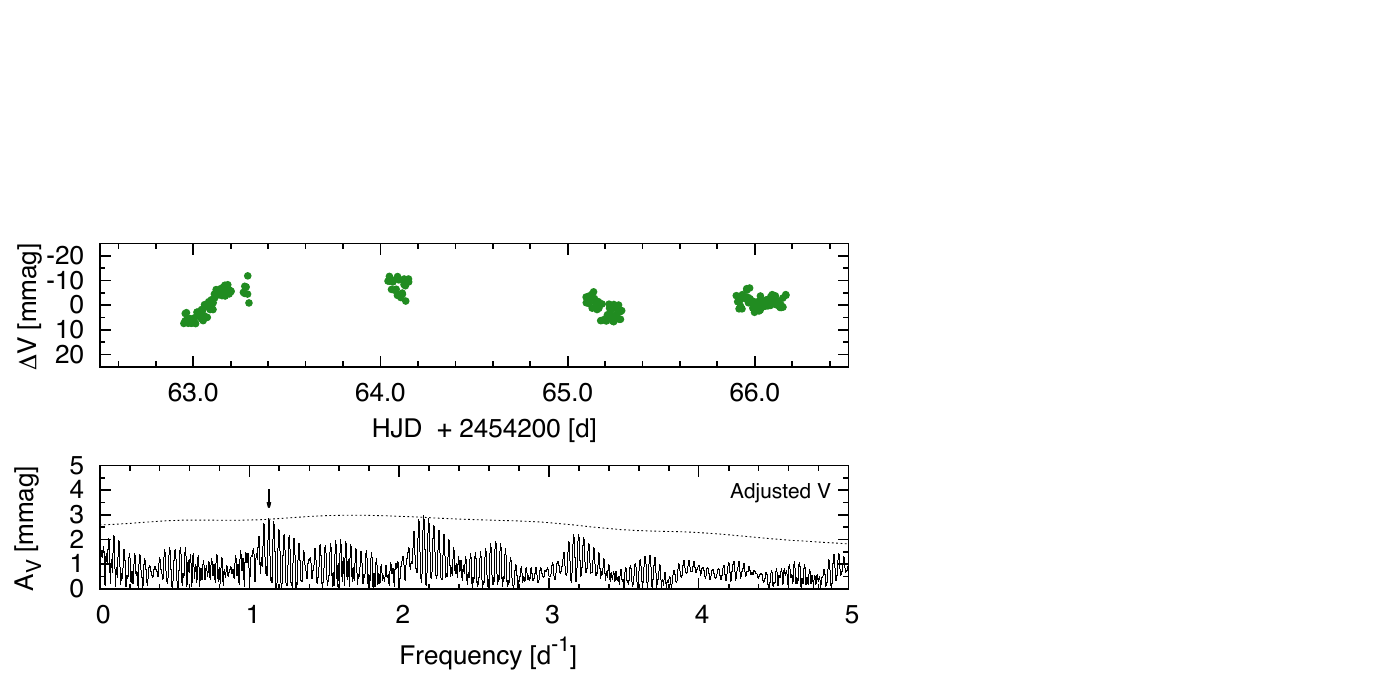}}
	\caption[SBL394]{Light curve in $V$ and frequency spectrum of the SPB star SBL394. Only one signal was significant at $f_{1} = 1.1298 \pm 0.0005 \; \mathrm{d^{-1}}$.} 
	\label{img:SBL394}
\end{figure}

\begin{figure*}
	\resizebox{\hsize}{!}{\includegraphics{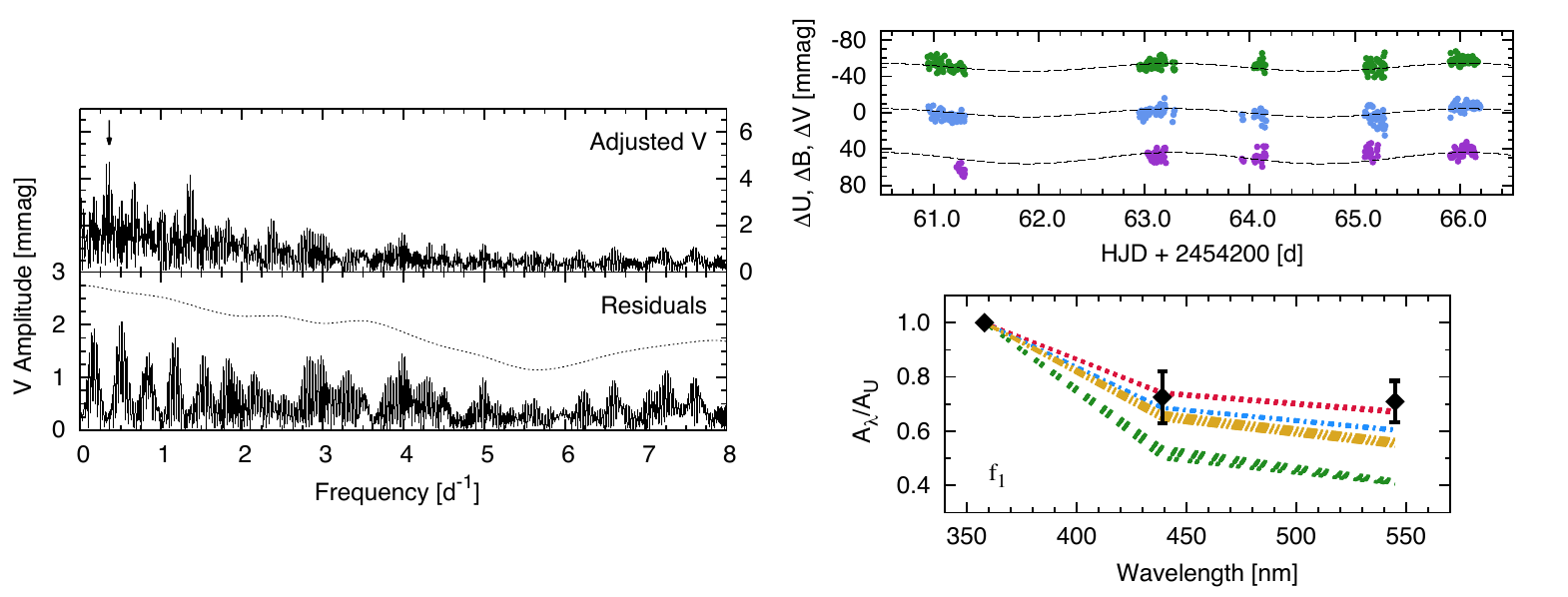}}
	\caption[SBL461]{Same as Fig. \ref{img:SBL226} where applicable, but for the SPB star SBL461. Only one frequency was found to be significant at $f_{1} = 0.3597 \pm 0.0005 \; \mathrm{d^{-1}}$. The theoretical amplitude ratios were calculated for models with masses ranging from 6.5 to $7.5 \; M_{\odot}$.} 
	\label{img:SBL461}
\end{figure*}

\begin{figure*}
	\resizebox{\hsize}{!}{\includegraphics{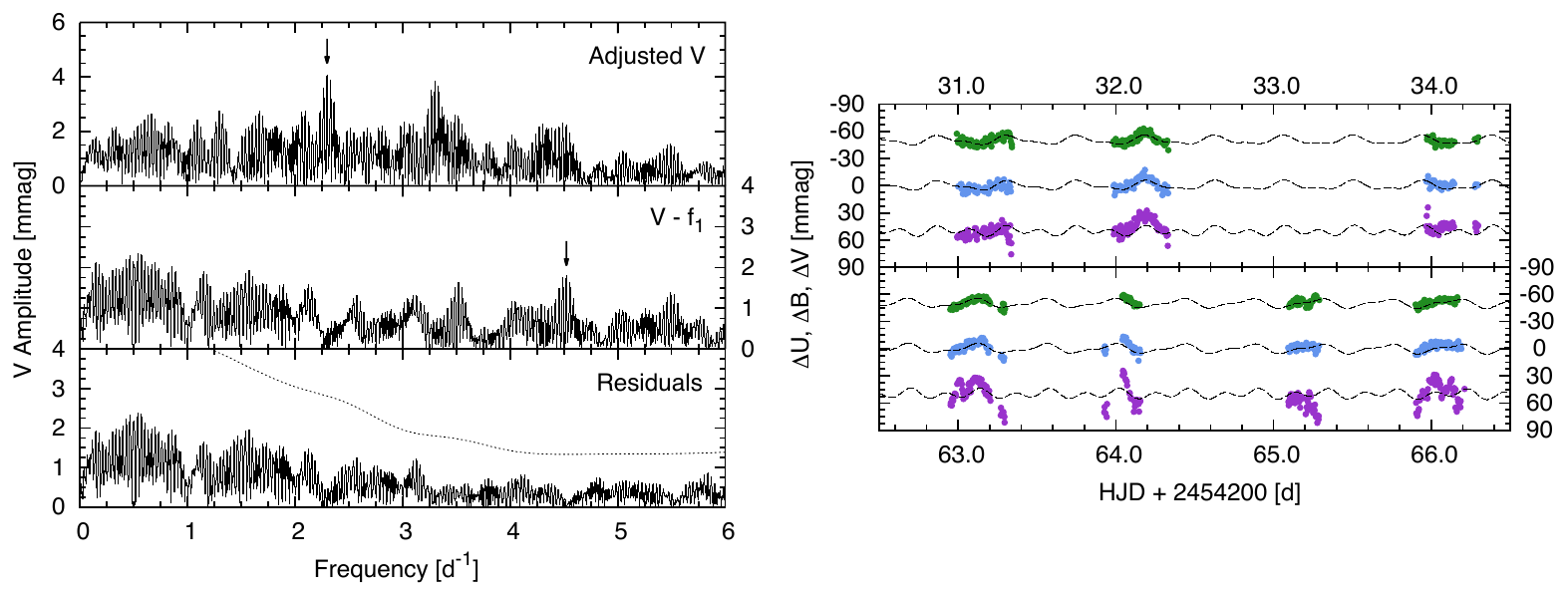}}
	\caption[SBL480]{Same as Fig. \ref{img:SBL226}, but for the newly detected SPB candidate SBL480. $f_{1} = 2.2981 \pm 0.0004 \; \mathrm{d^{-1}}$, $f_{2} = 4.5224 \pm 0.0008 \; \mathrm{d^{-1}}$.} 
	\label{img:SBL480}
\end{figure*}

\textbf{SBL461} is another SPB candidate as stated by ASK01. This star is found to be a member of NGC 6231 by RCB97 and BVF99 and has spectral type B1.5V according to PHC91. The question about its binary nature has yet to be settled since GM01 find it to be a single star, RM98, however, list it as a binary. It is interesting to mention here that PHC91 found a significantly different value for the $V$ magnitude when compared to work by BVF99, SBL98 and BL95 which might be due a blended PSF of SBL461 and SBL459. We also checked SBL459 for photometric variability and found signals near $2.4 \; \mathrm{d}^{-1}$ and $14.8 \; \mathrm{d}^{-1}$. However, due to the blending of these objects and changing ambient conditions/camera positions we cannot conclude that SBL459 is indeed a variable star on its own. For SBL461 ASK01 find evidence for two frequencies at $0.4 \; \mathrm{d^{-1}}$ and $0.5 \; \mathrm{d^{-1}}$ respectively but explicitly mention that their results need to be checked. The Fourier analysis with our data set resulted in only one significant frequency. Even though the light curves are well described by this single oscillation, the residuals are not clean suggesting additional variability at the mmag level. Due to the small amplitudes involved, the identification has been done by analyzing a monthly adjusted light curve to suppress very low-frequency variation. In all three bands the oscillation with the largest amplitude is found at $f_{1} = 0.3597 \; \mathrm{d^{-1}}$ which is reasonably close to the solution of ASK01 when aliasing is taken into account. An additional significant frequency is present in the $V$ data which can be found at $0.17 \; \mathrm{d^{-1}}$ but does not fit $f_{1}/2$ well. However, this might have been introduced by the length of a run in one month. Since variability at $0.17 \; \mathrm{d^{-1}}$ is not often seen in the light curves of other stars, the possibility remains, that it is an intrinsic effect. Because no significant improvement of the fit to the light curve is seen when this frequency is included, we do not consider it as an independent oscillation. Note here, that this frequency becomes significant only after pre-whitening and is therefore not above the $\mathrm{SNR} = 4$ threshold in the residual amplitude spectrum shown in Fig. \ref{img:SBL461}. Possible additional frequencies can be found at $0.5 \; \mathrm{d^{-1}}$ (which is the second frequency of ASK01) and interestingly in the range $7 - 8 \; \mathrm{d^{-1}}$ with the most prominent peak at $7.64 \; \mathrm{d^{-1}}$ when looking at daily adjusted data or at about $7.24 \; \mathrm{d^{-1}}$ for the the monthly adjustment. Figure \ref{img:SBL461} shows this monthly adjusted set of SBL461 and also our calculated frequency spectra. Clearly visible is the low frequency variability. Note that the noise in the light curve is unusually high for this relatively bright star. This can be explained by its close position to another cluster star introducing additional errors. Furthermore, for some nights the noise level is higher than for others which is attributed to the defocused state of the camera, where the distinction of the PSFs becomes more difficult. Together with its position in the HR diagram, our results suggest that SBL461 is indeed an SPB type pulsator. Even though only one frequency can be found above our significance level, the different amplitudes in the three bands and the potential presence of additional variability point to the conclusion that the variability is due to pulsation. The binary nature, however, still needs to be clarified. If the amplitude in $U$ is artificially increased due to unreliable data and if this star is in fact a binary it is possible that the variability is not originating from pulsations. Also visible in Fig. \ref{img:SBL461} are the amplitude ratios on top of predicted values for models with masses between 6.5 and $7.5 \; M_{\odot}$ within the error box of $T_{\mathrm{eff}}$ and $\mathrm{log} \, g$. A visual inspection suggests an $l=1$ mode for $f_{1}$.

\textbf{SBL480} has not been subject to a variability study so far. It was reported to be a member of the cluster by BVF99 and RCB97, has spectral type B2IVn according to PHC91 
and does not seem to be a binary following RM98. The frequency analysis with our data set turned out to be very difficult since the amplitudes of potential oscillations are very small. To minimize the effects of different zero points during the entire observations an adjustment was performed on a monthly basis. The strongest signal is found at $f_{1} = 2.2981 \; \mathrm{d^{-1}}$ and after prewhitening the periodogram is dominated by signals peaking at about $0.5 \; \mathrm{d^{-1}}$. Due to the characteristics of the Fourier method this region is very crowded in frequency space and no signal is found to be above the detection threshold. An additional significant frequency could be found at $f_{2} = 4.5224 \; \mathrm{d^{-1}}$ in the $V$ data. We investigated the possibility of $f_{2}$ being a harmonic of $f_{1}$ when taking aliasing into account. Either pair ($f_{2}/2 \approx f_{1} - 0.03 \; \mathrm{d^{-1}}$, $f_{2}$), or ($f_{1}$, $2f_{1} \approx f_{2} + 2\cdot 0.03 \; \mathrm{d^{-1}}$) resulted in a worse fit than the chosen one. Still, this does not rule out the possibility that $f_{2}$ is a harmonic of $f_{1}$. The variation around $0.5 \; \mathrm{d^{-1}}$ is difficult to interpret since some other stars also show variability in this region with about the same amplitude suggesting that this is due to instrumental or atmospheric effects. The data quality in $U$ is above average, but even so no significant oscillation was visible in the frequency spectra. Figure \ref{img:SBL480} displays sample light curves where the variability is clearly visible in both the frequency spectrum for the adjusted data set and the light curves. No Str\"omgren photometry is available for this target and efforts to calibrate effective temperature and luminosity on the basis of $UBV$ photometry using the same calibrations as SBL98 from \cite{1981ARA&A..19..295B} were not successful since they did not yield realistic results. However, the time scales of the variability and the multi-periodicity point towards pulsation as the origin of the light curve modulation. Since $f_{2}$ might be a harmonic of $f_{1}$ and because the amplitude differences between the passbands are very small the possibility remains that the variability is not caused by pulsations. Therefore SBL480 is considered to be an SPB candidate. 

\begin{figure*}
	\resizebox{\hsize}{!}{\includegraphics{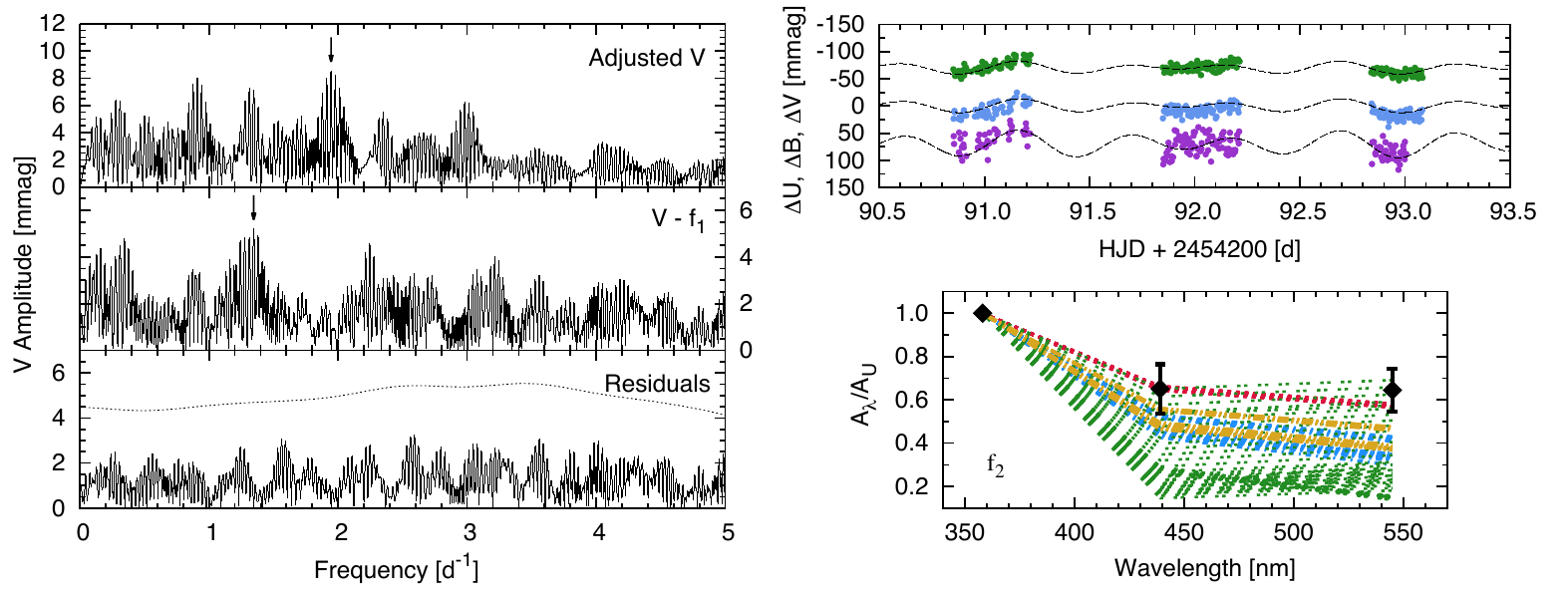}}
	\caption[SBL482]{Same as Fig. \ref{img:SBL226}, but for the newly discovered SPB star SBL482. The detected frequencies are $f_{1} = 1.9461 \; \mathrm{d^{-1}}$ and $f_{2}= 1.3466 \; \mathrm{d^{-1}}$ with the errors $\sigma_{f_{1}} = 0.0004 \; \mathrm{d^{-1}}$ and $\sigma_{f_{2}} = 0.0006 \; \mathrm{d^{-1}}$. The amplitude ratios were compared to models with masses ranging from 3.5 to $4 \; M_{\odot}$.}
	\label{img:SBL482}
\end{figure*}

\textbf{SBL482} was mentioned as a possible variable star by BVF99 for the first time. They arrived at this conclusion on the basis of the differences in their 
photometry to the values of PHC91. Since then, no variability studies included this star. Following BVF99, SBL482 is considered to be a cluster member even though RCB97 list it as a non-member. This decision was supported by the de-reddend $b-y$ color resulting from the Str\"omgren calibration which gave a similar color excess when compared to other cluster members. The first step in our frequency analysis was to adjust the light curves so that the average values in magnitude for each month matched. We detected two significant frequencies in the $V$ data which also show up in the other bands ($f_{1} = 1.9461 \; \mathrm{d^{-1}}$ and $f_{2}= 1.3466 \; 
\mathrm{d^{-1}}$). However, aliasing and some minor differences between $B$ and $V$ data do not allow an unambiguous determination of the oscillation frequencies. 
Nevertheless, the data are sufficient to detect clear amplitude differences in both $f_{1}$ and $f_{2}$ which we interpret as a clear sign of pulsation. Sample light curves and the steps in the frequency analysis are shown in Fig. \ref{img:SBL482}. The calculation of stellar parameters places this star in the SPB instability strip and together with the multiperiodicity, the time scales of the variability and the amplitude differences in the three bands, SBL482 can be declared an SPB star. We also attempted mode identification. To this end we considered models with masses ranging from 3.5 to $4 \; M_{\odot}$ as suggested by the star's position in the theoretical HR diagram. For $f_{1}$ no reliable statement can be given since too many modes were excited. The results for $f_{2}$ are somewhat better and are displayed in Fig. \ref{img:SBL482} among the frequency spectra and sample light curves. We found a large range of oscillations for $l=3$, but for spherical degrees $l \not= 3$ we found very similar behaviour across the mass range. Considering all models that pass through our error box we find compatibility with $l=1$ and $l=3$. If one considers only models with a mass of $3.5 \; M_{\odot}$ then a dipole mode is clearly favoured.

\begin{figure*}
	\resizebox{\hsize}{!}{\includegraphics{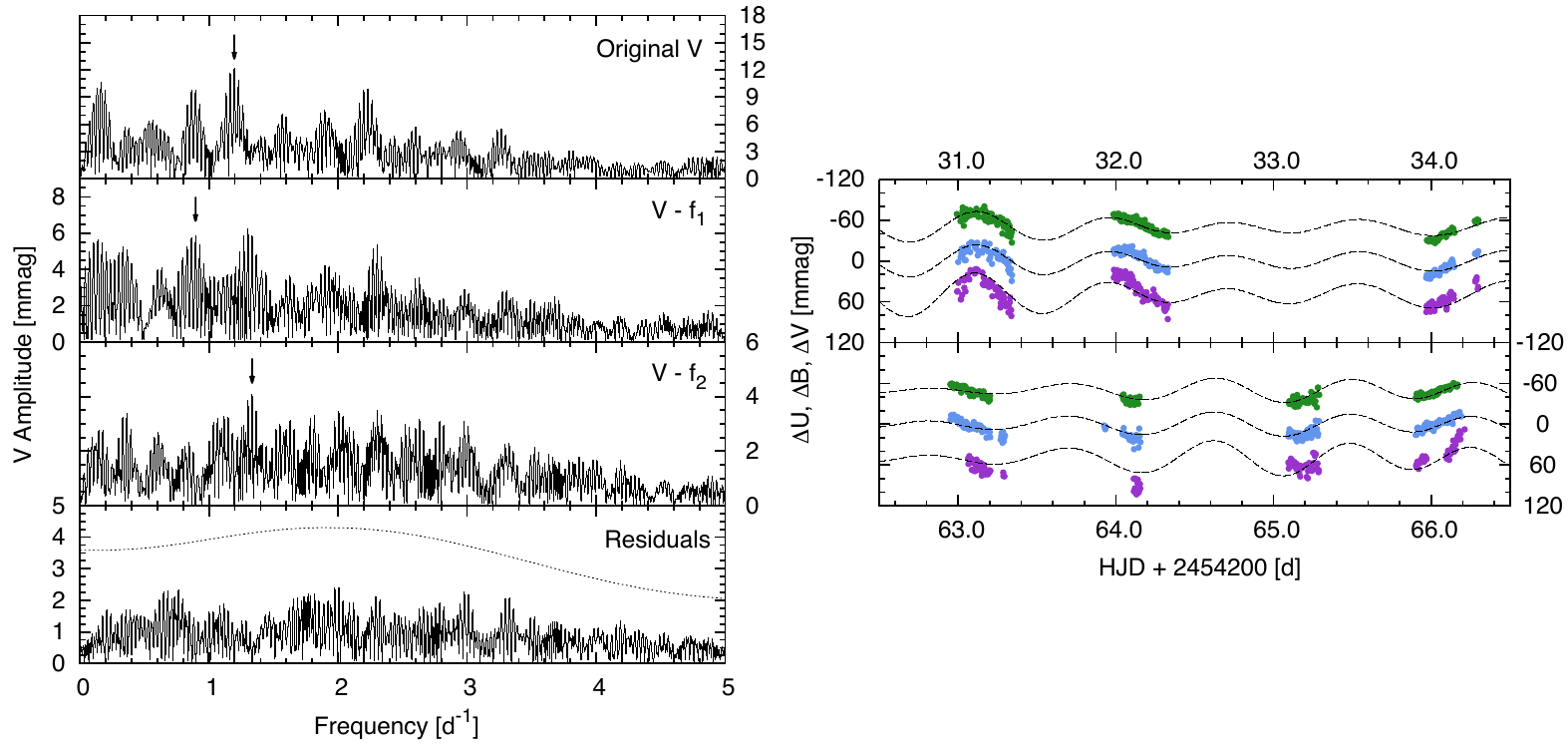}}
	\caption[SBL528]{Same as Fig. \ref{img:SBL226}, but for the newly detected SPB variable SBL528. We found the frequencies $f_{1} = 1.1959 \; \mathrm{d^{-1}}$, $f_{2}= 0.8962 \; \mathrm{d^{-1}}$, and $f_{3}= 1.3335 \; \mathrm{d^{-1}}$ with errors ranging from $\sigma_{f_{1}} = 0.0002 \; \mathrm{d^{-1}}$ to $\sigma_{f_{3}} = 0.0004 \; \mathrm{d^{-1}}$.} 
	\label{img:SBL528}
\end{figure*}

\textbf{SBL528} is a new variable star in NGC 6231. It is classified as a member of NGC 6231 by BVF99 and RCB97, has spectral type B6Vn according to PHC91 and is also a possible binary following RM98. Other sources, however, do not find indications for binarity. Since minor zero-point offsets were present our first step was to adjust the data on a monthly basis to suppress very low-frequency variability. In $U$ several measurements had to be deleted, but in general the quality of the light curve in this filter is above average. Our analysis resulted in three distinct oscillations ($f_{1} = 1.1959 \; \mathrm{d^{-1}}$, $f_{2}= 0.8962 \; \mathrm{d^{-1}}$, and $f_{3}= 1.3335 \; \mathrm{d^{-1}}$), where $V$ and $B$ data agree well in general but $f_{3}$ disappears into the noise in $B$. Only $f_{1}$ exceeded the SNR threshold in $U$. This can be attributed to the significant daily offsets visible in this light curve. Since we study variability of the order of $1 \; \mathrm{d^{-1}}$, we did not adjust the data to remove these offsets. The calculated multi-periodic fit matches well with the observed values, but a look at the residuals of the data suggests that the solution is not complete. Furthermore, since aliasing is a problem for this object the chosen frequencies  are those which minimize the residuals. The determination of effective temperature and luminosity is based on photometry from BL95 and puts SBL528 in the SPB instability domain. Over-plotted evolutionary tracks suggest a mass of about $5 \; M_{\odot}$. RM98 find two components each with a mass of about $5 \; M_{\odot}$. The multi-periodicity, the different amplitudes in $U$, $B$ and $V$, its position in the HR diagram, and the time scales of the variability clearly point towards the nature of an SPB star. Therefore it can safely be declared as such.  We were not able to set constraints on the mode of $f_{1}$ since in the given error box of $T_{\mathrm{eff}}$ and $\mathrm{log} \, g$ with masses ranging from 4.5 to $5.5 \; M_{\odot}$ far too many models showed pulsations. Restricting the mass range to include only $5 \; M_{\odot}$ models also does not solve this problem. Figure \ref{img:SBL528} therefore only shows the results of our frequency analysis and sample light curves for SBL528.

\subsection{Other variable stars}

\subsubsection{Possible cluster members}

\begin{figure*}
	\resizebox{\hsize}{!}{\includegraphics{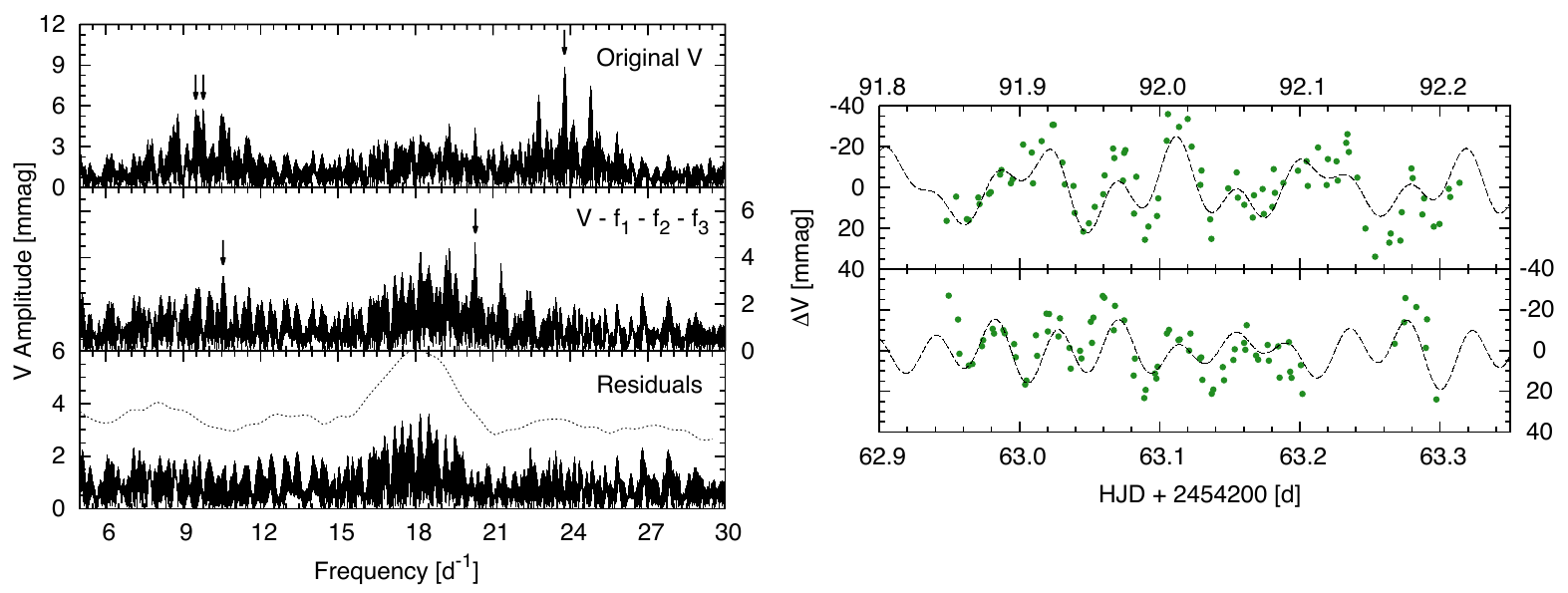}}
	\caption[SBL306]{Same as Fig. \ref{img:SBL226}, but for the $\delta$ Scuti star SBL306. The determined frequencies are: $f_{1} = 23.7713 \; \mathrm{d^{-1}}$, $f_{2}= 9.7797 \; \mathrm{d^{-1}}$, $f_{3}= 9.4853 \; \mathrm{d^{-1}}$, $f_{4} = 20.3195 \; \mathrm{d^{-1}}$, and  $f_{5} = 10.542 \; \mathrm{d^{-1}}$ with errors ranging from $\sigma_{f_{1}} = 0.0005 \; \mathrm{d^{-1}}$ to $\sigma_{f_{5}} = 0.001 \; \mathrm{d^{-1}}$.} 
	\label{img:SBL306}
\end{figure*}

\textbf{SBL306} was identified as a $\delta$ Scuti variable by ASK01. Since it is a rather faint object it was only possible to study this object with the $V$ data. No 
membership information is available, but Str\"omgren photometry suggests a slightly smaller color excess $(E_{b-y} \approx 0.26 \; \mathrm{mag})$ and distance modulus than the average cluster members. SBL306 is located in the southern parts of NGC6231 where SBL98 find larger differential reddening in general. Therefore it might be a foreground star. Broadband photometry puts it near SBL455, another $\delta$ Scuti variable which might be a member of NGC 6231. SBL306 is found inside the cool border the $\delta$ Scuti instability strip and is also located on the ZAMS in the theoretical HRD. ASK01 find three oscillations, two of which are readily visible in our data. Interestingly, the strongest signal they found at $15.6 \; \mathrm{d^{-1}}$ is barely seen in our analysis and cannot be distinguished from noise. Three additional frequencies are found for this star which were not detected before. The frequencies are $f_{1} = 23.7713 \; \mathrm{d^{-1}}$, $f_{2}= 9.7797 \; \mathrm{d^{-1}}$, $f_{3}= 9.4853 \; \mathrm{d^{-1}}$, $f_{4} = 20.3195 \; \mathrm{d^{-1}}$, and $f_{5} = 10.5421 \; \mathrm{d^{-1}}$. Due to strong aliasing our frequency solution must be seen as preliminary; the frequencies determined are the values that minimize the residuals. Furthermore, $f_{3}$ could be a combination of $f_{4}$ and $f_{5}$. A sample light curve in $V$, and the determined fit, as well as frequency spectra are displayed in Fig. \ref{img:SBL306}.

\begin{figure}
	\resizebox{\hsize}{!}{\includegraphics{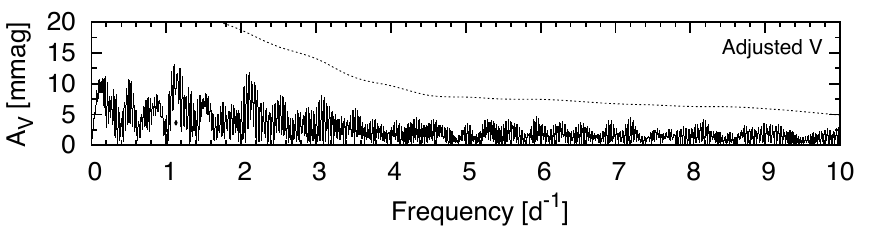}}
	\caption[SBL417]{Frequency spectrum for the potential PMS $\gamma$ Dor pulsator SBL417. No signal is significant, but the largest amplitude its found at the position where ASK01 detected an oscillation ($1.1 \; \mathrm{d^{-1}}$).} 
	\label{img:SBL417}
\end{figure}

\textbf{SBL417} has been identified as a PMS star by SBL98. ASK01 find this star to be variable with only one detected frequency at $1.1 \; \mathrm{d^{-1}}$ and even speculate that it might be a PMS $\gamma$ Dor pulsator. They also note that the variability might be irregular due to its PMS nature. With a visual magnitude of $14.62 \; \mathrm{mag}$ SBL417 depicts one of the faintest stars in our sample. A visual inspection of the light curve did not show any distinguishable regular variability, but a calculated frequency spectrum using monthly adjusted data showed an insignificant peak at the position where ASK01 found a significant signal. The frequency spectrum is shown in Fig. \ref{img:SBL417}. 

\begin{figure*}
	\resizebox{\hsize}{!}{\includegraphics{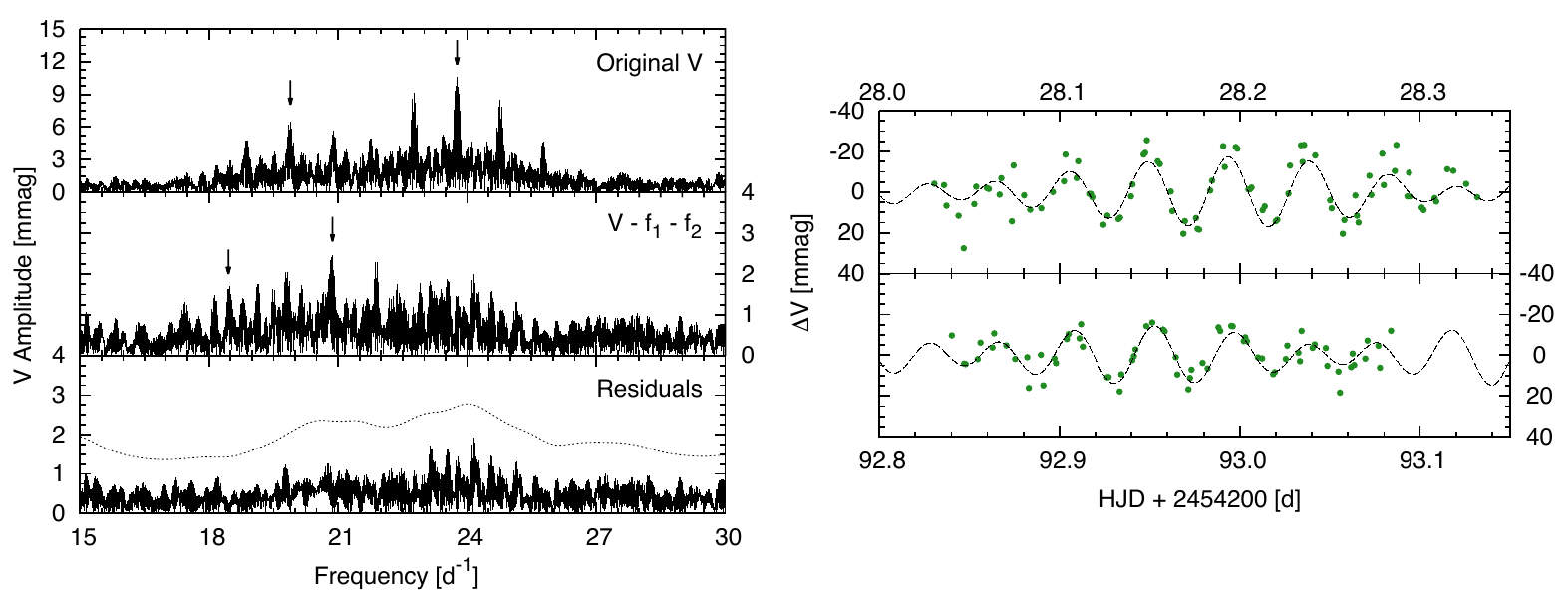}}
	\caption[SBL455]{Same as Fig. \ref{img:SBL226}, but for the $\delta$ Scuti variable SBL455. The significant frequencies are: $f_{1} = 23.7651 \; \mathrm{d^{-1}}$, $f_{2}= 19.8906 \; \mathrm{d^{-1}}$, $f_{3} = 20.869 \; \mathrm{d^{-1}}$, and $f_{4} = 18.449 \; \mathrm{d^{-1}}$. The errors range from $\sigma_{f_{1}} = 0.0003 \; \mathrm{d^{-1}}$ to $\sigma_{f_{4}} = 0.002 \; \mathrm{d^{-1}}$.} 
	\label{img:SBL455}
\end{figure*}

\begin{figure*}
	\resizebox{\hsize}{!}{\includegraphics{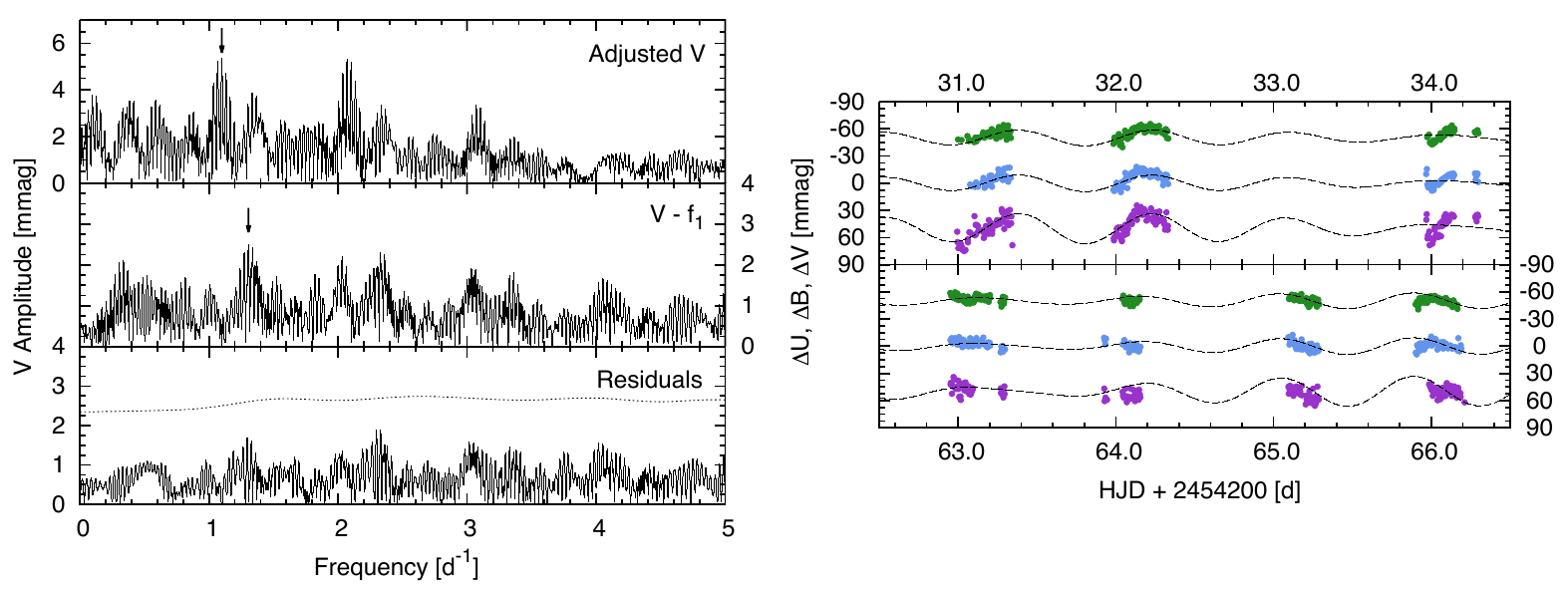}}
	\caption[SBL464]{Same as Fig. \ref{img:SBL226}, but for the newly discovered SPB variable SBL464. The determined frequencies are: $f_{1} = 1.1001 \pm 0.0003 \; \mathrm{d^{-1}}$, $f_{2}= 1.3059 \pm 0.0006 \; \mathrm{d^{-1}}$.} 
	\label{img:SBL464}
\end{figure*}

\textbf{SBL455} has been suggested as a $\delta$ Scuti variable by BL95 who performed a variability analysis and found two oscillations. ASK01 confirmed the variability time scales and also its nature as a $\delta$ Scuti star. They also mention that this star is positioned off the main sequence and therefore - if it is a cluster member - is a PMS star. However, no membership information is available from the main sources used in this work and SBL98 do not classify it as a PMS star with their criteria. Our variability analysis resulted in four significant frequencies and is displayed in Fig. \ref{img:SBL455}: $f_{1} = 23.7651 \; \mathrm{d^{-1}}$, $f_{2}= 19.8906 \; \mathrm{d^{-1}}$, $f_{3}= 20.869 \; \mathrm{d^{-1}}$, and $f_{4} = 18.4495 \; \mathrm{d^{-1}}$. All of them except $f_{4}$ match with the frequencies determined by either BL95 or ASK01, when aliasing is taken into account. The second oscillation detected by BL95 at $25.13 \; \mathrm{d^{-1}}$ is also visible in the residual frequency spectrum but does not exceed the SNR threshold. All frequencies and amplitudes were determined using an unadjusted data set since zero point offsets were negligible and did not affect the rapid $\delta$ Scuti variability. Str\"omgren photometry of BL95 can be used to determine the color excess, effective temperature and luminosity. The color excess is very similar to that of cluster members $(E_{b-y} \approx 0.3 \; \mathrm{mag})$ which supports the idea of it indeed being located in NGC 6231. The stellar parameters also position it in the $\delta$ Scuti instability strip away from the ZAMS, but within the errors SBL455 could also be very close to the ZAMS. In the color-color diagram SBL455 can be found on the ZAMS and also near another potential cluster member $\delta$ Scuti star (SBL306). GS79 classify it as spectral type A5III which in turn would mean that it already has evolved off the ZAMS and is therefore unlikely a member in such a young stellar aggregate as NGC 6231. However, since PMS stars also exhibit larger luminosities compared to their main-sequence counterparts it is possible that the spectral type given by GS79 is a misinterpretation. The $\delta$ Scuti nature of SBL455 could clearly be confirmed in our analysis. Whether this star is indeed a pre main sequence object and/or a cluster member still has to be verified, but according to our analysis it is likely a member.

The membership status of \textbf{SBL464} with respect to NGC 6231 is unclear since RCB97 list it as a member, whereas BVF99 reported it to be a non-member of the stellar aggregate. RCB97 also estimated its spectral type to be B8-9V and RM98 find evidence for possible binarity. Since no Str\"omgren photometry is available, a discussion of its possible membership must rely on broadband and Geneva photometry. In both the color-magnitude and the color-color diagram as shown in Fig. \ref{img:HRD} SBL464 is located among cluster member SPB variables which can be taken as an argument in favor of its membership. Our differential photometry clearly indicates variability of the order of $1 \; \mathrm{d^{-1}}$. Since the amplitudes of the variations are very small, an adjustment on a monthly basis took place to enhance possible SPB type variability. We found two significant frequencies for this star, $f_{1} = 1.1001 \; \mathrm{d^{-1}}$ and $f_{2}= 1.3059 \; \mathrm{d^{-1}}$. $f_{2}$ is only barely significant in the $V$ data. Also an unambiguous determination of the frequencies was not possible due to strong aliasing, the small amplitudes and the complex structure of the periodogram in the region around $1 \; \mathrm{d^{-1}}$. The values we chose here are those which minimize the residuals after fitting the two oscillations. Variability on shorter time scales is also indicated in the amplitude spectra in the range $12 - 13 \; \mathrm{d^{-1}}$ with some peaks coming close to the SNR threshold. The amplitude of such a variation would clearly be below $1 \; \mathrm{mmag}$. Figure \ref{img:SBL464} shows the frequency spectra for the adjusted data set and for the residuals after prewhitening the two detected frequencies. Also the light curve with the multi-periodic fit is displayed. Since our determined amplitude values in $U$,$B$, and $V$ are not significantly different from each other since the data in $U$ is unreliable our classification as an SPB type variable is based upon multi-periodicity and the color information from SBL98. Furthermore, we consider SBL464 to be likely a member of NGC 6231.

\begin{figure}
	\resizebox{\hsize}{!}{\includegraphics{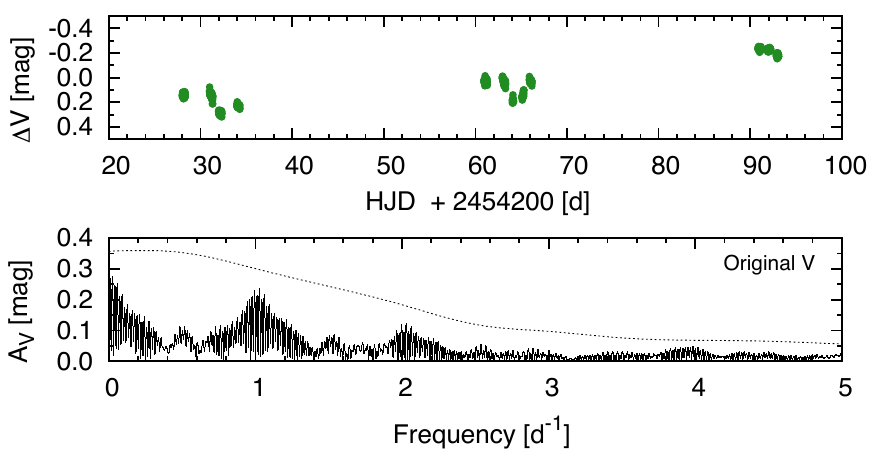}}
	\caption[SBL705]{Same as Fig. \ref{img:SBL226} where applicable, but for the newly discovered variable star SBL705. Note here that in both panels the y-axis scale has changed from mmag to mag.} 
	\label{img:SBL705}
\end{figure}

\textbf{SBL705} has only been studied by SBL98 so far. Therefore we could not calculate stellar parameters or rely on spectral type information. The photometric measurements of SBL98 place this star near the other possible member $\delta$ Scuti stars at the ZAMS and it does not show unusual colors with respect to other cluster stars. The light curve of SBL705 clearly shows long term variability with peak-to-peak amplitudes of at least $0.6 \; \mathrm{mag}$. However, we could not determine the period of this variability due to the large gaps between the observing runs. Referring to the complete data set displayed in Fig. \ref{img:SBL705}, it is also possible that the long term variability has a period exceeding our data set which covers about 2 months. Superimposed on the long term variation is variability in the range between $0.5$ and $2 \; \mathrm{d^{-1}}$.

\subsubsection{Non-members}

\begin{figure*}
	\resizebox{\hsize}{!}{\includegraphics{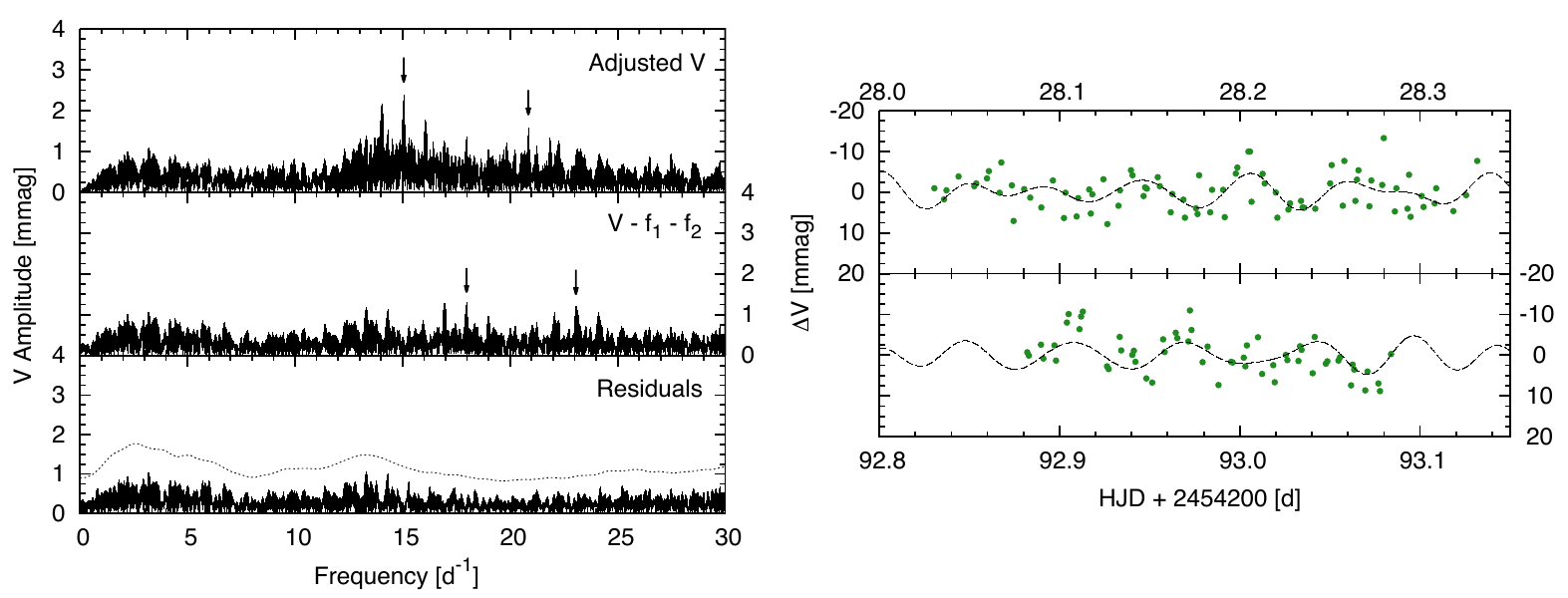}}
	\caption[SBL444]{Same as Fig. \ref{img:SBL226} where applicable, but for the newly detected non-member $\delta$ Scuti variable SBL444. We determined four oscillation frequencies at $f_{1} = 15.0539 \; \mathrm{d^{-1}}$, $f_{2}= 20.8483 \; \mathrm{d^{-1}}$, $f_{3}= 17.9737 \; \mathrm{d^{-1}}$, and $f_{4} = 23.056 \; \mathrm{d^{-1}}$, with errors ranging from $\sigma_{f_{1}} = 0.0007 \; \mathrm{d^{-1}}$ to $\sigma_{f_{4}} = 0.001 \; \mathrm{d^{-1}}$.} 
	\label{img:SBL444}
\end{figure*}

\begin{figure*}
	\resizebox{\hsize}{!}{\includegraphics{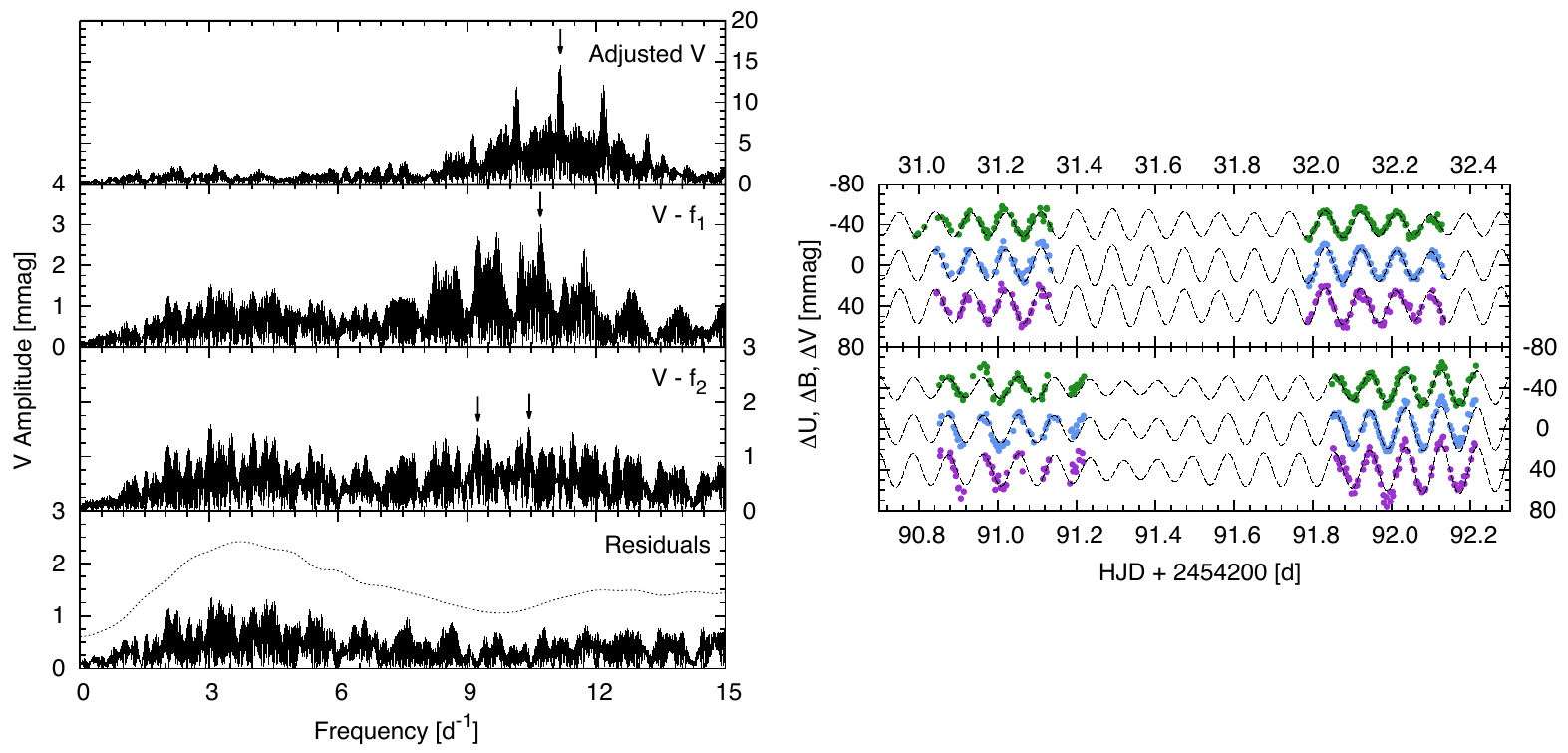}}
	\caption[SBL612]{Same as Fig. \ref{img:SBL226} where applicable, but for the non-member $\delta$ Scuti variable SBL612. $f_{1} = 11.1654 \; \mathrm{d^{-1}}$, $f_{2}= 10.7018 \; \mathrm{d^{-1}}$, $f_{3}= 10.4427 \; \mathrm{d^{-1}}$, and $f_{4} = 9.2525 \; \mathrm{d^{-1}}$. Errors range from $\sigma_{f_{1}} = 0.0001 \; \mathrm{d^{-1}}$ to $\sigma_{f_{4}} = 0.0008 \; \mathrm{d^{-1}}$.} 
	\label{img:SBL612}
\end{figure*}

\begin{figure*}
	\resizebox{\hsize}{!}{\includegraphics{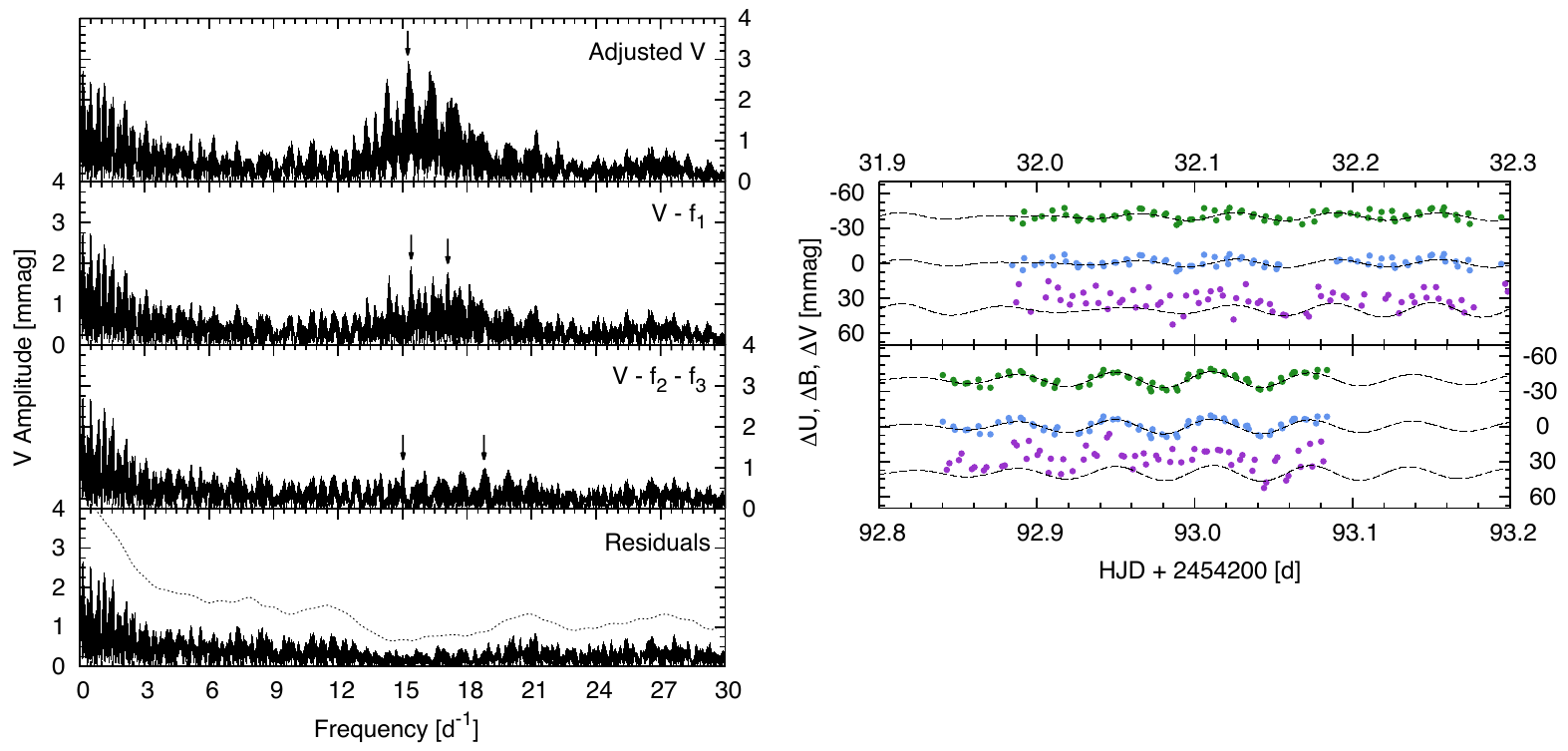}}
	\caption[SBL752]{Same as Fig. \ref{img:SBL226}, but for the newly discovered $\delta$ Scuti star SBL752. The determined frequencies are: $f_{1} = 15.2413 \; \mathrm{d^{-1}}$, $f_{2}= 15.3989 \; \mathrm{d^{-1}}$, $f_{3}= 17.1006 \; \mathrm{d^{-1}}$, $f_{4} = 15.0161 \; \mathrm{d^{-1}}$, and $f_{5}= 18.7892 \; \mathrm{d^{-1}}$ with errors ranging from $\sigma_{f_{1}} = 0.0006 \; \mathrm{d^{-1}}$ to $\sigma_{f_{5}} = 0.005 \; \mathrm{d^{-1}}$.} 
	\label{img:SBL752}
\end{figure*}

\textbf{SBL444} has not been determined to be variable so far. No information can be found on spectral type and BVF99 do not find it to be a member of NGC 6231. In addition, SBL444 is located near other non-member $\delta$ Scuti stars in the color-magnitude and color-color diagrams. Furthermore, no Str\"omgren photometry is available. Since this star is not a member of NGC 6231 and reaches only a reasonable SNR in $V$, our analysis was limited to frequency determination only. Our frequency analysis - which is based on a daily adjusted data set - resulted in four significant oscillations, all in the range of typical $\delta$ Scuti pulsation: $f_{1} = 15.0539 \; \mathrm{d^{-1}}$, $f_{2}= 20.8483 \; \mathrm{d^{-1}}$, $f_{3}= 17.9737 \; \mathrm{d^{-1}}$, and $f_{4} = 23.056 \; \mathrm{d^{-1}}$. Additional oscillations are likely present at the mmag level, for instance around $13 \; \mathrm{d^{-1}}$ but are not detected with our adopted SNR threshold criterion. Figure \ref{img:SBL444} shows selected data including the light curves and the calculated multiperiodic fit for two nights and the frequency spectra of our analysis. We classify SBL444 as a newly detected $\delta$ Scuti variable on the basis of a comparison with photometric data of other pulsators and the time scales of the variability.

\textbf{SBL612} has been classified as a $\delta$ Scuti variable by BL95 who also suspected this star to be a non-member of NGC6231 and list it as a foreground star since the observed colour excess is much lower when compared to cluster members. BVF99 confirm the non-membership photometrically as well. Our calibration of effective temperature and luminosity with measurements from PHC91 also yields a significantly lower color excess ($E_{b-y} = 0.19 \; \mathrm{mag}$) and furthermore positions this star in the $\delta$ Scuti instability strip. BL95 found only one dominant oscillation with a period of about $2.15 \; \mathrm{h}$. The analysis with our new data was done with daily adjusted data and resulted in four significant frequencies ($f_{1} = 11.1654 \; \mathrm{d^{-1}}$, $f_{2}= 10.7018 \; \mathrm{d^{-1}}$, $f_{3}= 10.4427 \; \mathrm{d^{-1}}$, and $f_{4} = 9.2525 \; \mathrm{d^{-1}}$), where the strongest signal, $f_{1}$, is a perfect match with the value of BL95. The shape of the residual frequency spectrum, which is shown in Fig. \ref{img:SBL612} besides sample light curves, suggests that there are very likely other oscillations present with amplitudes in the mmag range. None of these peaks, however, are significant. 

\textbf{SBL752} has not been identified as a variable star so far and is listed as a member of NGC 6231 by RCB97 who also find spectral type Amp based on a photometric estimation. On the other hand, \citetads{SG05} assigned spectral type B to this star. Our frequency analysis resulted in five detected oscillations ($f_{1} = 15.2413 \; \mathrm{d^{-1}}$, $f_{2}= 15.3989 \; \mathrm{d^{-1}}$, $f_{3}= 17.1006 \; \mathrm{d^{-1}}$, $f_{4} = 15.0161 \; \mathrm{d^{-1}}$, and $f_{5}= 18.7892 \; \mathrm{d^{-1}}$), but since the amplitudes are altogether very small, the light curve is dominated by zero-point offsets. The detected oscillations, however, are little affected by this trend. Furthermore, aliasing posed a problem during the analysis so that the frequencies were selected to minimize the residuals. The $U$ and $B$ data showed larger average single data point errors compared to $V$ making it harder to determine accurate amplitudes. Figure \ref{img:SBL752} shows frequency spectra and two nights of data with the calculated fit. For other nights the fit does not match the data similarly well since offsets are present which were not accounted for. The residual frequency spectrum, after prewhitening all five frequencies, still showed some excess power around $20 \; \mathrm{d^{-1}}$ but no peak was significant. No Str\"omgren photometry is available for this star, but looking at the broadband colors in Fig. \ref{img:HRD} SBL752 is not found in a common region with other variable stars: it shares the same apparent magnitude with some cluster SPB stars, but in the color-color diagram it is closer to other $\delta$ Scuti variables suggesting that it was misclassified as a member. However, looking at the reddening vector, SBL752 might be a foreground star but a comparison to SBL612 does not clarify this situation. SBL612, being a confirmed foreground $\delta$ Scuti pulsator is more than two magnitudes brighter in $V$, suggesting that it is closer. On the other hand, the $U-B$ color excess suggests that SBL752 lies closer to us compared to SBL612 if it is also a $\delta$ Scuti type variable. The available Geneva photometry for SBL752 can also be interpreted in both ways. RM98 assumed it to be a B-type cluster member and consequently determined a mass of $4.08 \; M_{\odot}$. On the other hand, assuming a reddening value consistent with that of a foreground $\delta$ Scuti star, the Geneva colours are also compatible with such an interpretation. In conclusion, given the detected stellar pulsation periods, we tend to believe that SBL752 is not a cluster member, but a foreground $\delta$ Scuti star.

\subsubsection{Eclipsing binaries}

\begin{figure}
	\resizebox{\hsize}{!}{\includegraphics{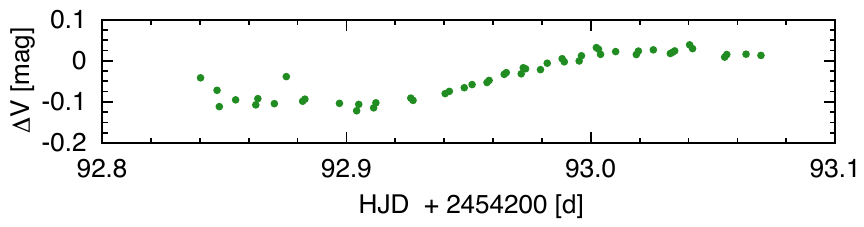}}
	\caption[SBL521]{The observed part of the minimum of SBL521 in the $V$ passband.} 
	\label{img:SBL521}
\end{figure}

\begin{figure}
	\resizebox{\hsize}{!}{\includegraphics{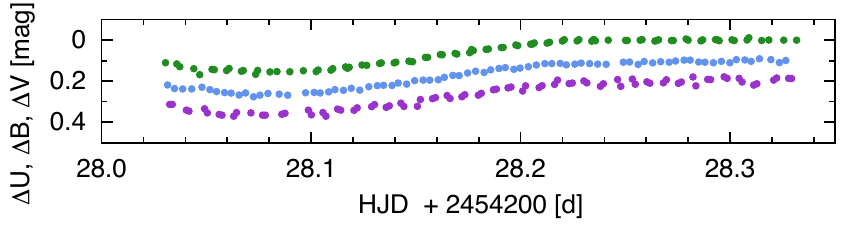}}
	\caption[SBL574]{The single observed minimum of the binary SBL574. The light curves were shifted to match in apparent brightness outside the minimum. $B$ was then displaced by additional $0.1 \; \mathrm{mag}$, $U$ by $0.2 \; \mathrm{mag}$.} 
	\label{img:SBL574}
\end{figure}

\textbf{SBL521} is a known eclipsing binary (BE85) for which we could observe a part of a minimum with a depth of about $m_V = 0.15 \; \mathrm{mag}$ (see Fig. \ref{img:SBL521}).

\textbf{SBL574} has already been classified as a binary by \cite{2004RMxAC..20...79S}. BVF99 do not find this star to be a member of NGC 6231. We can confirm its nature as an eclipsing binary, but we only observed one minimum - visible in Fig. \ref{img:SBL574} - which does not allow us to set further constraints on its orbital parameters.

\begin{figure}
	\resizebox{\hsize}{!}{\includegraphics{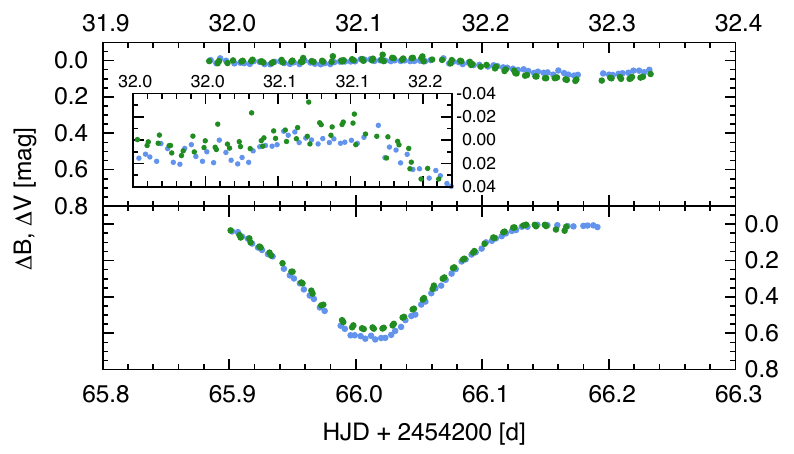}}
	\caption[3UC097]{The newly discovered eclipsing binary 097-240594. Visible are the secondary minimum in the top panel, the primary minimum in the bottom panel and the inlay shows a zoom to visualize the rising light curve towards the secondary eclipse.} 
	\label{img:3UC097}
\end{figure}

\textbf{097-240594} has not been included in any of the above mentioned previous observations of NGC 6231 since it is situated far away from the core of the cluster with a distance of about $13'$. The designation we use here is taken from the UCAC3 catalog. The light curves in $B$ and $V$ of this star are shown in Fig. \ref{img:3UC097} and clearly reveal its nature as an eclipsing binary. Two primary eclipses and one secondary eclipse have been observed which unfortunately are each about a month apart so that the orbital period cannot be determined unambiguously. The shape of the light curves point towards an Algol type binary with a flat light maximum. A possible hot spot feature is visible just before the secondary minimum where the light curves show a slight but still well visible rise towards the eclipse. Also clearly visible is the different color (i.e.: surface temperature) of both components since the primary minimum is deeper in $B$ compared to $V$ and vice versa in the secondary minimum.

\end{document}